\documentclass[12pt]{article}
\usepackage[utf8]{inputenc}
\usepackage{cmap}
\usepackage[T1]{fontenc}
\usepackage[english]{babel}
\usepackage{indentfirst}
\usepackage[top=1.4cm,bottom=1.9cm,right=1.4cm,left=1.4cm]{geometry}
\usepackage{setspace}

\usepackage{etoolbox}  
\usepackage{fancyhdr}
\usepackage{footnote}
\usepackage{footmisc}
\usepackage{url}
\usepackage{hyperref}

\usepackage{amsmath}
\usepackage{amsfonts}
\usepackage{amssymb}
\usepackage{amsthm}
\usepackage{textcomp}
\usepackage{amssymb}

\usepackage{multicol}
\usepackage{multirow}
\usepackage{array}
\usepackage{booktabs}

\usepackage{listings} 
\usepackage{yfonts}
\usepackage[usenames]{color} 
\usepackage{enumitem}
\usepackage{fancyhdr}
\usepackage[square,numbers,sort&compress]{natbib}

\usepackage[usenames,dvipsnames,svgnames,table]{xcolor}

\usepackage{pdfpages}
\usepackage{epsfig}
\usepackage{graphicx}
\usepackage{tikz}
\usetikzlibrary{fit}
\usepackage{wrapfig}
\usepackage{subfigure}
\usepackage{epstopdf}

\usepackage{algorithmic}

\usepackage{makeidx}
\usepackage[title]{appendix} 
\makeindex

\usepackage[font={small}]{caption}
\usepackage{lmodern} \normalfont
\DeclareFontShape{T1}{lmr}{bx}{sc} { <-> ssub * cmr/bx/sc }{}
\usepackage{mathrsfs}

\usepackage[all]{xy}

\newcommand{\autor}{Deison Préve$^{1,}${\footnote[1]{Corresponding authors: Tel: +39 0103352556 – Fax: +39 0103352517\\ 
			E-mail addresses: deison.preve@imtlucca.it; andrea.bacigalupo@unige.it}}, Andrea Bacigalupo$^{2,}$\footnotemark[1], Marco Paggi$^{1}$}
\newcommand{\authors}{Deison Préve, Andrea Bacigalupo, Marco Paggi}
\newcommand\titulo{Variational-asymptotic homogenization of thermoelastic periodic materials with thermal relaxation}
\newcommand\institutions{$^{1}$ IMT School for Advanced Studies Lucca, Piazza San Francesco 19, 55100 Lucca, Italy\\$^{2}$ DICCA, Università degli Studi di Genova, Via Montallegro 1, 16145 Genova, Italy}

\appto\mainmatter{\pagestyle{plain}}

{\normalfont}{\normalfont}

\newcommand{\annexname}{Anexo}
\makeatletter
\newcommand\annex{\par
	\setcounter{chapter}{0}%
	\setcounter{section}{0}%
	\gdef\@chapapp{\annexname}%
	\gdef\thechapter{\@Roman\c@chapter}}
\makeatother

\newcommand{\R}{\mathbb{R}} 
\newcommand{\C}{\mathbb{C}} 
\newcommand{\N}{\mathbb{N}} 
\newcommand{\Z}{\mathbb{Z}} 

\begin{document}
\thispagestyle{plain}

\begin{center}
	{\LARGE{\titulo}}
\end{center}

\begin{center}
	\large{\autor}\\
	{\small{\institutions}}
\end{center}


\vspace{1cm}
%
%
%
%
%
%
\begin{center}
	\large{\textbf{Abstract}}
\end{center}

A multiscale asymptotic homogenization method for periodic microstructured materials in presence of thermoelasticity with periodic spatially dependent one relaxation time is introduced. The asymptotic expansions of the micro-displacement and the micro-temperature fields are rewritten on the transformed Laplace space and expressed as power series of the microstructural length scale, leading to a set of recursive differential problems over the periodic unit cell. The solution of such cell problems leads to the perturbation functions. Up-scaling and down-scaling relations are then defined, and the latter allow expressing the microscopic fields in terms of the macroscopic ones and their gradients. Average field equations of infinite order are also derived. The efficiency of the proposed technique was tested in relation to a bi dimensional orthotropic layered body with orthotropy axis parallel to the direction of the layers, where the mechanical and temperature constitutive properties were well stabilised. The dispersion curves of the homogenized medium, truncated at the first order are compared with the dispersion curves of the heterogeneous continuum obtained by the Floquet-Bloch theory. The results obtained with the two different approaches show a very good agreement.

\vspace{.5cm}

%
%
%
%
\section*{Introduction}
\addcontentsline{toc}{chapter}{Introduction}

The thermoelasticity phenomenon bonds together a set of theories as \cite{biot1956thermoelasticity} quoted, namely, the general theory of heat conduction, thermal stresses, and strains set up by thermal flow in elastic bodies, and the reverse effect of temperature distribution caused by the elastic deformation itself leading to thermoelastic energy dissipation. The main theory of thermoelasticity based on Fourier law of heat conduction, leaves thermal perturbations propagate at infinite velocity in a diffusive manner, once the coupled displacement-temperature governing equation is a parabolic-type partial differential equation \citep{biot1956thermoelasticity,kupradze1980three,carcione2019simulation}. In practical terms, this means that when a temperature gradient is suddenly produced in some region of the sample, this entails in an instantaneous disturbance at each point of the material \citep{joseph1989heat,liu1999new}. However, there are some experimental observations, as \cite{ignaczak2010thermoelasticity} pinpointed, where the temperature acts like a wave propagating through the body with finite speed, commonly referred to as second sound \citep{chandrasekharaiah1998hyperbolic,zamani2011second}. The pattern of the heat wave propagation has been observed in superfluids, inhomogeneous materials like sand and processed meat, \cite{kaminski1990hyperbolic,mitra1995experimental}, and in pure crystals, \cite{narayanamurti1972observation}. Moreover, this factor contradicts the physical affirmation that for a finite time interval a disturbance of bounded support may only generate the response of a bounded support \citep{hetnarski1999generalized,hetnarski2009thermal,ignaczak2010thermoelasticity}. 

In order to overcome such a paradox, recent developments were made as \cite{sherief2010fractional} mentions, and a couple of generalized thermoelastic theories have flourished. \cite{lord1967generalized} introduced the theory of generalized thermoelasticity with one relaxation time, originally proposed by \cite{maxwell1867iv} in the context of theory of gases, and later by \cite{cattaneo1948sulla} in the context of heat conduction in rigid bodies \citep{ignaczak2010thermoelasticity}. In this theory, a modified law of heat conduction including both the heat flux and its time derivative replaces the conventional Fourier law \citep{dhaliwal1980generalized}. The heat equation associated to this theory becomes hyperbolic and hence eliminates the paradox of the propagation of thermal signals with an infinite speed. The equations of motion and constitutive relations, remain the same for both theories of  thermoelasticity, the uncoupled theory and the coupled one \citep{lord1967generalized}. 

Modelling multiphase materials containing a periodic microstructure, where additionally elasticity and heat transfer are combined phenomena, is a topic of vast importance in modern applications, such as aerospace, aircraft, biomedical and electronic \citep{prunty1978dimensionally,dodson1989thermal,farmer1992thermal,kim2000dimensional,zhu2003effects}. Nonetheless, a plethora of problems are assumed on heterogeneous continuum with a small accuracy over the microstructural length scale. Deriving the solution of the governing thermoelastic partial differential equation with one relaxation time may be cumbersome both analytically and numerically, due the periodicity of the media \citep{nayfeh1971thermoelastic,hawwa1995general}. The multiscale asymptotic homogenization approaches, well stabilised in earlier works by \cite{SanchezPalencia1974,papanicolau1978asymptotic,Bakhvalov1984,SmyshlyaevCherednichenko2000}, show themselves as effective tools for determining the responses of the microscopic phases on the overall properties of the composites, by replacing an heterogeneous continuum by an equivalent homogenized model whose solutions are good approximations of the real ones, but are characterized by constitutive tensors not affected by the fast variable which gives rapid oscillations due to the underlying microstructure. Therefore, such procedure might be quite effective computationally \citep{bacigalupo2014computational,nassar2016asymptotic}. 

At this stage, it must be highlighted that various homogenization methods have being applied to study the overall properties of multiphase periodic materials, which may be classified in asymptotic technique \citep{gambin1989higher,AndrianovBolshakov2008,bacigalupo2010second,yang2016multiscale,fish2019second}, variational asymptotic schemes \citep{willis1981variational,Smyshlyaev2009,bacigalupo2014second,nassar2016asymptotic,kamotski2019bandgaps}, analytical paths \citep{bigoni2007analytical,milton2007modifications,bacca2013amindlin,bacca2013mindlin2,bacigalupo2018identification}, and computational approaches \citep{ostoja1999couple,kouznetsova2002multi,forest2002homogenization,lew2004homogenisation,scarpa2009effective,bacigalupo2010second,de2011cosserat,forest2011generalized,chen2013elasticity,monchiet2020strain}. In what concerns to periodic materials in presence of thermoelastic effects, as seen in \cite{temizer2011homogenization}, the homogenization method has been investigate therein, and developed in the context of thermoelastic periodic material with pertinent physical applications highlighted in \cite{l1998characterization}. Furthermore, a variational-asymptotic technique for thermoelastic periodic materials was brought by \cite{yu2007variational}. 

As a matter of fact, cutting edge research regarding asymptotic homogenization techniques, variational-asymptotic approaches and computational methods are being made along several multiphase materials in presence of certain phenomena. For instance, asymptotic homogenization technique over piezoelectric composite materials may be found in \cite{pettermann2000comprehensive,berger2005comprehensive,schroder2012two,zah2013computational} and on thermal-piezoelectric materials with a periodic microstructure via asymptotic schemes can be checked in \cite{fantoni2017multi}. Also, multiscale homogenization schemes have been applied to characterize the behaviour and the global constitutive properties of viscoelastic heterogeneous materials with periodic microstructure \cite{ohno2000homogenized,abdessamad2009memory,chen2017finite,del2019characterization}. Particularly, in the latter case, the variational-asymptotic homogenization method was proposed in order to characterize the propagation of dispersive waves in viscoelastic materials with periodic microstructure. Likewise, the homogenization approach for describing the elastic, thermal and diffusive properties of periodic materials (also on periodic layered materials) in presence of thermal-diffusion has been explored by \cite{salvadori2014computational,bacigalupo2016multiscale,fantoni2020wave}. Effective analysis on heterogeneous continuum in presence of thermal-mechanical and thermal-magneto-electro-elastic deformations have been studied by \cite{aboudi2001linear,kanoute2009multiscale,zhang2007thermo} and \cite{sixto2013asymptotic}, respectively.

Inspired by those studies where the heterogeneous continuum is rather bartered by some homogeneous model led by an homogenization procedure, the herein proposed original theoretical framework combines the generalized theory of thermoelasticity with periodic spatially dependent one time relaxation with the homogenization technique in periodic composite materials. Having said that, the goal of this work is to present a novel method to characterize the overall properties, namely elastic and thermal, of multiphase periodic materials governed by Lord-Shulman generalized thermoelasticity equation via asymptotic homogenization. In the following, the Lord-Shulman generalized thermoelastic governing equations at the micro-scale, which describe the non-homogeneous medium, are determined in the time domain and subsequently transformed by the bi-lateral Laplace transform once the relaxation time varies within the material phase, in order to separate it from the partial derivative in time contained in the differential operator induced by Maxwell-Cattaneo law. Such a procedure hence takes the input to the Laplace domain (complex frequency domain). With the aim of separating the fast from the slow variable, the micro-displacement and micro-temperature fields are rewritten as power series expansions of the microstructural length scale, such that a cascade of recursive non-homogeneous differential problems is defined over the periodic unit cell. From this point, the solvability conditions are imposed to these recursive differential problems, arising from the down-scaling relations which are written in terms of the perturbation functions. From the substitution of the down-scaling relations into the microscopic field equations, the governing equations at infinite order are given in closed form in terms of the microscopic constitutive properties and the perturbation functions. Finally, by truncating the governing equations, the first order homogenized field equations are derived from which the overall constitutive tensors are derived. The capability of the method is then assessed through the study of thermal wave propagation over a two-dimensional bi-layered media with a periodic microstructure with orthotropic phases and axis of orthotropy parallel to the direction of layering, where the overall mechanical and thermal properties are analytically determined. The outcome dispersion functions from the homogenized model are therefore set to be compared with those of the heterogeneous thermoelastic material obtained from the solutions of the Floquet–Bloch theory \citep{Floquet1883,Bloch1928,brillouin1953wave}.

The work is sorted out as follows: Sec. \ref{section2} recalls the fundamentals of thermoelasticity with one relaxation time in a periodic media, describes the microscopic governing equations in the time domain and in the Laplace space. Sec. \ref{sec:expansions} is dedicated in developing the cascade of differential problems as well as their solutions. In Sec. \ref{Sec::CellsProblems} is derived the cell problems and their respective perturbation functions. The down-scaling and up-scaling relations are defined in Sec. \ref{Sec::DownScalingUpScaling}. The variational-asymptotic to establish an equivalence between the equations at macro-scale and micro-scale procedure and the zeroth order truncation is attacked in Sec. \ref{variationalmacro}. Sec. \ref{waveprophomogenized} reports wave propagation and the related dispersion functions. Throughout Sec. \ref{exemplos}, the herein method is applied for studying the overall properties of two-dimensional bi-layered  orthotropic composites. Specifically, the overall elastic, thermal dilatation and thermal conduction tensors are determined in theirs analytic form in Sec. \ref{homogenizedprocess}, whereas benchmark and some analysis are pursued in Sec. \ref{benchmark}. Further mathematical details about (i) asymptotic expansions of the field equations ( Sec. \ref{sec::additionaldetailshomogenization}), (ii) recursive differential equations (Sec. \ref{sec:recursivesection}), (iii) the average field equation of infinite order (\ref{sec:infinitorder}), (iv) power-like functional (Secs. \ref{powerlikefunctionaltruncation}, \ref{Euler-Lagrangepower-likefunctional}), (v) frequency spectrum via Floquet-Bloch theory within the heterogeneous continuum (Secs. \ref{Wavepropagationheterogeneous}, \ref{heterogeneousapproach}, \ref{dispersivewaveheterogeneous}), are vastly exploited in the Supplementary Material.

\section{Derivation of thermoelasticity with one relaxation time on the Laplace domain}
\label{section2}

Throughout this Section, the generalized Lord-Shulman thermoelasticity theory is recalled, in relation to an heterogeneous periodic composite material, which will be taken to the Laplace domain.

\begin{figure}[h!]
	\centering
	\includegraphics[scale=0.45]{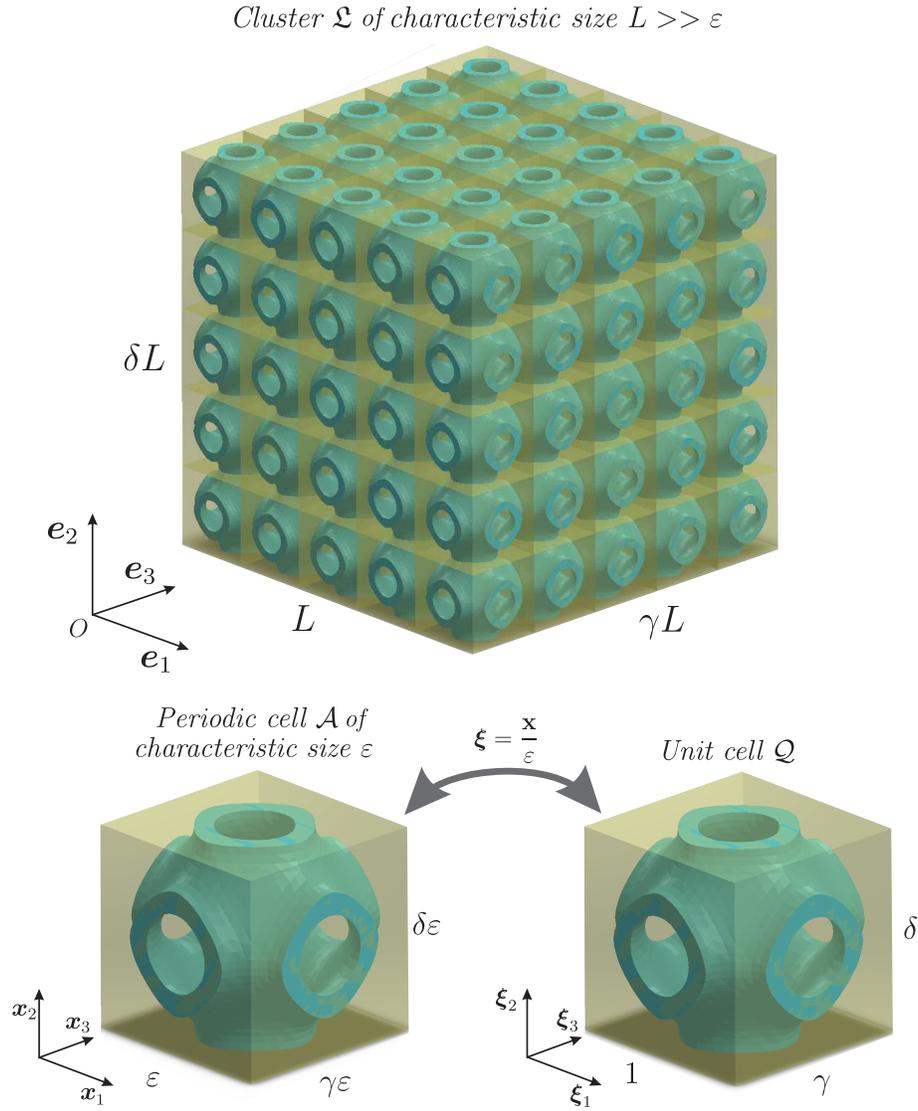}
	\caption{$3D$ continuum domain $\mathfrak{L}$ with periodic cell $\mathcal{A}$ and its dimensionless unit cell $\mathcal{Q}$.}
	\label{fig:stru3d}
\end{figure}

Let consider an heterogeneous composite material  $ \mathfrak{L}, $ Fig. \ref{fig:stru3d}, under the assumption of small strains (for instance a stretching load), which leads to a process of exchanging mechanical energy into thermal energy under the action of externally applied thermal-mechanical loadings. Such procedure is then followed by temperature variances and strains within the body, all of which vanish upon the removal of the mentioned thermal-mechanical loadings. The continuum $ \mathfrak{L} $ is described as a linear thermoelastic Cauchy medium \citep{hetnarski1999generalized,hetnarski2009thermal,ignaczak2010thermoelasticity} under stresses induced by body forces and temperature changes due to heat source.

On each point of the material is endowed with a displacement field $\boldsymbol{u}(\textbf{x},t) = u_i \boldsymbol{e}_{i}$ and a relative temperature field $\theta(\textbf{x}, t)=T(\textbf{x}, t)-T_0, $ where $ T(\textbf{x}, t) $ is the absolute temperature, $ T_0 $ is a reference stress-free temperature. The coupled constitutive relations link the stress tensor $\boldsymbol{\sigma}(\textbf{x},t)=\sigma_{ij}\boldsymbol{e}_{i}\otimes \boldsymbol{e}_{j}, $ the heat flux vector ${\mathbf{ q}}(\textbf{x},t)=q_{ij}\boldsymbol{e}_{i}\otimes \boldsymbol{e}_{j},  $  and the entropy per unit of volume $\boldsymbol{\eta}(\textbf{x},t)=\eta_{ij}\boldsymbol{e}_{i}\otimes \boldsymbol{e}_{j},$ to the
aforementioned relevant fields $ \boldsymbol{u}(\textbf{x},t), $ $ \theta(\textbf{x}, t), $ that is

\begin{subequations}
	\begin{align}	
		&\boldsymbol{\sigma}(\textbf{x},t)  = \textswab{C}(\textbf{x})\boldsymbol{\varepsilon}(\textbf{x},t) - {{ \boldsymbol{\alpha }}}(\textbf{x}) \theta(\textbf{x},t)\,,\label{sigma}\\
		&{\mathbf{ q}}(\textbf{x},t) + \tau(\textbf{x})\dfrac{\partial \mathbf{ q}(\textbf{x},t)}{\partial t}=  - {\bar{\mathbf{K}}}(\textbf{x}) \nabla  \theta(\textbf{x},t) \,,\label{q}\\
		&\boldsymbol{\eta}(\textbf{x},t) = \boldsymbol{\alpha}(\textbf{x}) \boldsymbol{\varepsilon}(\textbf{x},t)+\dfrac{C_E}{T_0}\theta(\textbf{x},t) \,,\label{eta}
	\end{align}
\end{subequations}

\noindent where  $\boldsymbol{\varepsilon}(\textbf{x},t)=\varepsilon_{ij}\boldsymbol{e}_{i}\otimes \boldsymbol{e}_{j}$ is the strain tensor, $\textswab{C}=C_{ijhk}\textbf{e}_i\otimes\textbf{e}_j \otimes\textbf{e}_h \otimes\textbf{e}_k$ is the fourth-order elastic tensor, ${\bar{\mathbf{K}}}=\bar{K}_{ij}\textbf{e}_i \otimes \textbf{e}_j$ is the second-order thermal conductivity tensor, ${{ \boldsymbol{\alpha }}}=\alpha_{ij}\textbf{e}_i \otimes \textbf{e}_j$ is the second-order stress-temperature tensor, $ C_E $ is the specific heat at zero strain, and for convenience we define $ p=C_E/T_0. $ Besides, as aforementioned, to avoid the physical paradox of infinite speed for the propagation of heat signal in the classical thermoelastic equations \citep{hetnarski2009thermal}, which are parabolic, a space dependent relaxation time $ \tau$ is introduced, known as the Maxwell-Cattaneo law replacing the Fourier law for thermal conduction, in order to transform them into hyperbolic equations.

As the material is under the effect of small displacements, then the micro-strain tensor is defined as

\begin{equation}
	\boldsymbol{\varepsilon}(\textbf{x},t) = \dfrac{1}{2}\left(\nabla \boldsymbol{u} (\textbf{x},t)+\nabla^{T} \boldsymbol{u}(\textbf{x},t)\right),\label{microstrain}
\end{equation}
where $\nabla \boldsymbol{u}$ is the gradient of the micro-displacement $\boldsymbol{u}(\textbf{x},t).$

The following balance equations hold

\begin{subequations}
	\begin{align}
		&\nabla\cdot(\boldsymbol{\sigma}(\textbf{x},t))  + \textbf{b}(\textbf{x},t)=\rho(\textbf{x}) \ddot{{\mathbf{ u}}}(\textbf{x},t),\label{bal1}\\
		&\nabla\cdot({\mathbf{q}}(\textbf{x},t)) - \bar{r}(\textbf{x},t)= -T_0\dot{\boldsymbol\eta}(\textbf{x},t)\,,\label{bal2}
	\end{align}
\end{subequations}
with $\textbf{b}=b_i\textbf{e}_i$ being the body forces, $\bar{r}$ the heat source per unit time per unit volume and $\rho$ is the mass density. Throughout, $t$ stands for the time coordinate, the superimposed dot denotes the derivative in relation to time, and consider $ \bar{r}=r/T_0. $

With the thermoelasticity constitutive equations and equilibrium equations well established, we may thus introduce the periodic microstructure of an heterogeneous composite material theory \citep{Mindlin1974}. Let a point be identified by the position vector $\textbf{x}=x_{1}\boldsymbol{e}_{1}+x_{2}\boldsymbol{e}_{2}+x_{3}\boldsymbol{e}_{3}\in\R^3$ related to a system of coordinates with origin at point $O$ and written in terms of the orthogonal base $\{\boldsymbol{e}_{1},\boldsymbol{e}_{2}, \boldsymbol{e}_{3}\},$ on the three-dimensional heterogeneous material $\mathfrak{L}$ characterized by a certain periodic microstructure. Let also $\mathcal{A}=[0,\varepsilon]\times [0,\delta \varepsilon]\times [0,\gamma\varepsilon]$ be a periodic cell with characteristic size $\varepsilon$ defined by three orthogonal periodic vectors $\boldsymbol{v}_{1}$, $\boldsymbol{v}_{2}$ and $\boldsymbol{v}_{3},$ written as $\boldsymbol{v}_{1} =  d_{1}\boldsymbol{e}_{1} = \varepsilon \boldsymbol{e}_{1} $,  $\boldsymbol{v}_{2}= d_{2}\boldsymbol{e}_{2} = \delta \varepsilon \boldsymbol{e}_{2}$ and $\boldsymbol{v}_{3} =  d_{3}\boldsymbol{e}_{3} =\gamma\varepsilon \boldsymbol{e}_{3}$. The material domain is set up by attaching three-dimensionally the cell $\mathcal{A}$ in accordance to the directions $\boldsymbol{v}_{1}$, $\boldsymbol{v}_{2}$ and $\boldsymbol{v}_{3}$, see Fig. \ref{fig:stru3d}.

Due to the periodicity of the material, throughout the work the superscript $ \emph{m} $ refers to the micro-scale. Therefore, $\textswab{C}^{m}(\textbf{x},t)=C_{ijhk}^{m}\textbf{e}_i\otimes\textbf{e}_j \otimes\textbf{e}_h \otimes\textbf{e}_k$ is the fourth-order micro-elastic tensor, $\bar{\mathbf{K}}^{m}(\textbf{x},t)=\bar{K}_{ij}^{m}\textbf{e}_i \otimes \textbf{e}_j$ is the second-order micro-thermal-conductivity tensor, $\boldsymbol{\alpha}^{m}(\textbf{x},t)=\alpha_{ij}^{m}\textbf{e}_i \otimes \textbf{e}_j$ is the second-order micro-stress temperature tensor, $\rho^{m}(\textbf{x})$ is the material density, $ p^m(\textbf{x}) $ is the specific heat at zero micro strain, and $ \tau^m(\textbf{x}) $ is the micro relaxation time. Furthermore, they obey the following conditions

\begin{subequations}
	\label{eq:pspecificheat} 
	\begin{align}
		&\textswab{C}^{m}(\textbf{x}+\boldsymbol{v}_i)         =  \textswab{C}^{m}(\textbf{x}),\\
		&\bar{\mathbf{K}}^{m}(\textbf{x}+\boldsymbol{v}_i)     =  \bar{\mathbf{K}}^{m}(\textbf{x}), \\
		&\boldsymbol{\alpha}^{m}(\textbf{x}+\boldsymbol{v}_i)  =  \boldsymbol{\alpha}^{m}(\textbf{x}), \\
		&\rho^{m}(\textbf{x}+\boldsymbol{v}_i)                 =  \rho^{m}(\textbf{x}),\\
		&p^{m}(\textbf{x}+\boldsymbol{v}_i)                    =  p^{m}(\textbf{x}),\\
		&\tau^{m}(\textbf{x}+\boldsymbol{v}_i)                 =  \tau^{m}(\textbf{x}),
	\end{align}
\end{subequations}
where $ i=1,2,3, $ for all $ \textbf{x}\in\mathcal{A}. $

In the following derivation, let us define $ L $ and $\varepsilon$ as the macroscopic length and the microstructural length, respectively, where $L>$$>\varepsilon,$ which gives the condition where the scales might be separated since the macroscopic length is taken to be much larger than the microstructural one. Let us also suppose that the heterogeneous material undergoes to $\mathfrak{L}$-periodic body forces $\boldsymbol{b}(\textbf{x},t),$ having zero mean values over the continuum $\mathfrak{L} = [0, L] \times [0, \delta L] \times [0, \gamma L],$ hence $\mathfrak{L}$ defined as above represents a portion of the continuum. Now, regarding the dimensionless cell, let us rescale the periodic cell $\mathcal{A}$ by a factor equals to the  characteristic length $\varepsilon,$ this implies that the non-dimensional cell model of the periodic microstructure is then $\mathcal{Q} = [0,1]\times [0,\delta]\times [0,\gamma].$ Moreover, from such rescaling process it arises two variables namely, the macroscopic (or slow) variable $\textbf{x}\in\mathcal{A}$ in charge of measuring the slow oscillations within the continuum, and the microscopic (or fast) variable $\boldsymbol{\xi}=\textbf{x}/\varepsilon\in\mathcal{Q}$ responsible in evaluating the fast heat propagation wave within the composite \citep{Bakhvalov1984,gambin1989higher,SmyshlyaevCherednichenko2000,Bacigalupo2014}. Thanks to cell $\mathcal{Q},$ the properties presented in \eqref{eq:pspecificheat} are now redefined as $\mathcal{Q}$-periodic over $\mathcal{Q}$ so the terms can be expressed by the microscopic variable $\boldsymbol{\xi}$ as  

\begin{subequations}
	\label{eq:MicroConstitutiveTensors1}
	\begin{align}
		&\medskip\textswab{C}^{m}(\textbf{x})         =  \textswab{C}^{m}\left(\mathbf{x},\boldsymbol{\xi}=\dfrac{\mathbf{x}}{\varepsilon}\right),\\
		&\medskip\bar{\mathbf{K}}^{m}(\textbf{x})     =  \bar{\mathbf{K}}^{m}\left(\mathbf{x},\boldsymbol{\xi}=\dfrac{\mathbf{x}}{\varepsilon}\right),\\
		&\medskip\boldsymbol{\alpha}^{m}(\textbf{x})  =  \boldsymbol{\alpha}^{m}\left(\mathbf{x},\boldsymbol{\xi}=\dfrac{\mathbf{x}}{\varepsilon}\right),\\
		&\medskip\rho^{m}(\textbf{x})                 =  \rho^{m}\left(\mathbf{x},\boldsymbol{\xi}=\dfrac{\mathbf{x}}{\varepsilon}\right),\\
		&\medskip p^{m}(\textbf{x})                   =  p^{m}\left(\mathbf{x},\boldsymbol{\xi}=\dfrac{\mathbf{x}}{\varepsilon}\right),\\
		&\medskip \tau^{m}(\textbf{x})                =  \tau^{m}\left(\mathbf{x},\boldsymbol{\xi}=\dfrac{\mathbf{x}}{\varepsilon}\right).
	\end{align}
\end{subequations}

Along with the $ \mathcal{Q} $-periodicity assumptions made for the relations \eqref{eq:pspecificheat}, regarding the micro-scale, the governing equations are finally obtained by plugging the micro-scale constitutive equations $ (\ref{sigma})$ to $(\ref{microstrain}) $ into the micro-scale balance equations $ (\ref{bal1})$ and $(\ref{bal2}),$ also hiding the arguments for a cleaner notation, one provides

\begin{subequations}
	\begin{align}
		&\nabla\cdot\left( \textswab{C}^{m}\nabla {\mathbf{u}} - {{\boldsymbol\alpha^{m}}\theta}\right)+{\mathbf{b}}=\rho^{m}{\mathbf{\ddot u}}\,,\label{gov1}\\
		&\mathbf{q} + \tau^m\mathbf{\dot{q}}=- (T_0)^{-1}\bar{\mathbf{K}}^{m} \nabla\theta\,,\label{gov2}\\
		&\nabla\cdot{\mathbf{{q}}} + {\boldsymbol\alpha^{m}} \nabla {\mathbf{\dot u}} + p^{m}\dot{\theta}  =  r\,.\label{gov3}		
	\end{align}
\end{subequations}
In the particular scenario where the relaxation time $ \tau^m(\mathbf{x},\boldsymbol{\xi})=\tau>0$ is assumed as a constant, the governing equations $ (\ref{gov1}) $ and  $ (\ref{gov2} )$ result into:

\begin{subequations}
	\begin{align}
		&\nabla\cdot\left( {\textswab{C}^{m}\nabla {\mathbf{u}}} - {\boldsymbol{\alpha}^{m}\theta}\right)+ {\mathbf{b}}= \rho^m\mathbf{\ddot u}\,,\label{eq:gov1}\\
		&\nabla\cdot\left( {{-(T_0)^{-1}\bar{\mathbf{K}}^{m}}}\nabla\theta\right) + {{\boldsymbol\alpha^{m}}}\mathcal{D}(\nabla\mathbf{\dot{u}})+p^{m}\mathcal{D}(\dot{\theta})=\mathcal{D}(r)\,,\label{eq:gov2}
	\end{align}
\end{subequations}
where the differential operator $\mathcal{D}=1+\tau^m\partial/\partial t$ depending on the relaxation time $ \tau^m $ is introduced.  

At this point it must be highlighted that, since the periodic cell $ \mathcal{Q} $ is composed by two different phase materials, the relaxation time $ \tau^m $ must be considered as space dependent, which means $ \tau(\mathbf{x},\boldsymbol{\xi}) $ is a $ \mathcal{Q} $-periodic function and hence varies within the composite material. For this reason, rather than continuing with the equation $ (\ref{gov2}),$ we apply the time bilateral Laplace transform, or analogously the Fourier transform with complex argument \citep{paley1934fourier,VanBre1947} to the equations $ (\ref{q}) ,$ $ (\ref{eta}), $ $ (\ref{gov1}) $ and  $ (\ref{gov3} ),$ in order to separate the time relaxation $ \tau^m $ from the partial derivative in time contained in the differential operator $ \mathcal{D}. $

To clarify ideas, the time bilateral Laplace transform of a function $ f:\R\rightarrow\R $ is defined as \cite{paley1934fourier}, 

\begin{equation}
	\label{eqn:lap}
	\mathcal{L}\left\{f(t)\right\}= \hat{f}(s) = \int_{-\infty}^{+\infty} f(t) \mathrm{e}^{-st}\mathrm{d}t, \quad s\in \mathbb{C}, 
\end{equation} 
where the Laplace argument $s$ and the Laplace transform $\hat{f}$ are complex valued (i.e. $\hat{f}:\mathbb{C}\to\mathbb{C}$ ). Whereas from an analytical standing point in \cite{paley1934fourier}, the inverse Laplace transform of a function $\hat{f}(s)$ is defined as

\begin{equation}
	\label{lapinverse}
	\mathcal{L}^{-1}\left\{\hat{f}(s)\right\}=f(t)=\dfrac{1}{2\pi \mathrm{i}}\int_{x-\mathrm{i} \infty}^{x+\mathrm{i} \infty} \mathrm{e}^{ts}\hat{f}(s)\mathrm{d}s.
\end{equation}
The Laplace transform of the $ n-$th derivative of $ f $ with respect to $ t $is given by the identity $ \mathcal{L}\left\{\partial^{n}f(t)/\partial t^{n}\right\} = s^{n}\hat{f}(s). $ The convolution between two functions $f_{1}$ and $f_{2}$ is defined as $ \left(f_{1}(t) \ast f_{2}(t)\right)=\int_{-\infty}^{\infty} f_1(u)f_2(t-\omega) \mathrm{d}\omega, $ and it follows that the Laplace transform convolution rule of two time dependent functions $f_{1}$ and $f_{2}$ is given as $ \mathcal{L}\left\{f_{1}(t) \ast f_{2}(t)\right\} = \mathcal{L}\left\{f_{1}(t)\right\} \mathcal{L}\left\{f_{2}(t)\right\}. $

Having established this, the governing equations on the transform space are

\begin{subequations}
	\begin{align}
		&\nabla\cdot\left(\textswab{C}^{m}\nabla\hat{{\mathbf{u}}} - {{\boldsymbol\alpha^{m}}\hat\theta}\right)+ \hat{{\mathbf{b}}} = \rho s^2\hat{ {\mathbf{u}}}\,,\label{eq:Bal1T}\\
		&\hat{\mathbf{{q}}} + \tau^m s\hat{\mathbf{ q}}=- (T_0)^{-1}\bar{\mathbf{K}}^{m} \nabla  \hat\theta\,,\label{eq:Bal2T}\\
		&\nabla\cdot\hat{\mathbf{{q}}} + {{\boldsymbol\alpha^{m} }}\nabla \hat{{\mathbf{u}}} + p^{m}s\hat{\theta} = \hat{r} \,,\label{eq:Bal3T}
	\end{align}
\end{subequations}
where $s$ is the unknown complex angular frequency $(s \in \mathbb{C}),$ 
and $\hat{\mathbf{u}}$, $\hat\theta,$ $\hat{{\mathbf{q}}},$ $ \hat{{\mathbf{b}}}, $ $ \hat{r}, $ are respectively, the bilateral Laplace transformed of the micro-displacement field, the micro-temperature field, the heat flux, the body forces and the heat source. 

Proceeding with the substitutions, Eqs. $ (\ref{gov1}), $ $ (\ref{gov2}) $ and $ (\ref{gov3}) $ become on the Laplace transform space,

\begin{subequations}
	\begin{align}
		&\nabla\cdot\left( {{\textswab{C}^{m}}\nabla\hat{{\mathbf{u}}}} - {{\boldsymbol\alpha^{m}}\hat\theta}\right)+ \hat{{\mathbf{b}}} = \rho^{m}s^2\hat{ {\mathbf{u}}}\,,\label{eq:Gov1T}\\
		&\nabla\cdot\left( {{-\mathbf{K}^{m}}}\nabla \hat\theta\right) + {s{\boldsymbol\alpha^{m}}}\nabla\hat{{\mathbf{u}}} + p^{m}s\hat{{\theta}}  =  \hat{r}\,,\label{eq:Gov2T}
	\end{align}
\end{subequations}
where, due the plugging we assume

\begin{equation}
	\mathbf{K}^{m}\left( {\mathbf{x}},\boldsymbol{\xi},s\right) =\dfrac{\bar{\mathbf{K}}^{m}(\mathbf{x},\boldsymbol{\xi})}{(1+\tau^m(\mathbf{x},\boldsymbol{\xi}) s)T_0},
	\label{plugging}
\end{equation}
with a $ \mathbf{K}^{m} $ well behaved over the defined domain.

The interface $\Sigma$ between two different phases $i$ and $j$ in the periodic cell $\mathcal{A},$ holds the jump of the values of function $f$ at it, written as $[[f]]=f^i(\Sigma)-f^j(\Sigma),$ follows that

\begin{subequations}
	\begin{align}
		&
		\left.
		\left[
		\left[
		\hat{\boldsymbol{u}}
		\right]
		\right]
		\right|_{\mathbf{x}\in\Sigma}=\mathbf{0},
		\label{eq:InterfaceConditionMicroDisplacement}\\
		&
		\left.
		\left[
		\left[
		\left(
		\textswab{C}^m\nabla\hat{\boldsymbol{u}}
		-
		\boldsymbol{\alpha}^m\hat\theta
		\right)\cdot\mathbf{n}
		\right]
		\right]
		\right|_{\mathbf{x}\in\Sigma}=\mathbf{0},
		\label{eq:InterfaceConditionMicroDisplacement1}\\
		&
		\left.
		\left[
		\left[
		\hat\theta
		\right]
		\right]
		\right|_{\mathbf{x}\in\Sigma}=\mathbf{0},
		\label{eq:InterfaceConditionMicrotemperature}\\
		&
		\left.
		\left[
		\left[
		-\mathbf{K}^m\nabla\hat\theta
		\cdot\mathbf{n}
		\right]
		\right]
		\right|_{\mathbf{x}\in\Sigma}=\mathbf{0},\label{eq:InterfaceConditionMicrotemperature1}
	\end{align}
\end{subequations}
representing the conditions of continuity for a bounded interface,
where the outward normal to the interface $\Sigma$ is indicated by the vector $\mathbf{n}.$ In case that $ \tau^m$ is constant within the phases of the material, and applying Laplace inverse transform we arrive in equations $ (\ref{eq:gov1}) $ and $ (\ref{eq:gov2}) $ again.

On behalf of the time domain $ t ,$ let us apply the inverse Laplace transform to the field Eqs. $ (\ref{eq:Gov1T}) $ and $ (\ref{eq:Gov2T} ),$ and due to the dependency on the variable $ s $ that $ \mathbf{K} $ holds on equation $ (\ref{plugging} ),$ which gives

\begin{equation}
	\label{eqn:convK}
	\mathcal{L}^{-1}\left\{\mathbf{K}^{m}\nabla\hat\theta\right\} = \mathcal{L}^{-1}\left\{\mathbf{K}^{m}\right\}\ast\nabla\theta,
\end{equation}
the field equations $(\ref{eq:gov1})$ and $(\ref{eq:gov2}),$ might be rewritten as an integral-differentiable form, therefore

\begin{subequations}
	\begin{align}
		&\nabla\cdot\left( \textswab{C}^{m}\nabla {\mathbf{u}} - {{\boldsymbol\alpha^{m}}\theta}\right)+{\mathbf{b}}=\rho^{m}{\mathbf{\ddot u}}\,,\label{eq:Gov1integraldifferenti}\\
		&\nabla\cdot\left(\mathcal{L}^{-1}\left\{\mathbf{K}^{m}\right\}\ast\nabla\theta-\boldsymbol{\alpha}^m\dot{\boldsymbol{u}}\right) + r = p^m\dot\theta\,\label{eq:Gov2integraldifferenti}
	\end{align}
\end{subequations}
known also as the field equations of a continuum with one time relaxation of first order.

We must point out here that, from the $\mathcal{Q}$-periodicity of microscopic constitutive tensors in Eqs. \eqref{eq:MicroConstitutiveTensors1}, \eqref{eq:Gov1T}, \eqref{eq:Gov2T}, and bearing in mind the $\mathcal{L}$-periodicity of the body forces, the micro-fields, displacement and temperature (either both Laplace transformed or not) are dependent on the fast variable $\boldsymbol\xi$ and the slow one $\mathbf{x},$ besides both fields may be written as $ \medskip\hat{\boldsymbol{u}} =  \hat{\boldsymbol{u}}\left(\mathbf{x},\boldsymbol\xi=\frac{\mathbf{x}}{\varepsilon}\right)$ and $ \hat\theta = \hat\theta\left(\mathbf{x},\boldsymbol\xi=\frac{\mathbf{x}}{\varepsilon}\right). $ In addition, given the fact of the $\mathcal{Q}$-periodicity of their coefficients, any attempt in deriving the solution of the set of PDE's $ (\ref{eq:Gov1T})$ and $(\ref{eq:Gov2T})$ might be rather cumbersome both analytically and numerically. Nevertheless, the homogenization technique arrives in this context to barter the microstructured continuum by an equivalent homogeneous one, where the solutions will be an approximation to those from $ (\ref{eq:Gov1T})$ and $(\ref{eq:Gov2T}),$ however on this scenario the coefficients will not be affected by the fast variable $ \boldsymbol\xi, $ which gives rapid oscillations led by the underlying microstructure previously held, and overall such procedure might be quite effective computationally.

Forward to the next Section, the homogenization technique on a bi-phase periodic microstructure composite subjected to thermal sources and periodic body forces will be applied over the equations of thermoelasticity with one time relaxation on the Laplace transformed domain, namely equations $ (\ref{eq:Gov1T})$ and $(\ref{eq:Gov2T}).$

\section{Asymptotic expansions of microscopic field equations on the Laplace domain}
\label{sec:expansions}

Inspired by the asymptotic approach developed in \cite{SanchezPalencia1974,Bakhvalov1984,SmyshlyaevCherednichenko2000,bacigalupo2014second}, the micro-displacement $\boldsymbol{u}$ and micro-temperature $\theta$ are expressed as power series in terms of $\varepsilon$, as well known as the asymptotic expansions in terms of $\varepsilon$ that separates the fast variable $\boldsymbol{\xi} = \textbf{x}/\varepsilon$ from the slow one $\textbf{x},$ i.e. in the hypothesis of scale separation, diving into the components, follows that

\begin{subequations}
	\begin{align}
		&u_{h}\left(\textbf{x},\dfrac{\textbf{x}}{\varepsilon},t\right) = \sum_{l=0}^{+\infty} \varepsilon^{l} u^{(l)}_{h} = u^{(0)}_{h}\left(\textbf{x},\dfrac{\textbf{x}}{\varepsilon},t\right) + \varepsilon u^{(1)}_{h}\left(\textbf{x},\dfrac{\textbf{x}}{\varepsilon},t\right) +\varepsilon^{2}u^{(2)}_{h}\left(\textbf{x},\dfrac{\textbf{x}}{\varepsilon},t\right)+\mathcal{O}(\varepsilon^{3}),\label{eqn:u}\\
		&\theta\left(\textbf{x},\dfrac{\textbf{x}}{\varepsilon},t\right) = \sum_{l=0}^{+\infty} \varepsilon^{l} \theta^{(l)} = \theta^{(0)}\left(\textbf{x},\dfrac{\textbf{x}}{\varepsilon},t\right) + \varepsilon \theta^{(1)}\left(\textbf{x},\dfrac{\textbf{x}}{\varepsilon},t\right) +\varepsilon^{2}\theta^{(2)}\left(\textbf{x},\dfrac{\textbf{x}}{\varepsilon},t\right)+\mathcal{O}(\varepsilon^{3}).\label{eqn:theta}
	\end{align}
\end{subequations}

The Laplace transform $\eqref{eqn:lap}$ is applied to equations $\eqref{eqn:u}$ and $\eqref{eqn:theta},$ which leads to

\begin{subequations}
	\begin{align}
		&\mathcal{L}\left(u_{h}\left(\textbf{x},\dfrac{\textbf{x}}{\varepsilon},t\right)\right)=\sum_{l=0}^{+\infty}\varepsilon^{l}\hat{u}^{(l)}_{h} = \hat{u}^{(0)}_{h}\left(\textbf{x},\dfrac{\textbf{x}}{\varepsilon},s\right)+\varepsilon\hat{u}^{(1)}_{h}\left(\textbf{x},\dfrac{\textbf{x}}{\varepsilon},s\right)+\varepsilon^{2}\hat{u}^{(2)}_{h}\left(\textbf{x},\dfrac{\textbf{x}}{\varepsilon},s\right)+\mathcal{O}(\varepsilon^{3}),\label{eqr:ulaplace}\\
		&\mathcal{L}\left(\theta\left(\textbf{x},\dfrac{\textbf{x}}{\varepsilon},t\right)\right)  =  \sum_{l=0}^{+\infty} \varepsilon^{l} \hat{\theta}^{(l)} = \hat{\theta}^{(0)}\left(\textbf{x},\dfrac{\textbf{x}}{\varepsilon},s\right) + \varepsilon \hat{\theta}^{(1)}\left(\textbf{x},\dfrac{\textbf{x}}{\varepsilon},s\right)+\varepsilon^{2}\hat{\theta}^{(2)}\left(\textbf{x},\dfrac{\textbf{x}}{\varepsilon},s\right)+ \mathcal{O}(\varepsilon^{3}).\label{eqr:thetalaplace}
	\end{align} 
\end{subequations}

It is noteworthy that both asymptotic expansions above are equivalent to the asymptotic expansions on the time domain $ t, $ $\eqref{eqn:u}$ and $\eqref{eqn:theta}.$ 

The homogenization procedure of the equations \eqref{eq:Gov1T}  and \eqref{eq:Gov2T} searches for solutions of the micro-displacement $\hat{u}_{h}$ and micro-temperature $\hat{\theta},$ as decompositions of increasing powers of the microscopic length $ \varepsilon.$ In order to do so, the replacement of the asymptotic expansions $~\eqref{eqr:ulaplace}$ and $~\eqref{eqr:thetalaplace}$ must be done into the microscopic field equations on the Laplace domain $ ~\eqref{eq:Gov1T} $ and $ \eqref{eq:Gov2T} ,$ respectively. Further mathematical details developing the asymptotic expansions in terms with equal power of $ \varepsilon $ of the field equations, among their interface conditions are stressed in the Sec. \ref{sec::additionaldetailshomogenization} of the Supplementary Material of the present work.

In fact, the arrived fields equations $ (\ref{eqn:uhomogenized})$ and $ (\ref{eqn:thetahomogenized}) $ in the Sec. \ref{sec::additionaldetailshomogenization} of the Supplementary Material, as power series obtained from the homogenization asymptotic procedure, have the following structure,

\begin{subequations}
	\begin{align}
		\label{fsourceterms}
		&\dfrac{1}{\varepsilon^2}f_i^{(0)}(\textbf{x})+\dfrac{1}{\varepsilon}f_i^{(1)}(\textbf{x})+\varepsilon^0f_i^{(2)}(\textbf{x})+\varepsilon f_i^{(3)}(\textbf{x})+\cdots+\varepsilon^lf_i^{(l+2)}(\textbf{x})+\hat{b}_i(\textbf{x})=0,\\
		\label{gsourceterms}
		&\dfrac{1}{\varepsilon^2}g^{(0)}(\textbf{x})+\dfrac{1}{\varepsilon}g^{(1)}(\textbf{x})+\varepsilon^0g^{(2)}(\textbf{x})+\varepsilon g^{(3)}(\textbf{x})+\cdots+\varepsilon^lg^{(l+2)}(\textbf{x})+\hat{r}(\textbf{x})=0,
	\end{align}
\end{subequations}
where the set of functions $ f_i^{(0)},\cdots,f_i^{(l+2)} $ and $ g^{(0)},\cdots,g^{(l+2)}, $ with $ l\in\N ,$ are such that the dependency goes only over the slow variable $ \textbf{x} ,$ and are determined imposing the solvability conditions (seen in Sec. \ref{sec:recursivesection}) on the class of the $ \mathcal{Q}$-periodic functions. Furthermore, the Sec. \ref{sec:recursivesection} of the Supplementary Material shows that the asymptotic field equations $\eqref{eqn:uhomogenized}$ and $\eqref{eqn:thetahomogenized}$ produce a cascade of recursive differential problems that determine sequentially the structure of the solutions of the displacements $\boldsymbol{\hat{u}}^{(0)},$ $\boldsymbol{\hat{u}}^{(1)}$...,\quad and also the structure of the solutions of the temperature $ \hat\theta^{(1)} ,$ $ \hat\theta^{(1)} $...,\quad respectively, in terms of suitable $\mathcal{Q}$-periodic perturbation functions which are determined by the solution of suitable cell problems detailed in the following sub-session.

\subsection{Cell problems and perturbation functions}
\label{Sec::CellsProblems}

The solutions of Eqs. $ (\ref{eq:Gov1T})$ and $(\ref{eq:Gov2T}) $ provide that the inhomogeneous cell problems at the different orders of $ \varepsilon $ can be expanded in terms of the perturbation functions. In regards to them, such perturbation functions depend exclusively on the microstructure features such as, material geometry and mechanical properties, where the last one influences the displacements and temperature due to the heterogeneity of the material. Despite the existence of the homogenized solution $\mathcal{Q}$-periodic holds without any other assumption \citep{papanicolau1978asymptotic,Bakhvalov1984}, it is noteworthy that imposing the normalization conditions $ (\ref{eq:normalization1}) - (\ref{eq:normalization3}) ,$ stated in the Sec. \ref{sec:recursivesection} of the Supplementary Material, to the cell problems, sooner stabilised on this Section, the uniqueness of the perturbation functions is also held.

At this stage, the structure of these cell problems is exploited at the different orders of $\varepsilon,$ for elasticity and thermal diffusion problems. For this purpose, the recursive differential problems of the elastic and thermal fields will be treated separately for each order of $ \varepsilon .$

On behalf of the elastic problem at the order $ \varepsilon^{-1},$ plugging the Eq. $ (\ref{eqn:solutiondisplac-1})$ , of the solution $ u_k^{(1)}, $ into the differential problem $ (\ref{eq:epsilon-1f}), $ both equation from the Sec. \ref{sec:recursivesection} of the Supplementary Material, lead to two cell problems:

\begin{subequations}
	\begin{align}
		&\medskip\left(C_{ijhk}^{m}\,N_{hpq_1,k}^{(1,0)}
		\right)_{,j}+C_{ijpq_1,j}^{m}=0,\label{eq:ElasticCellProblemOrder-1_1}\\
		&\left(C_{ijhk}^{m}\tilde{N}_{h,k}^{(1,0)}\right)_{,j}-\alpha_{ij,j}^{m}=0.
	\end{align}
\end{subequations}

Similarly, from Eq. $ (\ref{eq:epsilon-1g}) $ and in consideration of the solution $ (\ref{eqn:solutiontemp-1})$ reported in the Sec. \ref{sec:recursivesection} of the Supplementary Material, after substitution, the cell problem at the order $\varepsilon^{-1}$ takes the form

\begin{eqnarray}
	\label{eq:ThermalCellProblemOrder-1}
	\left(K_{ij}^{m}M_{q_1,j}^{(1,0)}\right)_{,i}+K_{iq_1,i}^{m}=0.
\end{eqnarray}
Therefore, for the terms at the order $ \varepsilon^{-1} ,$ we obtain the following set of cell problems

\begin{subequations}\label{eq:CellProblemOrder-1}
	\begin{align}
		&\left(C_{ijhk}^{m}\,N_{hpq_1,k}^{(1,0)}
		\right)_{,j}+C_{ijpq_1,j}^{m}=0,
		\\
		&\left(C_{ijhk}^{m}\tilde{N}_{h,k}^{(1,0)}\right)_{,j}-\alpha_{ij,j}^{m}=0,
		\\
		&\left(K_{ij}^{m}M_{q_1,j}^{(1,0)}\right)_{,i}+K_{iq_1,i}^{m}=0,
	\end{align}
\end{subequations}
with interface conditions expressed in terms of perturbation functions $N_{hpq_1}^{(1,0)},$ $\tilde{N}_{h}^{(1,0)},$ $M_{q_1}^{(1,0)}$ as

\begin{subequations}\label{eqns:InterfaceConditionsCellProblemOrder-1}
	\begin{align}
		&\left.
		\left[
		\left[
		N_{hpq_1}^{(1,0)}
		\right]
		\right]
		\right|_{\boldsymbol{\xi}\in\Sigma_1}=0,
		\\
		&
		\left.
		\left[
		\left[
		\left(
		C_{ijhk}^{m}
		\left(
		\delta_{hp}\,
		\delta_{kq_1}
		+
		N_{hpq_1,k}^{(1,0)}
		\right)
		\right)n_j
		\right]
		\right]
		\right|_{\boldsymbol\xi\in\Sigma_1}=0,
		\\
		&\left.
		\left[
		\left[
		\tilde{N}_{h}^{(1,0)}
		\right]
		\right]
		\right|_{\boldsymbol{\xi}\in\Sigma_1}=0,
		\\
		&
		\left.
		\left[
		\left[
		\left(
		C_{ijhk}^{m}
		\left(
		\tilde{N}_{h,k}^{(1,0)}
		-
		\alpha^m_{ij}
		\right)
		\right)n_j
		\right]
		\right]
		\right|_{\boldsymbol\xi\in\Sigma_1}=0,
		\\
		&\left.
		\left[
		\left[
		M_{q_1}^{(1,0)}
		\right]
		\right]
		\right|_{\boldsymbol{\xi}\in\Sigma_1}=0,
		\\
		&
		\left.
		\left[
		\left[
		\left(
		K_{ij}^m
		\left(
		M_{q_1,j}^{(1,0)}+\delta_{jq_1}
		\right)
		\right)n_i
		\right]
		\right]
		\right|_{\boldsymbol\xi\in\Sigma_1}=0,\label{eqns:InterfaceConditionsCellProblemOrder-1sexta}
	\end{align}
\end{subequations}

where the symmetries from the respective tensors are passed to the perturbation functions, for instance $ N_{hpq_1}^{(1,0)}=N_{hq_1p}^{(1,0)} ,$ since $ C_{ijpq_1,j}^{m}=C_{ijq_1p,j}^{m}. $

Once the perturbation functions $ N_{hpq_1}^{(1,0)} ,$ $ \tilde{N}_h^{(1,0)} $ and $M_{q_1}^{(1)}$ are determined, from the differential problem $ (\ref{eq:epsilon0f}) $ and recalling the solution $ u_k^{(2)} $ in Eq. $ (\ref{eqn:solutiondisplac0}) $ stated in the Sec. \ref{sec:recursivesection} of the Supplementary Material, one derives the three following cell problems, at the order $\varepsilon^{0},$ and symmetrizing with respect to indices $q_1$ and $q_2,$ leads to 

\begin{subequations}
	\begin{align}
		\label{eq:ElasticCellProblemOrder0_1}
		\begin{split}
			&\medskip\left(C_{ijhk}^{m}\,N_{hpq_1q_2,k}^{(2,0)}
			\right)_{,j}
			+
			\dfrac{1}{2}
			\left[
			\left(
			C_{ijhq_2}^{m}\,
			N_{hpq_1}^{(1,0)}
			\right)_{,j}
			+
			C_{iq_1pq_2}^{m}
			+
			C_{iq_2hj}^{m}\,
			N_{hpq_1,j}^{(1,0)}
			+
			\left(
			C_{ijhq_1}^{m}\,
			N_{hpq_2}^{(1,0)}
			\right)_{,j}
			+\right.
			\\
			&\medskip\left.+\,
			C_{iq_2pq_1}^{m}
			+
			C_{iq_1hj}^{m}\,
			N_{hpq_2,j}^{(1,0)}\right]
			=
			\dfrac{1}{2}
			\left\langle
			C_{iq_1pq_2}^m
			+
			C_{iq_2hj}^m\,
			N_{hpq_1,j}^{(1,0)}
			+
			C_{iq_2pq_1}^{m}
			+
			C_{iq_1hj}^m\,
			N_{hpq_2,j}^{(1,0)}
			\right\rangle,
		\end{split}\\
		\begin{split}
			&\medskip\left(
			C_{ijhk}^{m}\,
			\tilde{N}_{hq_1,k}^{(2,1)}
			\right)_{,j}
			+
			\left[
			\left(
			C_{ijhq_1}^{m}\,
			\tilde{N}_{h}^{(1,0)}
			\right)_{,j}
			+
			C_{iq_1hj}^{m}\,
			\tilde{N}_{h,j}^{(1,0)}
			-
			\left(
			\alpha_{ij}^{m}\,
			M_{q_1}^{(1,0)}
			\right)_{,j}
			-
			\alpha_{iq_1}^{m}
			\right]
			=
			\left\langle
			C_{iq_1hj}^m\,
			\tilde{N}_{h,j}^{(1,0)}
			-
			\alpha_{iq_1}^{m}
			\right\rangle
			\\
			&\left(
			C_{ijhk}^{m}\,
			N_{hp,k}^{(2,2)}
			\right)_{,j}
			-
			\rho^m\delta_{ip}
			=
			-\left\langle
			\rho^m
			\right\rangle\delta_{ip},
		\end{split}
	\end{align}
\end{subequations}
with interface conditions in terms of perturbation functions as

\begin{subequations}
	\begin{flalign}
		\label{eq:InterfaceConditionsElasticProblemOrder0}
		&
		\left.
		\left[
		\left[
		N_{hpq_1q_2}^{(2,0)}
		\right]
		\right]
		\right|_{\boldsymbol{\xi}\in\Sigma_1}=0,\\
		&
		\left.
		\left[
		\left[
		\left[
		C_{ijhk}^{m}
		N_{hpq_1q_2,k}^{(2,0)}
		+
		\frac{1}{2}
		\left(
		C_{ijhq_2}^m
		N_{hpq_1}^{(1,0)}
		+
		C_{ijhq_1}^m
		N_{hpq_2}^{(1,0)}
		\right)\right]n_j
		\right]
		\right]
		\right|_{\boldsymbol{\xi}\in\Sigma_1}=0,\\
		&
		\left.
		\left[
		\left[
		\tilde{N}_{hq_1}^{(2,1)}
		\right]
		\right]
		\right|_{\boldsymbol{\xi}\in\Sigma_1}=0,\\
		&
		\left.
		\left[
		\left[
		\left(
		C_{ijhk}^{m}
		\tilde{N}_{hq_1,k}^{(2,1)}
		+
		\left(
		C_{ijhq_2}^m\,
		\tilde{N}_{h}^{(1,0)}\delta_{kq_1}
		-
		\alpha_{ij}^m
		M_{q_1}^{(1,0)}
		\right)
		\right)n_j
		\right]
		\right]
		\right|_{\boldsymbol{\xi}\in\Sigma_1}=0,\\
		&
		\left.
		\left[
		\left[
		N^{(2,2)}_{hp}
		\right]
		\right]
		\right|_{\boldsymbol{\xi}\in\Sigma_1}=0,\\
		&
		\left.
		\left[
		\left[
		\left(
		C_{ijhk}^m
		N_{hp,k}^{(2,2)}
		\right)n_j
		\right]
		\right]
		\right|_{\boldsymbol{\xi}\in\Sigma_1}=0.
	\end{flalign}
\end{subequations}

Analogously, for the thermal homogenized field at the order $\varepsilon^0,$ replacing solution $ (\ref{eqn:solutiontemp0}) $ into the differential problem $ (\ref{eq:epsilon0g}) $ derived in the Sec. \ref{sec:recursivesection} of the Supplementary Material, one gives the next three cell problems

\begin{subequations}
	\begin{align}
		\begin{split}
			\label{eq:ThermalCellProblemOrder0_1}
			&\medskip\left(K_{ij}^{m}\,
			M_{q_1q_2,j}^{(2,0)}
			\right)_{,i}
			+
			\dfrac{1}{2}
			\left[
			\left(
			K_{iq_2}^{m}\,
			M_{q_1}^{(1,0)}
			\right)_{,i}
			+
			K_{q_1q_2}^{m}
			+
			K_{iq_2}^{m}\,
			M_{q_1,i}^{(1,0)}
			+
			\left(
			K_{iq_1}^{m}\,
			M_{q_2}^{(1,0)}
			\right)_{,i}
			+
			K_{q_2q_1}^{m}
			+\right.
			\\
			&\medskip\left.K_{q_1i}^{m}\,
			M_{q_2,i}^{(1,0)}
			\right]
			=
			\dfrac{1}{2}
			\left\langle
			K_{q_1q_2}^m
			+
			K_{q_2i}^m\,
			M_{q_1,i}^{(1,0)}
			+
			K_{q_2q_1}^{m}
			+
			K_{q_1i}^m\,
			M_{q_2,i}^{(1,0)}
			\right\rangle,
		\end{split}\\
		&\medskip\left(
		K_{ij}^{m}\,
		\tilde{M}_{pq_1,j}^{(2,1)}
		\right)_{,i}
		-
		\left[
		\alpha_{ij}^{m}\,
		N_{ipq_1,j}^{(1,0)}
		+
		\alpha_{pq_1}^{m}\,
		\right]
		=
		-\left\langle
		\alpha_{ij}^m\,
		N_{ipq_1,j}^{(1,0)}
		+
		\alpha_{pq_1}^{m}
		\right\rangle
		\\
		&\medskip\left(
		K_{ij}^{m}\,
		M_{,j}^{(2,1)}
		\right)_{,i}
		-
		\left[
		\alpha_{ij}^{m}\,
		\tilde{N}_{i,j}^{(1,0)}
		+
		p^{m}
		\right]
		=
		-\left\langle
		\alpha_{ij}^{m}\,
		\tilde{N}_{i,j}^{(1,0)}
		+
		p^{m}
		\right\rangle,
	\end{align}
\end{subequations}
and, in terms of perturbation functions, the interface conditions become

\begin{subequations}
	\label{eq:InterfaceConditionsThermalCellProblemOrder0}
	\begin{flalign}
		&
		\left.
		\left[
		\left[
		M_{q_1q_2}^{(2,0)}
		\right]
		\right]
		\right|_{\boldsymbol{\xi}\in\Sigma_1}=0,\\
		&
		\left.
		\left[
		\left[
		\left[
		K_{ij}^m
		M_{q_1q_2,j}^{(2,0)}
		+
		\frac{1}{2}
		\left(
		K_{iq_2}^m\,M_{q_1}^{(1,0)}+
		K_{iq_1}^m\,M_{q_2}^{(1,0)}
		\right)\right]n_i
		\right]
		\right]
		\right|_{\boldsymbol{\xi}\in\Sigma_1}=0.\\
		&
		\left.
		\left[
		\left[
		\tilde{M}_{pq_1}^{(2,1)}
		\right]
		\right]
		\right|_{\boldsymbol{\xi}\in\Sigma_1}=0,\\
		&
		\left.
		\left[
		\left[
		\left(
		K_{ij}^m(s)
		\tilde{M}_{pq_1,j}^{(2,1)}
		\right)n_i
		\right]
		\right]
		\right|_{\boldsymbol{\xi}\in\Sigma_1}=0.\\
		&
		\left.
		\left[
		\left[
		M^{(2,1)}
		\right]
		\right]
		\right|_{\boldsymbol{\xi}\in\Sigma_1}=0,\\
		&
		\left.
		\left[
		\left[
		\left(
		K_{ij}^m(s)
		M_{,j}^{(2,1)}
		\right)n_i
		\right]
		\right]
		\right|_{\boldsymbol{\xi}\in\Sigma_1}=0.
	\end{flalign}
\end{subequations}

\subsection{Down-scaling and up-scaling relations}
\label{Sec::DownScalingUpScaling}

On one hand, the feasibility of expressing the microscopic fields $\hat{u}_h(\mathbf{x},\boldsymbol{\xi},s)$ and $\hat\theta(\mathbf{x},\boldsymbol{\xi},s)$ as asymptotic expansions of powers of the microscopic length $\varepsilon$ in terms of the macroscopic fields $\hat{U}^M_h(\mathbf{x},s)$, $\hat\Theta^M(\mathbf{x},s),$ their gradients and in terms of the $\mathcal{Q}$-periodic perturbation functions arises from the solution of the cell problems at different orders of $\varepsilon$ developed in Sec. \ref{Sec::CellsProblems}. From the expansions $ (\ref{eqn:theta}) $ and $ (\ref{eqn:u}), $ and considering the solutions brought in the Sec. \ref{sec:recursivesection} of the Supplementary Material $ ~\eqref{eqn:solutiondisplac-2}, $ $ ~\eqref{eqn:solutiontemp-2}, $ $ ~\eqref{eqn:solutiondisplac-1}, $ $ ~\eqref{eqn:solutiontemp-1}, $ $ ~\eqref{eqn:solutiondisplac0}$ and $ ~\eqref{eqn:solutiontemp0} $ of cell problems at the different orders of $\varepsilon,$ one derives the down-scaling relations of the transformed micro-displacement field and the transformed micro-temperature field, respectively,

\begin{subequations}
	\begin{flalign}
		\label{downscalingu}
		\begin{split}
			&\hat{u}_{h}\left(\mathbf{x},\frac{\mathbf{x}}{\varepsilon},s\right)=
			\left[\hat{U}^{M}_{h}(\mathbf{x},s)
			+\varepsilon\left(N^{(1,0)}_{hpq_{1}}(\boldsymbol{\xi})\dfrac{\partial \hat{U}^{M}_{h}}{\partial x_{q_{1}}}+\tilde{N}^{(1,0)}_{h}(\boldsymbol{\xi})\hat{\Theta}^M\right)\right.+\\
			&+\left.\left.\varepsilon^{2}\left(N^{(2,0)}_{hpq_{1}q_{2}}(\boldsymbol{\xi})\dfrac{\partial^{2}\hat{U}^{M}_{h}}{\partial x_{q1}\partial x_{q2}}+\tilde{N}^{(2,1)}_{hq_{1}}(\boldsymbol{\xi})\dfrac{\partial \hat{\Theta}^{M}}{\partial x_{q_{1}}}+s^{2}N^{(2,2)}_{hp}(\boldsymbol{\xi})\hat{U}^{M}_{h}\right)+\mathcal{O}(\varepsilon^{3})\right]\right|_{\boldsymbol{\xi}= \frac{\mathbf{x}}{\varepsilon}},
		\end{split}
	\end{flalign}
	\begin{flalign}
		\label{downscalingtheta}
		\begin{split}
			&\hat{\theta}\left(\mathbf{x},\frac{\mathbf{x}}{\varepsilon},s\right)=
			\left[\hat{\Theta}^{M}(\mathbf{x},s)
			+\varepsilon\left(M^{(1,0)}_{q_{1}}(\boldsymbol{\xi})\dfrac{\partial \hat{\Theta}^{M}}{\partial x_{q_{1}}}\right)+\right.\\
			&+\left.\left.\varepsilon^{2}\left(M^{(2,0)}_{pq_{1}q_{2}}(\boldsymbol{\xi})\dfrac{\partial^{2}\hat{\Theta}^{M}}{\partial x_{q1}\partial x_{q2}}
			+s\tilde{M}^{(2,1)}_{pq_{1}}(\boldsymbol{\xi})\dfrac{\partial \hat{\Theta}^{M}}{\partial x_{q_{1}}}+s^{2}M^{(2,1)}_{hp}(\boldsymbol{\xi})\hat{\Theta}^{M}\right)+\mathcal{O}(\varepsilon^{3})\right]\right|_{\boldsymbol{\xi}= \frac{\mathbf{x}}{\varepsilon}}.
		\end{split}
	\end{flalign}
\end{subequations}

Noteworthy, the $\mathcal{Q}$-periodic perturbation functions $N_{hpq_1}^{(1,0)}, \tilde{N}_{h}^{(1,0)},N_{hpq_1q_2}^{(2,0)},\tilde{N}^{(2,0)}_{hq_1},N_{hp}^{(2,2)},M^{(1,0)}_{q_1},$ $M^{(2,0)}_{q_1q_2},$ $\tilde{M}^{(2,1)}_{pq_1},$ $M^{(2,1)},$ cast the effect of microstructural inhomogeneities of the material through their dependency on the fast variable $\boldsymbol{\xi}=\mathbf{x}/\varepsilon$, whereas the macro-fields $\hat{\boldsymbol{u}}(\mathbf{x},s),$ and $\hat\theta(\mathbf{x},s)$ are $\mathfrak{L}$-periodic functions and therefore rely only on the slow variable $\mathbf{x}$ \citep{fantoni2017multi}.

On the other hand, also known as the up-scaling relations, they allow the definition of the macroscopic fields in terms of the microscopic ones. This means that the macroscopic fields may be defined as the mean values of microscopic equations $ (\ref{downscalingu})$ and $ (\ref{downscalingtheta}) $ over the unit cell $\mathcal{Q},$ thus it follows that

\begin{subequations}
	\begin{eqnarray}
		\begin{split}
			\hat{U}_{h}(\mathbf{x},s)\doteq 
			\left\langle\hat{u}_h\left(\mathbf{x},\dfrac{\mathbf{x}}{\varepsilon}+{\boldsymbol{\zeta}},s\right)
			\right\rangle=&
			\dfrac{1}{|\mathcal{Q}|}\int_{\mathcal{Q}}\hat{u}_h\left(\mathbf{x},\dfrac{\mathbf{x}}{\varepsilon}+{\boldsymbol{\zeta}},s\right)d\boldsymbol{\zeta}=\\
			=&\dfrac{1}{|\mathcal{Q}|}\int_{\mathcal{Q}}\hat{u}_h\left(\mathbf{x},\boldsymbol{\xi},s\right)d\boldsymbol{\xi}=\left\langle\hat{u}_h\left(\mathbf{x},{\boldsymbol{\xi}},s\right)\right\rangle,\label{UpScalingLawu}
		\end{split}\\
		\begin{split}
			\hat\Theta(\mathbf{x},s)\doteq
			\left\langle\hat\theta\left(\mathbf{x},\dfrac{\mathbf{x}}{\varepsilon}+\boldsymbol{\zeta},s\right)
			\right\rangle=&
			\dfrac{1}{|\mathcal{Q}|}\int_{\mathcal{Q}}\hat{\theta}\left(\mathbf{x},\dfrac{\mathbf{x}}{\varepsilon}+{\boldsymbol{\zeta}},s\right)d\boldsymbol{\zeta}=\\
			=&\dfrac{1}{|\mathcal{Q}|}\int_{\mathcal{Q}}\hat{\theta}\left(\mathbf{x},\boldsymbol{\xi},s\right)d\boldsymbol{\xi}=\left\langle\hat\theta\left(\mathbf{x},\boldsymbol{\xi},s\right)
			\right\rangle,\quad\label{UpScalingLawtheta}
		\end{split}
	\end{eqnarray}
\end{subequations}
where the variable $\boldsymbol{\zeta}\in\mathcal{Q}$ has been introduced and the vector $\varepsilon\boldsymbol{\zeta}\in\mathcal{A}$ defines all the possible translations of the heterogeneous medium compared to a grid of cells having characteristic size $\varepsilon,$ with respect to the $\mathfrak{L}$-periodic body force $\hat{\mathbf{b}}(\mathbf{x})$ \citep{SmyshlyaevCherednichenko2000,Bacigalupo2014}. Besides, the mean value operator of the left hand side is taken over the variable $ \boldsymbol{\zeta} ,$ whereas the mean value operator of the right hand side has become over $\boldsymbol{\xi}. $

\section{Approximation of the power-like functional through truncation of its asymptotic expansion}
\label{variationalmacro}

A variational-asymptotic procedure \citep{SmyshlyaevCherednichenko2000,bacigalupo2014second} is exploited to furnish a finite order governing equation to establish an equivalence among the equations \eqref{eq:Gov1T} and \eqref{eq:Gov2T} at the macro-scale. For this matter, on the periodic domain $\mathfrak{L},$ let the power-like functional ${\Lambda}$ be expressed in terms of the energy-like density $\lambda_{m}$ at the micro-scale

\begin{align}
	{\Lambda}(\boldsymbol{u},\theta)&=\dfrac{\partial}{\partial t}\int_{\mathfrak{L}}\lambda_{m}\left(\boldsymbol{x},\frac{\boldsymbol{x}}{\varepsilon}\right) d\mathbf{x}=\nonumber\\
	&=\dfrac{\partial}{\partial t}\int_{\mathfrak{L}}\left( \dfrac{1}{2} \rho^{m} \dot{\boldsymbol{u}}\ast \dot{\boldsymbol{u}} +\dfrac{1}{2}\nabla \boldsymbol{u} \ast (\mathfrak{C}^{m}\nabla \boldsymbol{u})-\dfrac{1}{2}
	\nabla\boldsymbol{u}\ast(\boldsymbol\alpha^{m}\theta)-\boldsymbol{u}\ast\boldsymbol{b}\right)d\mathbf{x}\quad+\\
	&-\int_{\mathfrak{L}}\left( \dfrac{1}{2}\nabla\theta\ast\left(\mathcal{L}^{-1}\left\{\mathbf{K}^{m}\right\}\ast\nabla\theta\right)+\dfrac{1}{2}		\theta\ast\left(\boldsymbol\alpha^{m}\nabla\dot{\boldsymbol{u}}\right)+\dfrac{1}{2}\theta\ast(p^m\dot\theta)-\theta\ast r\right)d\mathbf{x}\,.\nonumber
\end{align} 
Now, applying the Laplace transform on $\mathcal{L}({\Lambda}),$ the functional above is taken to the Laplace domain and expressed  in terms of the power-like density $\hat{\lambda}_{m}$ as follows

\begin{align}
	\label{functionaltransformed}
	\mathcal{L}\{{\Lambda}\}&=\hat{\Lambda}(\hat{\boldsymbol{u}},\hat\theta)=\int_{\mathfrak{L}}\hat{\lambda}_{m}\left(\boldsymbol{x},\frac{\boldsymbol{x}}{\varepsilon}\right) d\mathbf{x}=\nonumber \\
	&=\int_{\mathfrak{L}}s\left(\dfrac{1}{2}\rho^{m}s^2\hat{\boldsymbol{u}}\cdot\hat{\boldsymbol{u}}+\dfrac{1}{2}\nabla\hat{\boldsymbol{u}}:(\mathfrak{C}^{m}\nabla\hat{\boldsymbol{u}})-\dfrac{1}{2}\nabla\hat{\boldsymbol{u}}:(\boldsymbol\alpha^{m}\hat\theta)-\hat{\boldsymbol{u}}\cdot\hat{\boldsymbol{b}}\right)d\mathbf{x}\quad+\\
	&-\int_{\mathfrak{L}}\left(\dfrac{1}{2}\nabla\hat\theta\cdot\left(\mathbf{K}^{m}\nabla\hat\theta\right)+\dfrac{1}{2}s\hat\theta\left(\boldsymbol\alpha^{m}:\nabla\hat{\boldsymbol{u}}\right)+\dfrac{1}{2}s\hat\theta(p^m\hat\theta)-\hat\theta\hat{r}\right)d\mathbf{x}\,,\nonumber
\end{align}
where the symbol $:$ indicates the tensor double inner product. Note that the Euler-Lagrange equation from the functional \eqref{functionaltransformed} agrees with the demonstration presented in the Sec. \ref{Euler-Lagrangepower-likefunctional} which leads to the field equations at the micro-scale $ (\ref{eq:Gov1T})$ and $(\ref{eq:Gov2T}).$

As it was introduced at the equations \eqref{UpScalingLawu} and \eqref{UpScalingLawtheta}, the transformed power-like functional $\hat{\Lambda}$ and its corresponding power-like density $\hat{\lambda}_{m}$ are dependent by the translation parameter $\boldsymbol{\zeta}\in\mathcal{Q}.$ Which means, the perturbation functions $N_{hpq_1}^{(1,0)}, \tilde{N}_{h}^{(1,0)},N_{hpq_1q_2}^{(2,0)},\tilde{N}^{(2,0)}_{hq_1},N_{hp}^{(2,2)},M^{(1,0)}_{q_1},$ $M^{(2,0)}_{q_1q_2},$ $\tilde{M}^{(2,1)}_{pq_1},$ $M^{(2,1)},$ determined from the cell problems in Sec. \ref{Sec::CellsProblems}, also depend on the translation variable $\boldsymbol{\zeta}.$ Besides, the power-like density $\hat{\lambda}_{m}$ in the Laplace domain complies with the property 

\begin{equation}
	\hat{\lambda}^{\zeta}_{m}\left(\textbf{x},\dfrac{\textbf{x}}{\varepsilon}\right)=\hat{\lambda}_{m}\left(\textbf{x},\frac{\textbf{x}}{\varepsilon}+\boldsymbol{\zeta}\right),
\end{equation}
implying that the Laplace transformed of the power-like functional $\hat\Lambda,$ depends on the parameter $\boldsymbol\zeta,$ one has

\begin{flalign}
	\label{beli}
	&\hat{\Lambda}^{\zeta}=\hat{\Lambda}(\boldsymbol{\zeta})=\int_{\mathfrak{L}}\hat{\lambda}^{\zeta}_{m}\left(\textbf{x},\dfrac{\textbf{x}}{\varepsilon}\right)d\textbf{x}=\int_{\mathfrak{L}}\hat{\lambda}_{m}\left(\textbf{x},\frac{\textbf{x}}{\varepsilon}+\boldsymbol{\zeta}\right)d\textbf{x}.
\end{flalign}

Let $\hat{\Lambda}_{m}$ be the average transformed power-like functional at the micro-scale
\begin{flalign}
	\begin{split}
		\label{mediafuncional}
		&\hat{\Lambda}_{m}\dot{=}\langle\hat{\Lambda}^{\zeta}\rangle=\dfrac{1}{|\mathcal{Q}|}\int_{\mathcal{Q}}\left[\int_{\mathfrak{L}}\hat{\lambda}_{m}\left(\textbf{x},\frac{\textbf{x}}{\varepsilon}+\boldsymbol{\zeta}\right)d\textbf{x}\right]d\boldsymbol{\zeta}=\\
		&=\int_{\mathfrak{L}}\left[\dfrac{1}{|\mathcal{Q}|}\int_{\mathcal{Q}}\hat{\Lambda}(\boldsymbol{\zeta})d\boldsymbol{\zeta}\right]d\textbf{x} = \int_{\mathfrak{L}}\left\langle\hat{\lambda}_{m}\left(\textbf{x},\dfrac{\textbf{x}}{\varepsilon}\right)\right\rangle d\textbf{x},
	\end{split}
\end{flalign} 
where the Fubini's theorem was applied in the third equality. 

Since the power-like functional $\hat{\Lambda}^{\zeta}$ is averaged by the translated realizations of the microstructure, which entails that the transformed power-like density at the micro-scale satisfies  

\begin{flalign}
	\label{sou}
	&\Big \langle \hat{\lambda}_{m}\Big ( \boldsymbol{x},\frac{\boldsymbol{x}}{\varepsilon}+\boldsymbol{\zeta}\Big ) \Big \rangle = \frac{1}{|\mathcal{Q}|}\int_{\mathcal{Q}} \hat{\lambda}_{m}\Big ( \boldsymbol{x},\frac{\boldsymbol{x}}{\varepsilon}+\boldsymbol{\zeta}\Big ) d\boldsymbol{\zeta}=\frac{1}{|\mathcal{Q}|}\int_{\mathcal{Q}} \hat{\lambda}_{m}\Big ( \boldsymbol{x},\boldsymbol{\xi}\Big ) d\boldsymbol{\xi}=\langle \hat{\lambda}_{m}(\boldsymbol{x},\boldsymbol{\xi})\rangle,
\end{flalign}
consequently, this ensures that the average transformed power-like functional $\langle \hat{\Lambda}^{\zeta} \rangle$ at the micro-scale does not rely on the translation variable $\boldsymbol\zeta.$

Through a similar variational approach made in the Sec. \ref{Euler-Lagrangepower-likefunctional}, we will determine the governing field equations at the macro-scale and the overall constitutive tensors, finding the Euler-Lagrange equation of the power-like function from the down-scaling relations \eqref{downscalingu} and \eqref{downscalingtheta}. So, let us once again introduce the down-scaling relations related to the transformed micro-displacement $\boldsymbol{\hat{u}}(\textbf{x},\boldsymbol{\xi},s),$ and to the transformed micro-temperature $ \hat\theta(\textbf{x},\boldsymbol{\xi},s) $, i.e.

\begin{subequations}
	\begin{flalign}
		&\hat{u}_{h}\left(\mathbf{x},\frac{\mathbf{x}}{\varepsilon},s\right)=\hat{U}^{M}_{h}(\mathbf{x},s)
		+\varepsilon\left(N^{(1,0)}_{hpq_{1}}(\boldsymbol{\xi})\dfrac{\partial \hat{U}^{M}_{h}}{\partial x_{q_{1}}}+\tilde{N}^{(1,0)}_{h}(\boldsymbol{\xi})\hat{\Theta}^M\right)+\mathcal{O}(\varepsilon^{2}),\label{downscalingu1order}\\
		&\hat{\theta}\left(\mathbf{x},\frac{\mathbf{x}}{\varepsilon},s\right)=\hat{\Theta}^{M}(\mathbf{x},s)+\varepsilon\left(M^{(1,0)}_{q_{1}}(\boldsymbol{\xi})\dfrac{\partial \hat{\Theta}^{M}}{\partial x_{q_{1}}}\right)+\mathcal{O}(\varepsilon^{2})\label{downscalingtheta1order},
	\end{flalign}
\end{subequations}
and  applying the gradients \eqref{eqn:utotalderivative} and \eqref{eqn:thetatotalderivative} to the down-scaling relations above, one gives

\begin{flalign}
	\dfrac{D \hat{u}_{h}}{D x_{k}}&=\dfrac{\partial\hat{U}^{M}_{h}}{\partial x_{k}}+N^{(1,0)}_{hpq_{1},k}\dfrac{\partial\hat{U}^{M}_{p}}{\partial x_{q_{1}}}+\tilde{N}^{(1,0)}_{h,k}\hat\Theta^M+ \varepsilon \left(N^{(1,0)}_{hpq_{1}}\dfrac{\partial^{2}\hat{U}^{M}_{p}}{\partial x_{q_{1}}\partial x_{k}}+\tilde{N}^{(1,0)}_{h}\dfrac{\partial\hat\Theta^{M}}{\partial x_k}\right)+\mathcal{O}(\varepsilon^{2})=\nonumber\\
	&=B^{(1,0)}_{hkpq_{1}}\dfrac{\partial\hat{U}^{M}_{h}}{\partial x_{q_{1}}}+\tilde{B}^{(1,0)}_{hk}\hat\Theta^M+\mathcal{O}(\varepsilon),\label{downscalingu1orderderivative}
\end{flalign}
where $ B^{(1,0)}_{hkpq_{1}}=\delta_{hp}\delta_{kq_1}+N^{(1,0)}_{hpq_{1},k} $ and $ \tilde{B}^{(1,0)}_{hk}=\tilde{N}^{(1,0)}_{h,k}, $ and

\begin{flalign}
	\dfrac{D \hat{\theta}_{h}}{D x_{j}}&=\dfrac{\partial\hat{\Theta}^{M}}{\partial x_{j}}+M^{(1,0)}_{q_{1},j}\dfrac{\partial\hat\Theta^{M}}{\partial x_{q_{1}}}+ \varepsilon M^{(1,0)}_{q_{1}}\dfrac{\partial^{2}\hat\Theta^{M}}{\partial x_{q_{1}}\partial x_{j}}+\mathcal{O}(\varepsilon^{2})=\nonumber\\
	&=A^{(1,0)}_{jq_{1}}\dfrac{\partial\hat{\Theta}^{M}}{\partial x_{q_{1}}}+\mathcal{O}(\varepsilon),\label{downscalingtheta1orderderivative}
\end{flalign}
where $ A^{(1,0)}_{jq_{1}}=\delta_{jq_1}+M^{(1,0)}_{q_{1},j}. $ The tensors $B^{(1,0)}_{hkpq_{1}},$ $\tilde{B}^{(1,0)}_{hk}$ and $ A^{(1,0)}_{jq_{1}} $ are called localization tensors and are also $\mathcal{Q}$-periodic functions in relation to the fast variable $\boldsymbol{\xi},$ once the perturbation functions and their gradients are $\mathcal{Q}$-periodic functions.

As previously mentioned, such first variation must vanish for all admissible $ \delta\hat{U}^{M}_{h}, $ $ \delta\hat\Theta^M $ i.e,\\ $ \delta\hat{\Lambda}_{m}(\hat{U}^{M}_{h},\delta\hat{U}^{M}_{h})=0, $ and hence the Euler-Lagrange differential equation associated to the variational problem \eqref{firstvariation} leads to the following two equations

\begin{subequations}
	\begin{align}
		\label{eulerlagrange1}
		&s^{2}\langle\rho^{m}\rangle\hat{U}^{M}_{l}-\left\langle B^{(1,0)}_{ijlr_{1}}C^{m}_{ijhk}B^{(1,0)}_{hkpq_{1}}\right\rangle\dfrac{\partial^2\hat{U}^{M}_{p}}{\partial x_{q_{1}}\partial x_{r_{1}}}
		+\left\langle B^{(1,0)}_{ijlq_{1}}\alpha^{m}_{ij}-B^{(1,0)}_{hklq_{1}}C^{m}_{ijhk}\tilde{B}^{(1,0)}_{ij}\right\rangle\dfrac{\partial\hat{\Theta}^{M}}{\partial x_{q_{1}}}-\hat{b}_l=0,\\
		\begin{split}\label{eulerlagrange2}
			&\left\langle A^{(1,0)}_{ir_{1}}K^{m}_{ij}A^{(1,0)}_{jq_{1}}\right\rangle\dfrac{\partial^2\hat{\Theta}^{M}}{\partial x_{r_{1}}x_{q_{1}}}	
			+s\left(\left\langle B^{(1,0)}_{hkpq_{1}}C^{m}_{ijhk}\tilde{B}^{(1,0)}_{ij}\right\rangle\dfrac{\partial\hat{U}^{M}_{l}}{\partial x_{q_{1}}}-\left\langle B^{(1,0)}_{ijpq_{1}}\alpha^{m}_{ij}\right\rangle\dfrac{\partial\hat{U}^{M}_{l}}{\partial x_{q_{1}}}\right)\quad+\\
			&-s\langle p^m+2\tilde{B}^{(1,0)}_{ij}\alpha^{m}_{ij}+\tilde{B}^{(1,0)}_{ij}C^{m}_{ijhk}\tilde{B}^{(1,0)}_{hk}\rangle\hat\Theta^M+\hat{r}=0,
		\end{split}
	\end{align}
\end{subequations}
which are clearly developed in terms of the Laplace transformed macro-temperature and macro-displacement among with their gradients. Since the equations \eqref{eulerlagrange1} and \eqref{eulerlagrange2} are set by positive definite constitutive tensors, one implies that the existence and the uniqueness of their solutions are endorsed by the Legendre-Hadamard condition.

Rearranging and rewriting the equations \eqref{eulerlagrange1} and \eqref{eulerlagrange2} in terms of the constitutive tensors, one gives

\begin{subequations}
	\begin{align}
		&C_{lr_{1}pq_{1}}\dfrac{\partial^2\hat{U}^{M}_{p}}{\partial x_{q_{1}}\partial x_{r_{1}}}-\alpha_{lq_{1}}\dfrac{\partial\hat{\Theta}^{M}}{\partial x_{q_{1}}}+\hat{b}_l=\rho s^{2}\hat{U}^{M}_{l},\label{eulerlagrange1re}\\
		&K_{r_{1}q_{1}}\dfrac{\partial^2\hat{\Theta}^{M}}{\partial x_{r_{1}}x_{q_{1}}}-s\alpha_{lq_{1}}\dfrac{\partial\hat{U}^{M}_{l}}{\partial x_{q_{1}}}+\hat{r}= ps\hat\Theta^M,\label{eulerlagrange2re}
	\end{align}
\end{subequations}
where the overall constitutive tensors are defined as

\begin{subequations}
	\begin{eqnarray}
		\label{C}
		&&
		C_{lr_{1}pq_{1}}=\left\langle B^{(1,0)}_{ijlr_{1}}C^{m}_{ijhk}B^{(1,0)}_{hkpq_{1}}\right\rangle,\\
		\label{K}
		&&
		K_{r_{1}q_{1}}=\left\langle A^{(1,0)}_{ir_{1}}K^{m}_{ij}A^{(1,0)}_{jq_{1}}\right\rangle,\\
		\label{alpha}
		&&
		\alpha_{lq_{1}}=\left\langle B^{(1,0)}_{ijlq_{1}}\alpha^{m}_{ij}-B^{(1,0)}_{hklq_{1}}C^{m}_{ijhk}\tilde{B}^{(1,0)}_{ij}\right\rangle,\\
		&&
		\label{p}
		p=\langle p^m+2\tilde{B}^{(1,0)}_{ij}\alpha^{m}_{ij}+\tilde{B}^{(1,0)}_{ij}C^{m}_{ijhk}\tilde{B}^{(1,0)}_{hk}\rangle,\\
		&&
		\rho=\langle\rho^{m}\rangle.\label{rho}
	\end{eqnarray}
\end{subequations}

Note that, at this stage, the equations \eqref{eulerlagrange1re} and \eqref{eulerlagrange2re} are the macro-scale equivalent to the field equations \eqref{eq:Gov1T} and \eqref{eq:Gov2T} on the transformed Laplace space. Moreover, by applying the inverse Laplace transform \eqref{lapinverse} to them, and recalling the identity \eqref{eqn:convK}, the field equations at the macro-scale corresponding to the equations \eqref{eq:gov1} and \eqref{eq:gov2} are recast on the time domain as

\begin{subequations}
	\begin{align}
		&C_{lr_{1}pq_{1}}\dfrac{\partial^2\ddot{U}^{M}_{p}}{\partial x_{q_{1}}\partial x_{r_{1}}}-\alpha_{lq_{1}}\dfrac{\partial\dot{\Theta}^{M}}{\partial x_{q_{1}}}+\hat{b}_l=\rho \ddot{U}^{M}_{l},\label{eulerlagrange1macrotempo}\\
		&\mathcal{L}^{-1}\left\{K_{r_{1}q_{1}}\right\}\ast\dfrac{\partial^2{\Theta}^{M}}{\partial x_{r_{1}}x_{q_{1}}}-\alpha_{lq_{1}}\dfrac{\partial\dot{U}^{M}_{l}}{\partial x_{q_{1}}}+r= p\dot\Theta^M.\label{eulerlagrange2macrotempo}
	\end{align}
\end{subequations}

\section{Wave propagation in homogenized continuum}
\label{waveprophomogenized}

In this Section, we exploit the wave propagation along the homogenized continuum, which is equivalent to the started thermoelastic periodic material. Such procedure consists in applying the Fourier transform to equations \eqref{eulerlagrange1re} and \eqref{eulerlagrange2re}, with respect to the macroscopic variable $\textbf{x}\in\R^3,$ in order to derive these field equations at the macro-scale defined over the complex frequency $ s ,$ and also defined over the wave vector $ \boldsymbol{k}=(k_1,k_2,k_3)^T\in\C^3$ ($ k_1, k_2,k_3 $ are the wave numbers and $ T $ identifies transpose vector), in other terms, the thermoelastic field equations \eqref{eq:gov1} and \eqref{eq:gov2}, initially defined on the space-time domain $ (\textbf{x},t)\in\R^3\times\R ,$ will be taken to the wave vector-frequency domain $ (\boldsymbol{k},s)\in\C^3\times\C $ \citep{laude2015phononic}. 

Proceeding the same way as made in Sec. \ref{section2}, we begin defining the space Fourier transform and its properties that will be used throughout the work \citep{paley1934fourier}. Let $ f:\R^3\rightarrow\R $ be an arbitrary function defined over the domain $ \textbf{x}\in\R^3,$ the complex space Fourier transform is defined as,

\begin{equation}
	\label{transformadafourier}
	\mathcal{F}(f(\textbf{x}))= \check{f}(\boldsymbol{k}) = \int_{\R^3}f(\textbf{x})\mathrm{e}^{-\mathrm{i}\boldsymbol{k}\cdot\textbf{x}}\mathrm{d}\textbf{x},\quad \boldsymbol{k}\in \C^{3},
\end{equation} 
where $\check{f}:\C^3\to\C. $ 

The Fourier transform of the derivative of order $ n+m$ of  $ f $ with respect to $ x^{n}_j$ and $ x^{m}_r$ is given by the identity$ \mathcal{F}\left\{\partial^{n+m}f(\textbf{x})\partial x^{n}_j\partial x^{m}_r\right\}=(\mathrm{i}{k}_j)^{n}(\mathrm{i}{k}_r)^{m}\check{f}(\boldsymbol{k}). $

Transforming the macro-scale field equations \eqref{eulerlagrange1re} and \eqref{eulerlagrange2re} into the Fourier transformed space, one has

\begin{subequations}
	\begin{align}
		&C_{lr_{1}pq_{1}}(\mathrm{i}{k}_{q_{1}})(\mathrm{i}{k}_{r_{1}})\check{\hat{U}}^{M}_{p}-\alpha_{lq_{1}}(\mathrm{i}{k}_{q_{1}})\check{\hat{\Theta}}^{M}+\check{\hat{b}}_l=\rho s^{2}\check{\hat{U}}^{M}_{l},\label{eulerlagrangefourier1re}\\
		&K_{r_{1}q_{1}}(\mathrm{i}{k}_{q_{1}})(\mathrm{i}{k}_{r_{1}})\check{\hat{\Theta}}^{M}-s\alpha_{lq_{1}}(\mathrm{i}{k}_{q_{1}})\check{\hat{U}}^{M}_l+\check{\hat{r}}= ps\check{\hat{\Theta}}^M.\label{eulerlagrangefourier2re}
	\end{align}
\end{subequations}

In case of free wave propagation, we take the external influences (source terms) being zero, such as  the heat source equivalent  $ \check{\hat{r}} ,$ and the body force equivalent  $ \check{\hat{\boldsymbol{b}}}, $ so the equations above might be rephrased to the compact form as

\begin{subequations}
	\begin{align}
		&\left[\bar{\mathfrak{C}}(\boldsymbol{k}\otimes\boldsymbol{k})+\rho s^{2}\boldsymbol{I}\right]\cdot\check{\hat{\boldsymbol{U}}}^{M}+\mathrm{i}(\boldsymbol\alpha\boldsymbol{k})\check{\hat{\Theta}}^{M}=\boldsymbol{0},\label{sistema1}\\
		&\left[\mathbf{K}:(\boldsymbol{k}\otimes\boldsymbol{k})+ps\right]\check{\hat{\Theta}}^{M}+\mathrm{i}s(\boldsymbol\alpha\boldsymbol{k})\cdot\check{\hat{\boldsymbol{U}}}^{M}= 0,\label{sistema2}
	\end{align}
\end{subequations}
where the tensor $ \bar{\mathfrak{C}} $ corresponds to the shift $ \bar{\mathfrak{C}}=\bar{C}_{lpr_{1}q_{1}}\textbf{e}_l\otimes\textbf{e}_p\otimes\textbf{e}_{r_{1}}\otimes\textbf{e}_{q_{1}},$ with $ \bar{C}_{lpr_{1}q_{1}}=C_{lpr_{1}q_{1}}. $ Note that, the system of linear equations formed by \eqref{sistema1} and \eqref{sistema2} may be rewritten in a matricial form as 

\begin{equation}
	\label{matriz}
	\left[\begin{array}{cc}
		\bar{\mathfrak{C}}(\boldsymbol{k}\otimes\boldsymbol{k})+\rho s^{2}\boldsymbol{I} & \mathrm{i}\boldsymbol{k}\boldsymbol\alpha\\
		\mathrm{i}s\boldsymbol{k}\boldsymbol\alpha                                               & \mathbf{K}:(\boldsymbol{k}\otimes\boldsymbol{k})+ps
	\end{array}\right]
	\left[\begin{array}{c}
		\check{\hat{\boldsymbol{U}}}^{M}\\\check{\hat{\Theta}}^M
	\end{array}\right]
	=
	\left[\begin{array}{c}
		\boldsymbol{0}\\0
	\end{array}\right].
\end{equation}

The matricial system \eqref{matriz}, is an eigenvector-eigenvalue problem that provides the frequency spectrum as an implicit function of the wave vector $ \boldsymbol{k}$ and the complex frequency $ s ,$ which is given by the frequency equation $F(\boldsymbol{k},s)=\textrm{det}([\boldsymbol{H}(\boldsymbol{k},s)])=0 ,$ also known as implicit dispersion relation, where the matrices in  the system \eqref{matriz} are named as $ [\boldsymbol{H}(\boldsymbol{k},s)][\boldsymbol{V}]=[\boldsymbol{0}] .$ Since the implicit function $ F(\boldsymbol{k},s)=0 $ is a complex function, it can be rephrased in terms of its real part and imaginary part as $ F(\boldsymbol{k},s)=\Re e(F(\boldsymbol{k},s))+\mathrm{i}\Im m(F(\boldsymbol{k},s))=0, $ or simply 

\begin{equation}
	\label{hypersurface}
	\left\{\begin{array}{lcr}
		\Re e(F(\boldsymbol{k},s)) & = & 0\\
		\Im m(F(\boldsymbol{k},s)) & = & 0
	\end{array}\right.,
\end{equation}
which represents the collection of branches (frequency spectrum) by the hyper-surfaces $ \Re e(F(\boldsymbol{k},s)) $ and $ \Im m(F(\boldsymbol{k},s)) $ into the space $(\boldsymbol{k},s)\in\C^4,$ such hyper-surface $ F(\boldsymbol{k},s) $ is also seen in mathematics as a level set of the complexed-valued function $ F $. As \cite{laude2015phononic} quotes, from a mathematical standpoint, there is no reason to choose real over complex values for either the frequency or the wave vector. What matters is actually the physical meaning that might be associated to them. Having described the implicit dispersion function, from this stage on we emphasize two possible lanes that might provide us the dispersion spectrum, and finally analyse the wave propagation either with spatial damping or with the damping in time, \cite{carcione2007wave}.

In the first case, let us consider the complex angular frequency as $ s=\mathrm{i}\omega $ with $ \omega\in\R $ in the implicit dispersion relation \eqref{hypersurface}. Thus the dispersion relation dependence goes over to the complex wave vector $\boldsymbol{k}$ and the angular frequency $\omega.$ Therefore, the dispersion relation associated to generic inhomogeneous harmonic waves holds

\begin{equation}
	\label{surfacedispersives(k)}
	\left\{\begin{array}{lcr}
		\Re e(F(\Re e(\boldsymbol{k} ),\Im m(\boldsymbol{k} ),\omega)) & = & 0\\
		\Im m(F(\Re e(\boldsymbol{k} ),\Im m(\boldsymbol{k} ),\omega)) & = & 0
	\end{array}\right.,
\end{equation}
where the intersection between these two hyper-surfaces defines the frequency spectrum of the material as an hyper-curve immersed in $ \R^7. $ More specifically, such hyper-curve describes the relation between the angular frequency $ \omega $ and the complex vector $ \boldsymbol{k} ,$ which can be expressed as $\boldsymbol{k}=\Re e(\boldsymbol{k})+\mathrm{i}\Im m(\boldsymbol{k})=\|\Re e(\boldsymbol{k})\|\boldsymbol{n}_{r}+\mathrm{i}\|\Im m(\boldsymbol{k})\|\boldsymbol{n}_{i},$ where $ \boldsymbol{n}_{r},$ $ \boldsymbol{n}_{i} $ are versors (i.e $\|\boldsymbol{n}_{r}\|=\|\boldsymbol{n}_{i}\|=1 $ and $ \boldsymbol{n}_{r},\boldsymbol{n}_{i}\in\R^3 $), representing the direction of the normal to planes of constant phase and planes of constant amplitude of the propagating wave, respectively. Moreover, in the particular case of the homogeneous harmonic waves characterized by $ \boldsymbol{n}_{r}=\boldsymbol{n}_{i}=\boldsymbol{n} $ and for the complex wave expressed in the form $ \boldsymbol{k}=(\|\Re e(\boldsymbol{k})\|+i\|\Im m(\boldsymbol{k})\|)\boldsymbol{n}=\kappa\boldsymbol{n}, $ the dispersion relation is obtained by specializing \eqref{surfacedispersives(k)} in the form






\begin{equation}
	\label{surfacedispersiveheterogeneous}
	\left\{\begin{array}{lcr}
		\Re e(F(\Re e(\kappa),\Im m(\kappa),\omega)) & = & 0\\
		\Im m(F(\Re e(\kappa),\Im m(\kappa),\omega)) & = & 0
	\end{array}\right.,
\end{equation}
a three dimensional hyper-surface whose intersection is a curve immersed in $ \R^3. $

Nonetheless, in the second case of wave propagation with attenuation in time, let us assume in the set of equations \eqref{hypersurface} the complex frequency being as $ s=\Re e(s)+\mathrm{i}\Im m(s),$ the wave vector such that $ \boldsymbol{k}\in\R^3,$ and let us also fix a direction of the wave vector $ \boldsymbol{k}=\kappa\boldsymbol{n},$ where $ \boldsymbol{n} $ is its versor and $ \kappa=||\boldsymbol{k}||\in\R ,$ thus \eqref{hypersurface} becomes

\begin{equation}
	\label{surfacedispersive}
	\left\{\begin{array}{lcr}
		\Re e(F(\kappa,\Re e(s),\Im m(s))) & = & 0\\
		\Im m(F(\kappa,\Re e(s),\Im m(s))) & = & 0
	\end{array}\right.,
\end{equation}
and hence taking the intersection between the two hyper-surfaces from \eqref{surfacedispersive}, one yields a hyper-curve $ s(\kappa)$-curves immersed in $ \R^3 ,$ given a certain propagation direction $ \boldsymbol{n}. $ 

\section{Illustrative example: homogenization procedure on a bi-phase orthotropic layered material}
\label{exemplos}

In order to contrast the results found by the analysis of the heterogeneous approach stressed in the Sec. \ref{Wavepropagationheterogeneous} of the Supplementary Material, the homogenized formulation obtained in the Sec. \ref{exemplos} is now applied for a layered two-dimensional infinite thermoelastic body with orthotropic phases having the orthotropy axes parallel to the layering direction $ \boldsymbol{e_1}.$ Analytical precise formulae for the overall elastic, thermal dilatation and thermal conduction tensors are determined. Subsequently, we compare the results of the heterogeneous model developed in the Sec. \ref{Wavepropagationheterogeneous}, against those from the model obtained proceeding with the first order homogenization technique, both under the same hypothesis, by finding their dispersions spectrum seen in the Sec. \ref{waveprophomogenized}.

\subsection{Perturbation functions and overall constitutive tensors}
\label{homogenizedprocess}

\begin{figure}[h!]
	\centering
	\includegraphics[scale=0.83]{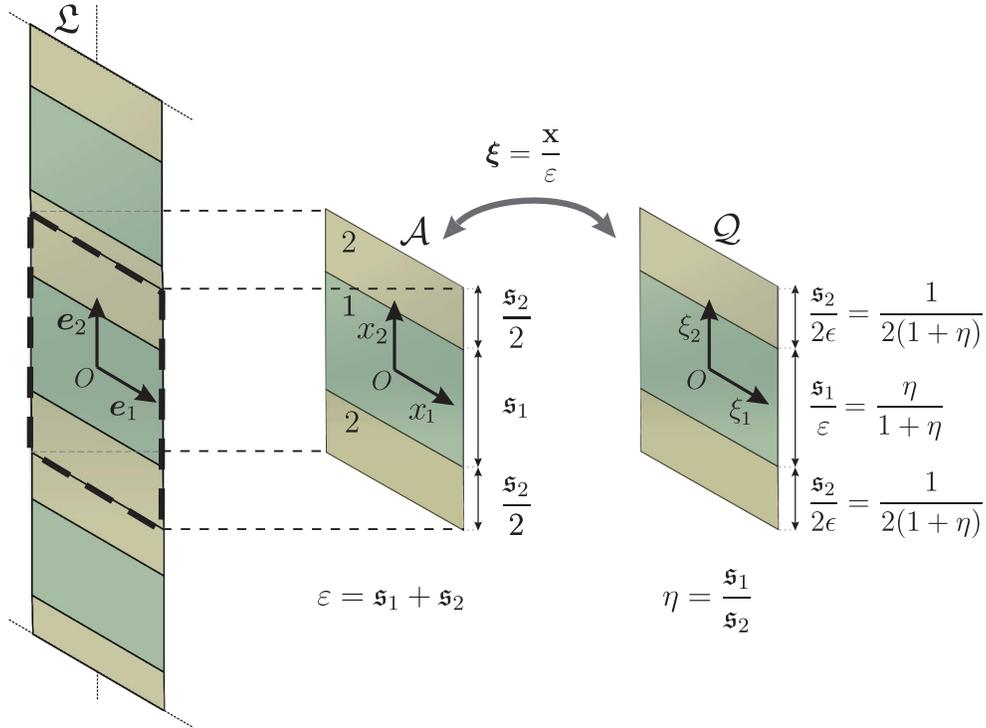}
	\caption{Heterogeneous bi-dimensional domain bi-phase layered periodic cell.}
	\label{fig:layeredmaterial2}
\end{figure}

Let us assume a layered body composed by two phases where the first order homogenization process will be applied on the system. As may seen in Fig. \ref{fig:layeredmaterial2}, let the domain $ \mathcal{A} $ made by two layered materials be defined having thickness $\mathfrak{s}_{1}$ and $\mathfrak{s}_{2},$ where $ d_2=\varepsilon=\mathfrak{s}_{1}+\mathfrak{s}_{2} ,$ and $ \eta=\mathfrak{s}_{1}/\mathfrak{s}_{2}.$ Also the domain displays orthotropic phases, which it coincides with the axis that determines the layering direction $\boldsymbol{e}_{1}.$ The perturbation functions at the order $ \varepsilon^{-1} ,$  $N_{hpq_1}^{(1,0)}$ $\tilde{N}_{h}^{(1,0)},$ $M_{q_1}^{(1,0)}$ are analytically obtained by the solutions of the three cell problems presented in \eqref{eq:CellProblemOrder-1} formulated in Sec. \ref{exemplos}, along with the interface conditions in \eqref{eqns:InterfaceConditionsCellProblemOrder-1}. Due to the microstructure symmetry, these functions rely exclusively on the microscopic (fast) component $ \xi_2,$ which is perpendicular to the layering direction $\boldsymbol{e}_{1}.$ Considering only non-zero perturbations functions, they are given by

\begin{subequations}
	\begin{align}
		N_{112}^{(1,0)_1}&=N_{121}^{(1,0)_1}=\dfrac{\left(C_{1212}^{_{(2)}}-C_{1212}^{_{(1)}}\right)\xi_2^{_{(1)}}}{C_{1212}^{_{(2)}}\eta+C_{1212}^{_{(1)}}}, 
		&N_{112}^{(1,0)_2}&=N_{121}^{(1,0)_2}=\dfrac{\eta\left(C_{1212}^{_{(2)}}-C_{1212}^{_{(1)}}\right)\xi_2^{_{(2)}}}{C_{1212}^{_{(2)}}\eta+C_{1212}^{_{(1)}}},\label{N112}\\
		N_{211}^{(1,0)_1}&=\dfrac{\left(C_{1122}^{_{(2)}}-C_{1122}^{_{(1)}}\right)\xi_2^{_{(1)}}}{C_{2222}^{_{(2)}}\eta+C_{2222}^{_{(1)}}},	&N_{211}^{(1,0)_2}&=\dfrac{\eta\left(C_{1122}^{_{(2)}}-C_{1122}^{_{(1)}}\right)\xi_2^{_{(2)}}}{C_{2222}^{_{(2)}}\eta+C_{2222}^{_{(1)}}},\label{N211}\\
		N_{222}^{(1,0)_1}&=\dfrac{\left(C_{2222}^{_{(2)}}-C_{2222}^{_{(1)}}\right)\xi_2^{_{(1)}}}{C_{2222}^{_{(2)}}\eta+C_{2222}^{_{(1)}}},	&N_{222}^{(1,0)_2}&=\dfrac{\eta\left(C_{2222}^{_{(2)}}-C_{2222}^{_{(1)}}\right)\xi_2^{_{(2)}}}{C_{2222}^{_{(2)}}\eta+C_{2222}^{_{(1)}}},\label{N222}\\
		\tilde{N}_{2}^{(1,0)_1}&=\dfrac{\left(\alpha_{22}^{_{(1)}}-\alpha_{22}^{_{(2)}}\right)\xi_2^{_{(1)}}}{C_{2222}^{_{(2)}}\eta+C_{2222}^{_{(1)}}},	&\tilde{N}_{2}^{(1,0)_2}&=\dfrac{\eta\left(\alpha_{22}^{_{(2)}}-\alpha_{22}^{_{(1)}}\right)\xi_2^{_{(2)}}}{C_{2222}^{_{(2)}}\eta+C_{2222}^{_{(1)}}},\label{Ntilde}\\
		M_{2}^{(1,0)_1}&=\dfrac{\left(K_{22}^{_{(2)}}-K_{22}^{_{(1)}}\right)\xi_2^{_{(1)}}}{K_{22}^{_{(2)}}\eta+K_{22}^{_{(1)}}},
		&M_{2}^{(1,0)_2}&=\dfrac{\eta\left(K_{22}^{_{(1)}}-K_{22}^{_{(2)}}\right)\xi_2^{_{(2)}}}{K_{22}^{_{(2)}}\eta+K_{22}^{_{(1)}}},\label{perturbationM2xi2}	
	\end{align}
\end{subequations}
where the superscript $i=\{1,2\}$ denotes for the phase 1 and the phase 2, respectively. Moreover, the dimensionless vertical variables $ \xi_2^{_{(1)}} $ and $ \xi_2^{_{(2)}} $ centred in each layer are such that, $ \xi_2^{_{(1)}}\in\left[-\eta/2(\eta+1),\eta/2(\eta+1)\right]$ and $\xi_2^{_{(2)}}\in\left[-1/2(\eta+1),1/2(\eta+1)\right],$ agreeing with the Fig. \ref{fig:layeredmaterial2}.

\begin{figure}[h!]
	\centering
	\includegraphics[scale=0.34]{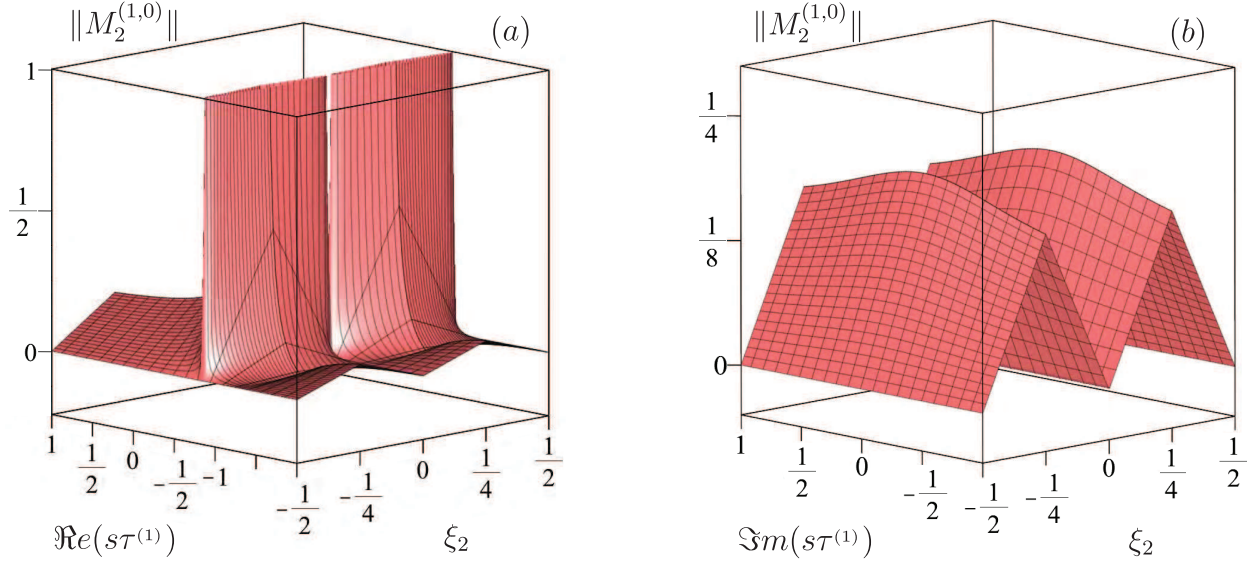}
	\caption{Dimensionless complex absolute value of the perturbation function $ M_2^{(1,0)}$ for the dimensionless parameters assumed $ \eta=1,$ $ \tau^{_{(2)}}/\tau^{_{(1)}}=3$ and $ \bar{K}_{22}^{_{(2)}}/\bar{K}_{22}^{_{(1)}}=10.$ $ (a) $ $ \|M_2^{(1,0)}\| $ vs. $ \Re e(s\tau^{_{(1)}}) $ and $ \xi_2 ;$ $ (b) $ $ \|M_2^{(1,0)}\| $ vs. $ \Im m(s\tau^{_{(1)}}) $ and $ \xi_2 .$ }
	\label{fig:|M2|}
\end{figure}

In particular, once the perturbation functions $ M_{q1}^{(1,0)} $ depend on the complex angular frequency $ s $ (see Eqs. \eqref{plugging} and \eqref{eqn:solutiontemp-1}), such dependency is evaluated, but dimensionless. As may be seen in Fig. \ref{fig:|M2|}, the complex absolute value of the dimensionless perturbation function, which is $\|M_2^{(1,0)}\|,$ is analytically computed by the equations in \eqref{perturbationM2xi2}, with respect to the phase $1$ and the phase $2,$ taking the following values for dimensionless parameters $ \eta=1,$ $ \tau^{_{(2)}}/\tau^{_{(1)}}=3$ and $ \bar{K}_{22}^{_{(2)}}/\bar{K}_{22}^{_{(1)}}=10.$ Such a function depends on the fast variable $\xi_{2},$ which is perpendicular to the transversal direction $\boldsymbol{e}_{1}$ and to the dimensionless complex parameter $s\tau^{_{(1)}}=\Re e(s\tau^{_{(1)}})+\mathrm{i}\Im m(s\tau^{_{(1)}}).$ It must be noted in Fig. \ref{fig:|M2|}$ (a) $ that the function $ \|M_2^{(1,0)}(\xi_{2},\Re e(s\tau^{_{(1)}}))\| $ presents a singularity along the variable $ \xi_{2}, $ 
whereas the function $ \|M_2^{(1,0)}(\xi_{2},\Im m(s\tau^{_{(1)}}))\| $ displays the absence of any sort of singularities regardless the variable.

As previously mentioned, we also derive the non-vanishing overall elastic, thermal dilatation and thermal conduction tensors corresponding to a first order equivalent continuum. Thus, from the perturbation functions \eqref{N112}, \eqref{N211} and \eqref{N222} are used into equation \eqref{C}, hence the overall elastic tensors $ C_{lr_{1}pq_{1}} $ are expressed as

\begin{subequations}\label{Coverall}
	\begin{align}
		&C_{1111} =\dfrac{\eta^2C_{1111}^{_{(1)}}C^{_{(2)}}_{2222} +\eta\left(C^{_{(2)}}_{1111}C^{_{(2)}}_{2222}+C^{_{(2)}}_{1111}C^{_{(1)}}_{2222}  -\left(C_{1122}^{_{(1)}}-C_{1122}^{_{(2)}}\right)^2\right)+C_{1111}^{_{(2)}}C^{_{(1)}}_{2222}}{(\eta+ 1)\left(C^{_{(1)}}_{2222} + \eta C^{_{(2)}}_{2222} \right)}, \\
		&C_{1122}=\dfrac{\eta C^{_{(1)}}_{1122}C^{_{(2)}}_{2222}+C^{_{(2)}}_{1122}C^{_{(1)}}_{2222}}{C^{_{(1)}}_{2222} + \eta C^{_{(2)}}_{2222}},\\
		&C_{1212}=\dfrac{(\eta+1)C^{_{(1)}}_{1212}C^{_{(2)}}_{1212}}{C^{_{(1)}}_{1212} + \eta C^{_{(2)}}_{1212}},\\
		&C_{2222} = \dfrac{(\eta+ 1) C^{_{(1)}}_{2222}C^{_{(2)}}_{2222}}{C^{_{(1)}}_{2222} + \eta C^{_{(2)}}_{2222}}. 
	\end{align}
\end{subequations}
Similarly, for the thermal dilatation components follows that from the perturbation functions \eqref{N112} to \eqref{Ntilde} into equations \eqref{C} and \eqref{alpha}, one provides

\begin{subequations}\label{alphaoverall}
	\begin{align}
		&\alpha_{11} = \dfrac{\eta^2 C^{_{(2)}}_{2222}\alpha^{_{(1)}}_{11} +\eta\left(C^{_{(1)}}_{2222}\alpha^{_{(1)}}_{11}+C^{_{(2)}}_{2222}\alpha^{_{(2)}}_{11}-\left(\alpha^{_{(1)}}_{22}-\alpha^{_{(2)}}_{22}\right)\left(C^{_{(1)}}_{1122}-C^{_{(2)}}_{1122}\right)\right)+C^{_{(1)}}_{2222}\alpha^{_{(2)}}_{11}}{(\eta+ 1)(C^{_{(1)}}_{2222} + \eta C^{_{(2)}}_{2222} )}, \\
		&\alpha_{22} =\dfrac{\eta C^{_{(2)}}_{2222}\alpha^{_{(1)}}_{22} +\alpha^{_{(2)}}_{22} C^{_{(1)}}_{2222} }{ \eta C^{_{(2)}}_{2222}+C^{_{(1)}}_{2222} } .
	\end{align}
\end{subequations}

\noindent The specific heat $ p $ and the mass density $ \rho $ in the equations \eqref{p} and \eqref{rho} respectively, are obtained following the same steps made latter for the elastic, and thermal dilatations tensors, giving

\begin{subequations}\label{prhooverall}
	\begin{align}
		&p =\dfrac{p^{_{(1)}}\eta^2C^{_{(2)}}_{2222}+\left(p^{_{(1)}}C_{2222}^{_{(1)}}+p^{_{(2)}}C_{2222}^{_{(2)}}+2\left(\alpha_{22}^{_{(1)}}-\alpha_{22}^{_{(2)}}\right)^2\right)\eta+p^{_{(2)}}C_{2222}^{_{(1)}}}{ \left(\eta C^{_{(2)}}_{2222}+C^{_{(1)}}_{2222}\right)\left(\eta+1\right) },\\
		&\rho = \dfrac{\eta\rho^{_{(1)}}+\rho^{_{(2)}}}{\eta+1}. 
	\end{align}
\end{subequations}

Still in this matter, from the perturbation functions \eqref{perturbationM2xi2} and equation \eqref{K}, the components of the overall thermal conduction tensor are

\begin{subequations}\label{Koverall}
	\begin{align}
		&K_{11}(s) = \dfrac{\eta K^{_{(1)}}_{11}(s) +K^{_{(2)}}_{11}(s)}{ \eta +1}, \\
		&K_{22}(s) =\dfrac{(\eta+1)K^{_{(1)}}_{22}(s)K^{_{(2)}}_{22}(s)}{\eta K^{_{(2)}}_{22}(s)+ K^{_{(1)}}_{22}(s) } .
	\end{align}
\end{subequations}
Consequently, from the Eq. \eqref{plugging}, the corresponding parts of the overall thermal conduction tensor namely $ K_{11}(s) $ and $ K_{22}(s), $ are also dependent on the complex angular frequency $ s,$ which ends up being rewritten in terms of the relaxation times $ \tau^{_(1)} ,$ $ \tau^{_(2)} ,$ the specific heat $T_0 $ and also the thermal conductivity tensors defined at the begining of the theory. After the respective substitutions of \eqref{plugging} to both equations in \eqref{Koverall}, the equivalent components for overall thermal conduction tensor are, respectively:

\begin{figure}[h!]
	\centering
	\includegraphics[scale=0.22]{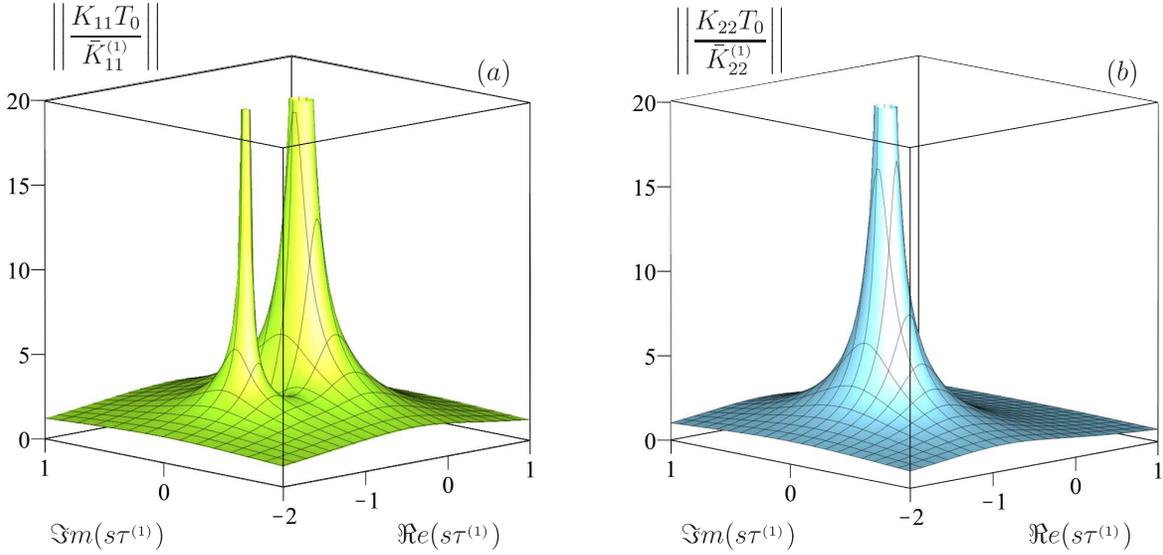}
	\caption{Dimensionless complex absolute value of the non-vanishing overall thermal conduction tensor components for the dimensionless parameters $ \eta=1,$ $ \tau^{_{(2)}}/\tau^{_{(1)}}=3$ and $ \bar{K}_{22}^{_{(2)}}/\bar{K}_{22}^{_{(1)}}=10.$ $ (a) $ $ \|{K}_{11}T_0/\bar{K}_{11}^{_{(1)}}\| $ vs. $ \Re e(s\tau^{_{(1)}})\times\Im m(s\tau^{_{(1)}}) ;$ $ (b) $ $ \|\bar{K}_{22}T_0/{K}_{22}^{_{(1)}}\| $ vs. $ \Re e(s\tau^{_{(1)}})\times\Im m(s\tau^{_{(1)}}) .$ }
	\label{fig:K11K22}
\end{figure}

\begin{subequations}\label{Ktoverall}
	\begin{align}
		&{K}_{11}(s)=\dfrac{\eta\bar{K}^{_{(1)}}_{11}(s\tau^{_{(2)}}+1)+\bar{K}^{_{(2)}}_{11}(s\tau^{_{(1)}}+1)}{T_0(\eta +1)(s\tau^{_{(1)}}+1)(s\tau^{_{(2)}}+1)}, \label{K11toverall}\\
		&{K}_{22}(s)=\dfrac{(\eta+1)\bar{K}^{_{(1)}}_{22}\bar{K}^{_{(2)}}_{22}}{T_0\left(\eta\bar{K}^{_{(2)}}_{22}(s\tau^{_{(1)}}+1)+\bar{K}^{_{(1)}}_{22}(s\tau^{_{(2)}}+1)\right)} .\label{K22toverall}
	\end{align}
\end{subequations}

The Fig. \ref{fig:K11K22} displays the complex absolute value of the dimensionless overall thermal conduction tensor components  $\|{K}_{11}T_0/\bar{K}_{11}^{_{(1)}}\|$ and $\|{K}_{22}T_0/\bar{K}_{22}^{_{(1)}}\|$ in terms of the dimensionless complex frequency $ (\Re e(s\tau^{_{(1)}}),\Im m(s\tau^{_{(1)}})), $ for the same dimensionless parameters $ \eta=1,$ $ \tau^{_{(2)}}/\tau^{_{(1)}}=3$ and $ \bar{K}_{22}^{_{(2)}}/\bar{K}_{22}^{_{(1)}}=10$ already assumed when it was analysed the perturbation function $ M_2^{(1,0)}. $ Still in the regard of the Fig. \ref{fig:K11K22}, note that the Eq. \eqref{K11toverall} presents a quadratic behaviour in its denominator and hence, as expected, it does display two poles, whereas the equation for $ K_{22} $, namely Eq. \eqref{K22toverall}, has a linear nature in its denominator, giving only one pole.

\subsection{Comparative analysis: heterogeneous material vs. homogenized solid}
\label{benchmark}

Under the same hypothesis assumed in the beginning of this Section, herein we proceed with a comparative analysis between the results obtained from the heterogeneous modelling procedure developed in the Sec. \ref{Wavepropagationheterogeneous} via Floquet-Bloch theory, and the homogenized first order model approached in the previous sections of the generalized thermoelastic problem with the derived overall elastic, thermal dilatation and thermal conduction constants.

For the considered two-dimensional body with the orthotropy axes perpendicular to the layering direction $ \boldsymbol{e_2}$ and the wave vector taken such as $ k_1=0 ,$ i.e the propagation axes is exclusively along the $ \boldsymbol{e_2}.$ For this scenario we consider an uncoupled problem which means that the dilatation tensor is assumed zero. Having said so, on the one hand by plugging overall constitutive tensors displayed in the set of Eqs. \eqref{Coverall}, \eqref{prhooverall} and \eqref{Ktoverall} into the homogenized specialized governing equations from \eqref{eulerlagrange1macrotempo}, \eqref{eulerlagrange2macrotempo} and following the procedure reported in Sec. \ref{waveprophomogenized}, the dispersive relation \eqref{surfacedispersive} (for $ \kappa=k_2 $) provides the explicit functions

\begin{equation}\label{}
	k_2(s)=\pm\dfrac{\mathrm{i}s\sqrt{C_{1212}^{_{(1)}}C_{1212}^{_{(2)}}\left(\eta\rho^{_{(1)}}+\rho^{_{(2)}}\right)\left(\eta C_{1212}^{_{(2)}}+C_{1212}^{_{(1)}}\right)}}{C_{1212}^{_{(1)}}C_{1212}^{_{(2)}}(\eta + 1)},
\end{equation}
or in its inverse form

\begin{equation}\label{shearwave}
	s(k_2)= \pm\dfrac{\mathrm{i}k_2(\eta+1)\sqrt{C_{1212}^{_{(1)}}C_{1212}^{_{(2)}}\left(\eta\rho^{_{(1)}}+\rho^{_{(2)}}\right)\left(\eta C_{1212}^{_{(2)}}+C_{1212}^{_{(1)}}\right)}}{\left(\eta\rho^{_{(1)}}+\rho^{_{(2)}}\right)\left(\eta C_{1212}^{_{(2)}}+C_{1212}^{_{(1)}}\right)},
\end{equation}
corresponding to the dispersion relation associated to the shear waves, giving rise to the standard non-dispersive behaviour. Similarly

\begin{equation}\label{}
	k_2(s)=\pm\dfrac{\mathrm{i}s\sqrt{C_{2222}^{_{(1)}}C_{2222}^{_{(2)}}\left(\eta\rho^{_{(1)}}+\rho^{_{(2)}}\right)\left(\eta C_{2222}^{_{(2)}}+C_{2222}^{_{(1)}}\right)}}{C_{2222}^{_{(1)}}C_{2222}^{_{(2)}}(\eta + 1)},
\end{equation}
with the inversion function as

\begin{equation}\label{compressionalwave}
	s(k_2)=\pm \dfrac{\mathrm{i}k_2(\eta+1)\sqrt{C_{2222}^{_{(1)}}C_{2222}^{_{(2)}}\left(\eta\rho^{_{(1)}}+\rho^{_{(2)}}\right)\left(\eta C_{2222}^{_{(2)}}+C_{2222}^{_{(1)}}\right)}}{\left(\eta\rho^{_{(1)}}+\rho^{_{(2)}}\right)\left(\eta C_{2222}^{_{(2)}}+C_{2222}^{_{(1)}}\right)},
\end{equation}
corresponding to the dispersion relation associated to the compressional waves, providing once again the standard non-dispersive behaviour. Finally, for dispersion relation associated to the thermal waves, 

\begin{equation}\label{thermalwavewavenumberk2}
	k_2(s)=\pm\dfrac{\mathrm{i}\sqrt{\mathscr{Z}_0\left[s^2\left(\eta\bar{K}_{22}^{_{(2)}}\tau^{_{(1)}}+\bar{K}_{22}^{_{(1)}}\tau^{_{(2)}}\right)+s\left(\eta\bar{K}_{22}^{_{(2)}}+\bar{K}_{22}^{_{(1)}}\right)\right]}}{\bar{K}^{_{(1)}}_{22}\bar{K}^{_{(2)}}_{22}(\eta+1)\left(\eta C_{2222}^{_{(2)}}+C_{2222}^{_{(1)}}\right)},
\end{equation}
where $\mathscr{Z}_0= T_0\left(\eta^2p^{_{(1)}}C_{2222}^{_{(2)}}+\eta\left(p^{_{(1)}}C_{2222}^{_{(1)}}+2\left(\alpha_{22}^{_{(1)}}-\alpha_{22}^{_{(2)}}\right)^2\right)+p^{_{(2)}}C_{2222}^{_{(1)}}\right) ,$ or in its inverse form

\begin{equation}\label{thermalwavecomplexfrequencys}
	s(k_2)=\dfrac{\mathscr{Z}_0\left(\eta\bar{K}_{22}^{_{(2)}}+\bar{K}_{22}^{_{(1)}}\right)\pm\sqrt{\mathscr{Z}_1(k_2)}}{2\mathscr{Z}_0\left(\eta\bar{K}_{22}^{_{(2)}}\tau^{_{(1)}}+\bar{K}_{22}^{_{(1)}}\tau^{_{(2)}}\right)},
\end{equation}
with the function $ \mathscr{Z}_1(k_2) $ defined as

\begin{equation}\label{thermalwaveausiliarioR}
	\mathscr{Z}_1(k_2)=-4k^2_2\mathscr{Z}_0\left(\eta\bar{K}_{22}^{_{(2)}}\tau^{_{(1)}}+\bar{K}_{22}^{_{(1)}}\tau^{_{(2)}}\right)(\eta+1)^2\left(\eta C_{2222}^{_{(2)}}+C_{2222}^{_{(1)}}\right)^2\left(\bar{K}^{_{(1)}}_{22}\bar{K}^{_{(2)}}_{22}\right)^2+\mathscr{Z}_0^2\left(\eta\bar{K}_{22}^{_{(2)}}+\bar{K}_{22}^{_{(1)}}\right)^2.
\end{equation}
It is important to notice here that, the dispersion relations \eqref{thermalwavewavenumberk2} and \eqref{thermalwavecomplexfrequencys} are expressed, as expected, in terms of both phases of the micro-thermal-conductivity tensor components $ \bar{K}^{_{(1)}}_{22} $ and $ \bar{K}^{_{(2)}}_{22} ,$ as well as the relaxation times $  \tau^{_{(1)}}$ and $ \tau^{_{(2)}} .$

\begin{figure}[h!]
	\centering
	\includegraphics[scale=0.31]{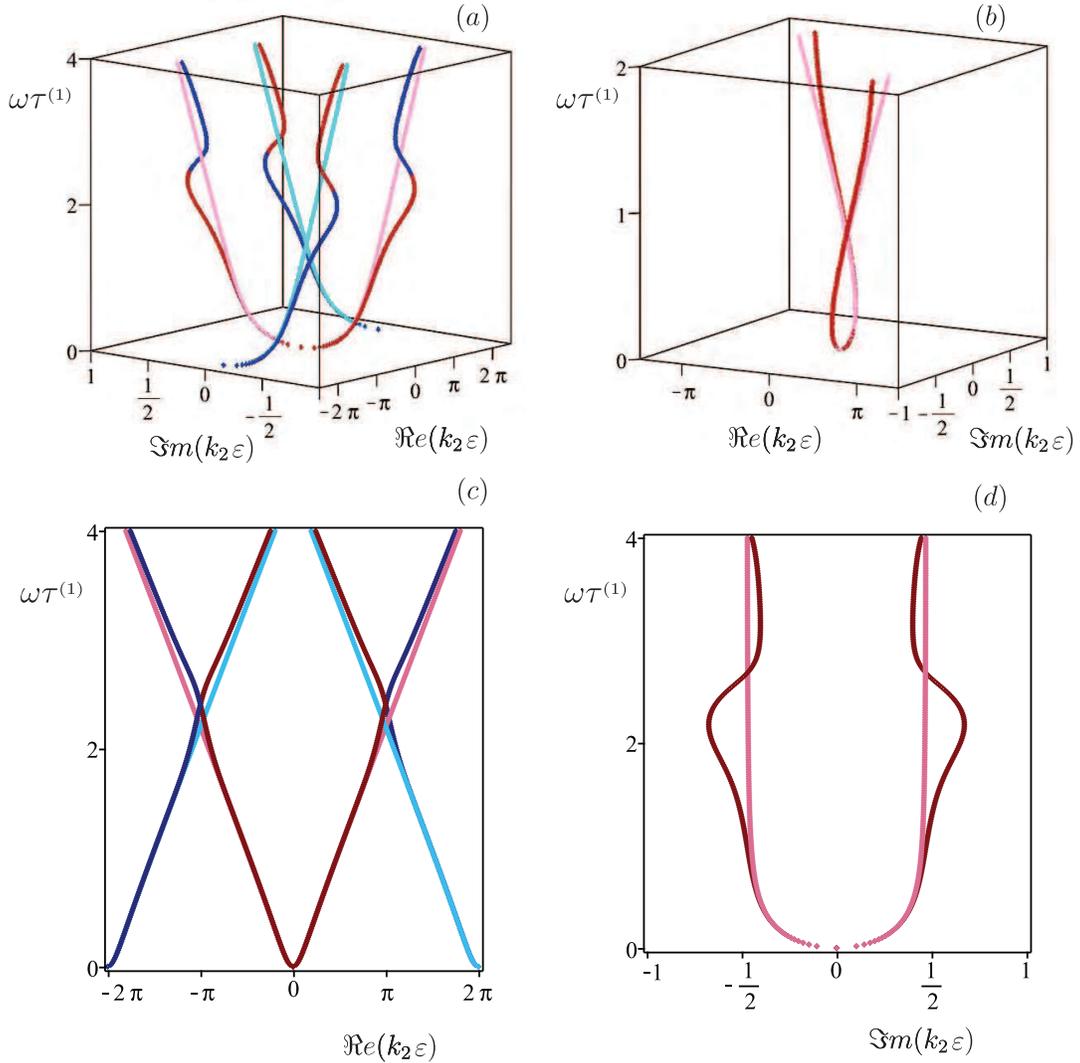}
	\caption{Dimensionless dispersion functions associated to thermal waves characterized when $ k_1=0. $ Comparison between a first order homogenized model (light blue curve and light red curve) with its respective heterogeneous one (dark blue curve and dark red curve), assuming the numerical values for the dimensionless parameters $ \eta=1,$ $ \tau^{_{(2)}}/\tau^{_{(1)}}=3,$ $ \bar{K}_{22}^{_{(2)}}/\bar{K}_{22}^{_{(1)}}=3,$ $ p^{_{(2)}}/p^{_{(1)}}=3 $ and $ p^{_{(1)}}T_0/\left(\tau^{_{(1)}}\bar{K}_{22}^{_{(1)}}\right)=1.$  $ (a) $ $ \omega\tau^{_{(1)}}$ vs. $\Re e(k_2\varepsilon)\times\Im m(k_2\varepsilon) ;$ $ (b) $ zoomed view of the angular frequency spectrum $ \omega\tau^{_{(1)}}$ vs. $\Re e(k_2\varepsilon)\times\Im m(k_2\varepsilon) ;$ $ (c) $ view of the plane $ \omega\tau^{_{(1)}}\times\Re e(k_2\varepsilon);$ $ (d) $ view of the plane $ \omega\tau^{_{(1)}}\times\Im m(k_2\varepsilon).$}
	\label{fig:bechmarkabcd}
\end{figure}

In respect of characterizing harmonic waves in the homogenized continuum, it is worth noting that for these above equations and their respective inverse forms allow us to either describe them by spatial damping or time damping as previously seen in Sec. \ref{waveprophomogenized}.

On the other hand, followed from the Secs. \ref{Wavepropagationheterogeneous} and \ref{heterogeneousapproach} of the Supplementary Material, the dispersion relations related to the heterogeneous bi-dimensional body with the orthotropy axes perpendicular to the layering direction $ \boldsymbol{e_2}$ and the wave vector taken such as $ k_1=0 ,$ with $ d_2=d^{_{(1)}}+d^{_{(2)}} $ are given by

\begin{figure}[h!]
	\centering
	\includegraphics[scale=0.31]{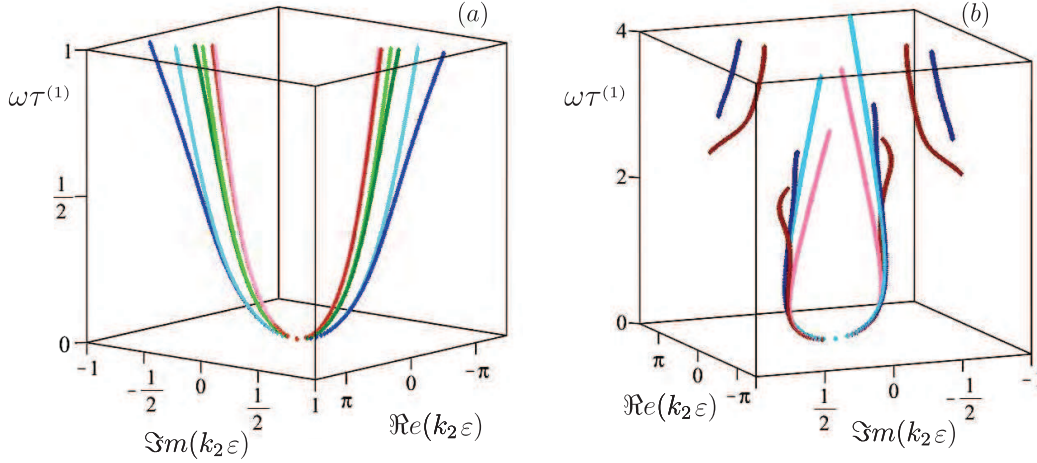}
	\caption{Dimensionless dispersion functions associated to thermal waves characterized when $ k_1=0.$ Comparison between homogenized models (light curves) and their respective heterogeneous ones (dark curves) given different values of the parameters. $ (a) $ setting $ \eta=1,$ $ p^{_{(1)}}T_0/\left(\tau^{_{(1)}}\bar{K}_{22}^{_{(1)}}\right)=1 $ and varying the parameters as $ \tau^{_{(2)}}/\tau^{_{(1)}}=3,$ $ \bar{K}_{22}^{_{(2)}}/\bar{K}_{22}^{_{(1)}}=3,$ $ p^{_{(2)}}/p^{_{(1)}}=3 $ (red curves), $ \tau^{_{(2)}}/\tau^{_{(1)}}=5,$ $ \bar{K}_{22}^{_{(2)}}/\bar{K}_{22}^{_{(1)}}=5,$ $ p^{_{(2)}}/p^{_{(1)}}=5 $ (green curves), $ \tau^{_{(2)}}/\tau^{_{(1)}}=10,$ $ \bar{K}_{22}^{_{(2)}}/\bar{K}_{22}^{_{(1)}}=10,$ $ p^{_{(2)}}/p^{_{(1)}}=10 $ (blue curves);
		$ (b) $ fixing the parameters $ \tau^{_{(2)}}/\tau^{_{(1)}}=3,$ $ \bar{K}_{22}^{_{(2)}}/\bar{K}_{22}^{_{(1)}}=3,$ $ p^{_{(2)}}/p^{_{(1)}}=3 ,$ $ p^{_{(1)}}T_0/\left(\tau^{_{(1)}}\bar{K}_{22}^{_{(1)}}\right)=1 ,$ and varying the thickness as $ \eta=1 $ (red curves), $ \eta=10 $ (blue curves).}
	\label{fig:bechmark2ab2}
\end{figure}

\begin{equation}\label{thermalwavehetero}
	k_2(s)=\dfrac{1}{\mathrm{i}d_2}\left[\ln\left(\dfrac{-\tilde{I}_1^{_{j}}(s)\pm\sqrt{\left(\tilde{I}_1^{_{j}}(s)\right)^2-4}}{2}\right)-2\pi \mathrm{i}n\right], \quad\textrm{with $ n\in\Z,\quad j={1,2,3},$}
\end{equation}
where the details about the invariants $ \tilde{I}_1^{_{j}}(s) $ are brought in the Sec. \ref{dispersivewaveheterogeneous} of the Supplementary Material.

For the case in which the wave number $ k_1 $ is assumed zero and when $ s=\mathrm{i}\omega ,$ the Fig. \ref{fig:bechmarkabcd} exhibits the behaviour of the thermal wave propagation given by Eqs. \eqref{thermalwavecomplexfrequencys} and \eqref{thermalwaveausiliarioR} for the homogenized material represented in light blue and light red curves, while the thermal wave function \eqref{thermalwavehetero} for the heterogeneous solid via Floquet-Bloch theory is presented in dark blue and light red curves, where the parameters have been set as $ \eta=1,$ $ \tau^{_{(2)}}/\tau^{_{(1)}}=3,$ $ \bar{K}_{22}^{_{(2)}}/\bar{K}_{22}^{_{(1)}}=3,$ $ p^{_{(2)}}/p^{_{(1)}}=3 $ and $ p^{_{(1)}}T_0/\left(\tau^{_{(1)}}\bar{K}_{22}^{_{(1)}}\right)=1.$ Both light and dark blue lines stand for the translations of their respective dispersion curves due to the periodicity of the material along $ \boldsymbol{e}_2 $ around $ \Re e(k_2\varepsilon)=2\pi n ,$ which they go all along the real wave number axis $ \Re e(k_2\varepsilon) .$ The dispersion relations of the homogenized model can also be determined seeking solutions from the governing equations \eqref{eulerlagrange1re} and \eqref{eulerlagrange2re} (with source terms assumed zero) of the forms ${\hat{\boldsymbol{U}}}^{M}=\check{\hat{\boldsymbol{U}}}^{M}\exp[\mathrm{i}(\boldsymbol{k}\cdot\textbf{x}+2\pi n)]$ and ${\hat{\Theta}}^{M}=\check{\hat{\Theta}}^{M}\exp[\mathrm{i}(\boldsymbol{k}\cdot\textbf{x}+2\pi n)]  $ (with $ n\in\Z $) for the displacement and temperature, respectively. When $ n=0 $ one arrives to the functions \eqref{shearwave}, \eqref{compressionalwave}, \eqref{thermalwavewavenumberk2}, while when $ n=-1 ,$ $ n=1 $ one produces the translated curves drawn in light blue displayed in the Figs. \ref{fig:bechmarkabcd}$ (a)$ and \ref{fig:bechmarkabcd}$(c) .$ Besides, since a first order homogenization was applied one may notice an accurate approximation between the light curves and dark ones along the thermal wave propagation, in other words the dispersion function derived from the homogenization process shows a very good agreement with the dispersion function obtained taking advantage of the Floquet–Bloch theory, seen in the Secs. \ref{Wavepropagationheterogeneous}, \ref{heterogeneousapproach} and \ref{dispersivewaveheterogeneous} of the Supplementary Material, for the interval $ 2\pi(1-n)/3< \Re e(k_2\varepsilon)<2\pi(1+n)/3,\forall n\in\Z .$ The Figs. \ref{fig:bechmarkabcd}$ (a)$ and \ref{fig:bechmarkabcd}$(c) $ show the dispersion functions associated to $ n=-1 ,$ $ n=0 $ and $ n=1, $ where may be observed that for a low frequency the curves of the heterogeneous continuum (dark red) are quite precise in respect to those obtained from the homogenization process of the material (light red). For a high frequency can be noticed that the curves of the heterogeneous continuum (dark red) are well approximated with a tiny deviation when compared to the translated curves ($ n=-1 $ and $ n=1 $) gotten from the homogenization process of the material (light blue). Fig. \ref{fig:bechmarkabcd}$ (b) $ shows a zoom of the thermal wave propagation by attenuation in space seen in Fig. \ref{fig:bechmarkabcd}$ (a) .$ The dimensionless angular frequency $ \omega\tau^{_{(1)}}$ by the dimensionless real wave number $ \Re e(k_2\varepsilon) $ plane and the dimensionless angular frequency $ \omega\tau^{_{(1)}}$ by the dimensionless attenuation factor $ \Im m(k_2\varepsilon)$ plane are represented in \ref{fig:bechmarkabcd}$ (c)$ and \ref{fig:bechmarkabcd}$ (d),$ respectively. 

\begin{figure}[h!]
	\centering
	\includegraphics[scale=0.31]{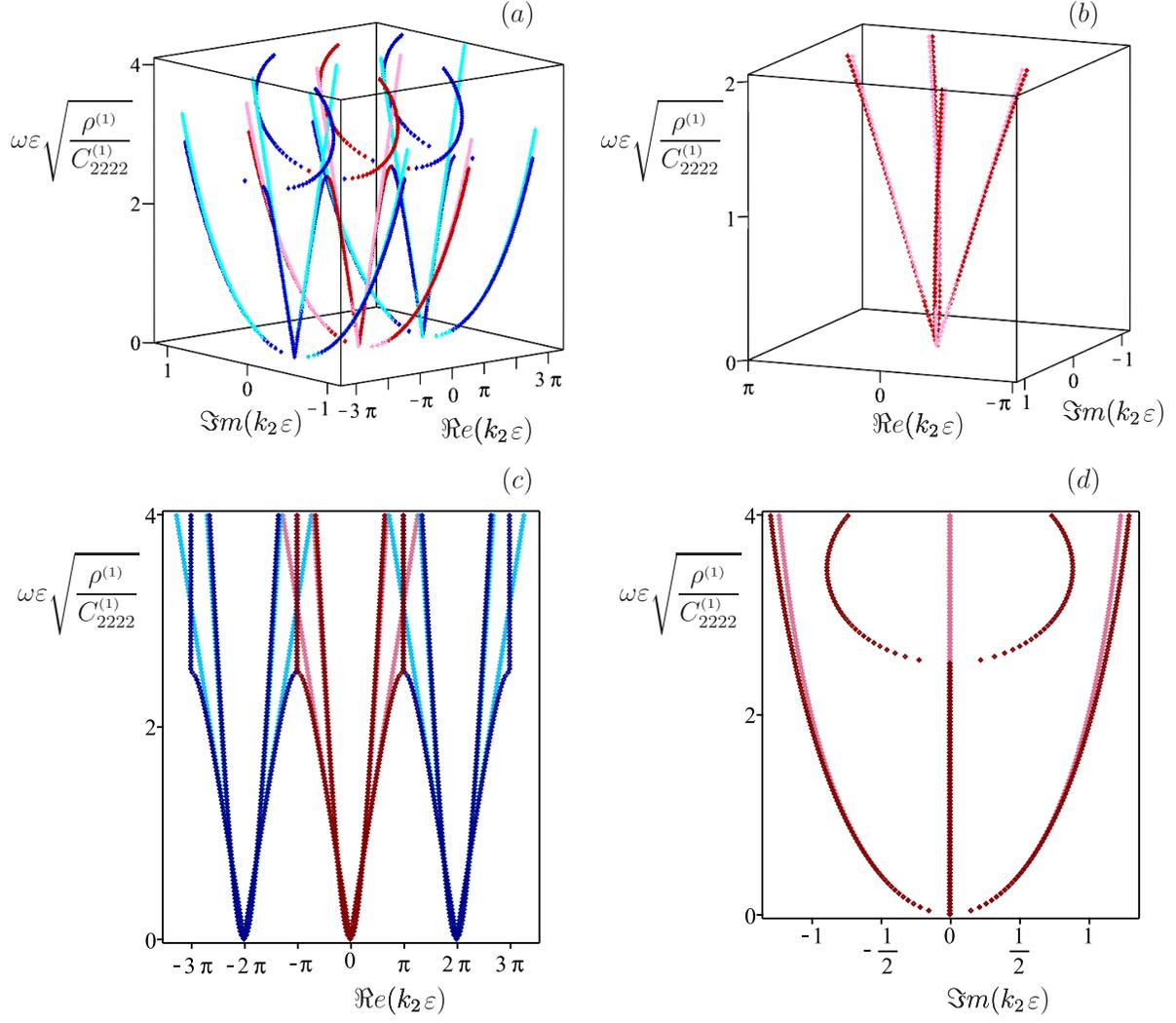}
	\caption{Dimensionless dispersion functions associated to compressional-thermal waves characterized when $ k_1=0. $ Comparison between a first order homogenized model (light blue curve and light red curve) with its respective heterogeneous one (dark blue curve and dark red curve), assuming the numerical values for the dimensionless parameters $ \eta=1,$ $ \tau^{_{(2)}}/\tau^{_{(1)}}=3,$ $ {C}_{2222}^{_{(2)}}/{C}_{2222}^{_{(1)}}=3,$ $ \bar{K}_{22}^{_{(2)}}/\bar{K}_{22}^{_{(1)}}=3,$ $ p^{_{(2)}}/p^{_{(1)}}=3 ,$ $ \rho^{_{(2)}}/\rho^{_{(1)}}=2,$ $(\alpha_{22}^{_{(1)}}T_0)/{C}_{2222}^{_{(1)}}=1/100,  $ $\alpha_{22}^{_{(2)}}T_0/{C}_{2222}^{_{(2)}}=1/10,  $ $\alpha_{22}^{_{(1)}}\eta\sqrt{C_{2222}^{_{(1)}}/\rho^{_{(1)}}}/\bar{K}_{22}^{_{(1)}}=1/100,$ $\alpha_{22}^{_{(2)}}\eta\sqrt{C_{2222}^{_{(1)}}/\rho^{_{(1)}}}/\bar{K}_{22}^{_{(2)}}=1/10,$ $p^{_{(1)}}T_0\eta\sqrt{C_{2222}^{_{(1)}}/\rho^{_{(1)}}}/\bar{K}_{22}^{_{(1)}}=1$ and $\tau^{_{(1)}}\sqrt{C_{2222}^{_{(1)}}/\rho^{_{(1)}}}/\varepsilon=1/10.$ $ (a) $ $ \omega\varepsilon\sqrt{\rho^{_{(1)}}/C_{2222}^{_{(1)}}}$ vs. $\Re e(k_2\varepsilon)\times\Im m(k_2\varepsilon) ;$ $ (b) $ zoomed view of the angular frequency spectrum $ \omega\varepsilon\sqrt{\rho^{_{(1)}}/C_{2222}^{_{(1)}}}$ vs. $\Re e(k_2\varepsilon)\times\Im m(k_2\varepsilon) ;$ $ (c) $ view of the plane $ \omega\varepsilon\sqrt{\rho^{_{(1)}}/C_{2222}^{_{(1)}}}\times\Re e(k_2\varepsilon);$ $ (d) $ view of the plane $ \omega\varepsilon\sqrt{\rho^{_{(1)}}/C_{2222}^{_{(1)}}}\times\Im m(k_2\varepsilon).$}
	\label{fig:COUPLED1}
\end{figure}

For a better understanding of the first order homogenization facing its heterogeneous version, the Fig. \ref{fig:bechmark2ab2} has been made varying some parameters, where dark curves represent the heterogeneous material obtained via Floquet-Bloch theory, whereas light curves are for the first order homogenized model. Essentially, Fig. \ref{fig:bechmark2ab2}$(a)$ displays three zoomed different situations for the thermal wave propagation when $ \eta=1$ and $ p^{_{(1)}}T_0/\left(\tau^{_{(1)}}\bar{K}_{22}^{_{(1)}}\right)=1 ,$ namely the curves in red were plotted setting  $ \tau^{_{(2)}}/\tau^{_{(1)}}=3,$ $ \bar{K}_{22}^{_{(2)}}/\bar{K}_{22}^{_{(1)}}=3,$ $ p^{_{(2)}}/p^{_{(1)}}=3 ,$ for the second scenario the curves in green were generated choosing $ \tau^{_{(2)}}/\tau^{_{(1)}}=5,$ $ \bar{K}_{22}^{_{(2)}}/\bar{K}_{22}^{_{(1)}}=5,$ $ p^{_{(2)}}/p^{_{(1)}}=5, $ and for the third situation the curves in blue were plotted choosing the parameters $ \tau^{_{(2)}}/\tau^{_{(1)}}=10,$ $ \bar{K}_{22}^{_{(2)}}/\bar{K}_{22}^{_{(1)}}=10,$ $ p^{_{(2)}}/p^{_{(1)}}=10 .$ Furthermore, regarding the behaviour led by the different values of the dimensionless parameters taken, it can be noted in the Fig. \ref{fig:bechmark2ab2}$ (a) $ that an increase in the parameters $ \tau^{_{(2)}}/\tau^{_{(1)}},$ $ \bar{K}_{22}^{_{(2)}}/\bar{K}_{22}^{_{(1)}}$ and $ p^{_{(2)}}/p^{_{(1)}} ,$ produces curves that bend less sharply, and hence have smaller curvatures. With an analogous idea, Fig. \ref{fig:bechmark2ab2}$(b)$ displays two scenarios varying the thickness that are $ \eta=1 $ represented by the red curves and $ \eta=10 $ represented by the blue curves, while the parameters $ \tau^{_{(2)}}/\tau^{_{(1)}}=3,$ $ \bar{K}_{22}^{_{(2)}}/\bar{K}_{22}^{_{(1)}}=3,$ $ p^{_{(2)}}/p^{_{(1)}}=3 $ and $p^{_{(1)}}T_0/\left(\tau^{_{(1)}}\bar{K}_{22}^{_{(1)}}\right)=1,$ are fixed. Analysing the Fig. \ref{fig:bechmark2ab2}$ (b) ,$ it is notable that a decrease in the dimensionless thickness induce to bigger curvatures along the wave propagation. From these comparisons between the wave for the heterogeneous solid and their respectives for the homogenized model illustrated in Fig. \ref{fig:bechmark2ab2}, it may be verified a good matching for the wave propagation between the two models exemplified considering the interval $ 2\pi(1-n)/3< \Re e(k_2\varepsilon)<2\pi(1+n)/3,\forall n\in\Z .$  

\begin{figure}[h!]
	\centering
	\includegraphics[scale=0.31]{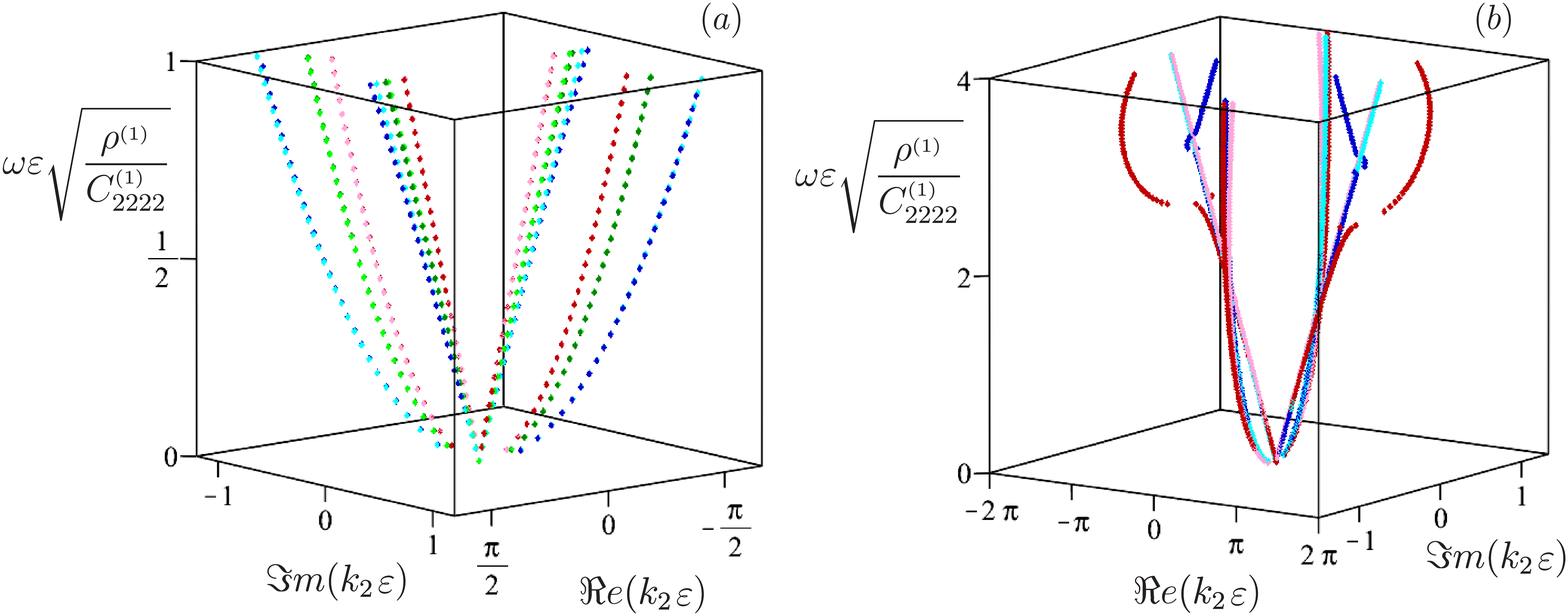}
	\caption{Dimensionless dispersion functions associated to compressional-thermal waves characterized when $ k_1=0.$ Comparison between homogenized models (light curves) and their respective heterogeneous ones (dark curves) given different values of the parameters. In both the images the dimensionless parameters are fixed as $(\alpha_{22}^{_{(1)}}T_0)/{C}_{2222}^{_{(1)}}=1/100,  $ $\alpha_{22}^{_{(2)}}T_0/{C}_{2222}^{_{(2)}}=1/10,  $ $\alpha_{22}^{_{(1)}}\eta\sqrt{C_{2222}^{_{(1)}}/\rho^{_{(1)}}}/\bar{K}_{22}^{_{(1)}}=1/100,$ $\alpha_{22}^{_{(2)}}\eta\sqrt{C_{2222}^{_{(1)}}/\rho^{_{(1)}}}/\bar{K}_{22}^{_{(2)}}=1/10,$ $p^{_{(1)}}T_0\eta\sqrt{C_{2222}^{_{(1)}}/\rho^{_{(1)}}}/\bar{K}_{22}^{_{(1)}}=1$ and $\tau^{_{(1)}}\sqrt{C_{2222}^{_{(1)}}/\rho^{_{(1)}}}/\varepsilon=1/10.$  $ (a) $ setting $ \eta=1,$ and varying the parameters as $ \tau^{_{(2)}}/\tau^{_{(1)}}=3,$ $ {C}_{2222}^{_{(2)}}/{C}_{2222}^{_{(1)}}=3,$ $ \bar{K}_{22}^{_{(2)}}/\bar{K}_{22}^{_{(1)}}=3,$ $ p^{_{(2)}}/p^{_{(1)}}=3 ,$ $ \rho^{_{(2)}}/\rho^{_{(1)}}=2,$ (red curves), $ \tau^{_{(2)}}/\tau^{_{(1)}}=5,$ $ {C}_{2222}^{_{(2)}}/{C}_{2222}^{_{(1)}}=5,$ $ \bar{K}_{22}^{_{(2)}}/\bar{K}_{22}^{_{(1)}}=5,$ $ p^{_{(2)}}/p^{_{(1)}}=5 ,$ $ \rho^{_{(2)}}/\rho^{_{(1)}}=4,$  (green curves), $ \tau^{_{(2)}}/\tau^{_{(1)}}=10,$ $ {C}_{2222}^{_{(2)}}/{C}_{2222}^{_{(1)}}=10,$ $ \bar{K}_{22}^{_{(2)}}/\bar{K}_{22}^{_{(1)}}=10,$ $ p^{_{(2)}}/p^{_{(1)}}=10 ,$ $ \rho^{_{(2)}}/\rho^{_{(1)}}=6,$  (blue curves);	$ (b) $ fixing the parameters $ \tau^{_{(2)}}/\tau^{_{(1)}}=3,$ $ {C}_{2222}^{_{(2)}}/{C}_{2222}^{_{(1)}}=3,$ $ \bar{K}_{22}^{_{(2)}}/\bar{K}_{22}^{_{(1)}}=3,$ $ p^{_{(2)}}/p^{_{(1)}}=3 ,$ $ \rho^{_{(2)}}/\rho^{_{(1)}}=2,$ and varying the thickness as $ \eta=1 $ (red curves), $ \eta=20 $ (blue curves).}
	\label{fig:COUPLED2}
\end{figure}

\begin{figure}[h!]
	\centering
	\includegraphics[scale=0.31]{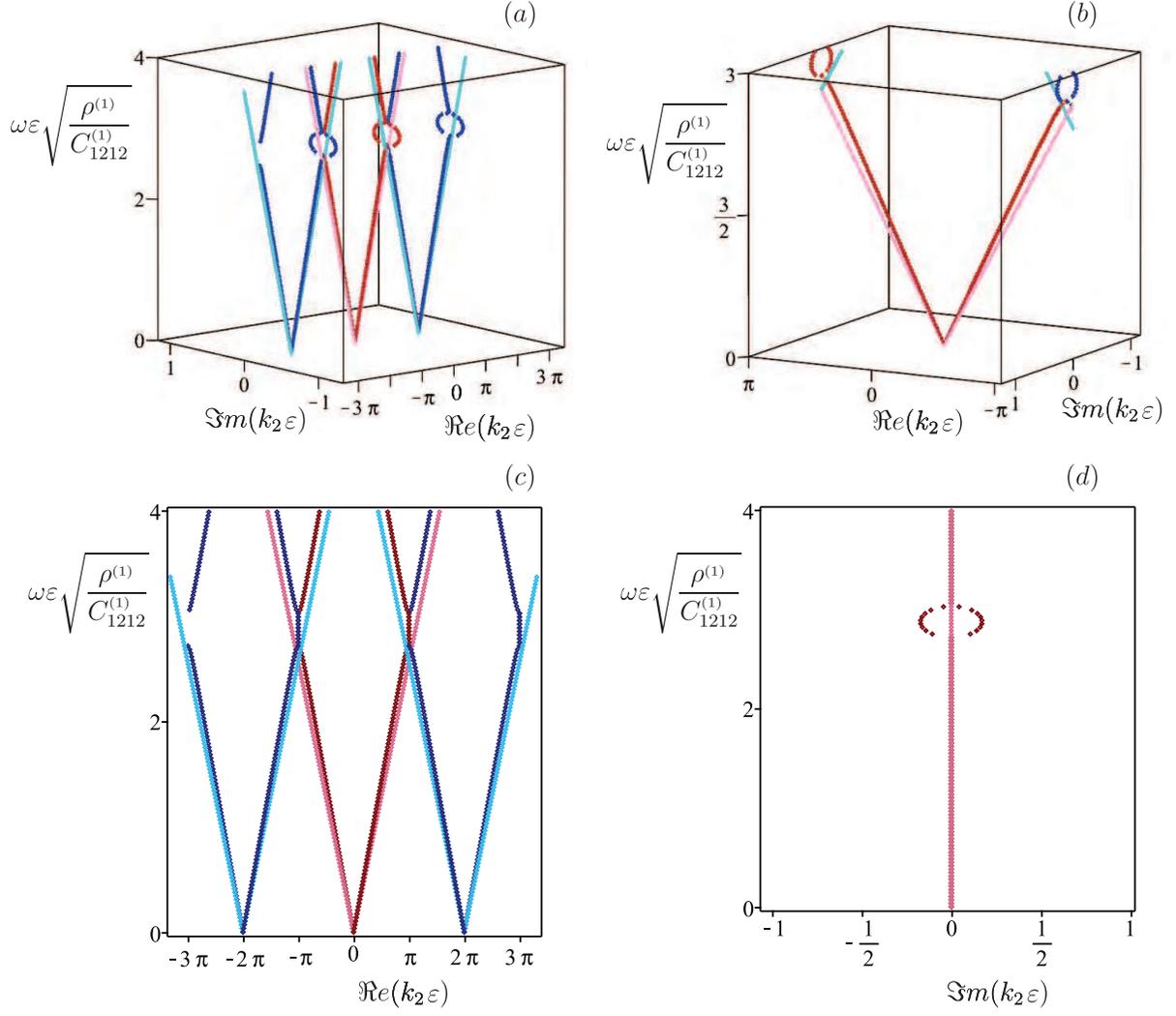}
	\caption{Dimensionless dispersion functions associated to shear waves characterized when $ k_1=0. $ Comparison between a first order homogenized model (light blue curve and light red curve) with its respective heterogeneous one (dark blue curve and dark red curve), assuming the numerical values for the dimensionless parameters $ \eta=1,$ $ {C}_{1212}^{_{(2)}}/{C}_{1212}^{_{(1)}}=1$ and $ \rho^{_{(2)}}/\rho^{_{(1)}}=2.$ $ (a) $ $ \omega\varepsilon\sqrt{\rho^{_{(1)}}/C_{1212}^{_{(1)}}}$ vs. $\Re e(k_2\varepsilon)\times\Im m(k_2\varepsilon) ;$ $ (b) $ zoomed view of the angular frequency spectrum $ \omega\varepsilon\sqrt{\rho^{_{(1)}}/C_{1212}^{_{(1)}}}$ vs. $\Re e(k_2\varepsilon)\times\Im m(k_2\varepsilon) ;$ $ (c) $ view of the plane $ \omega\varepsilon\sqrt{\rho^{_{(1)}}/C_{1212}^{_{(1)}}}\times\Re e(k_2\varepsilon);$ $ (d) $ view of the plane $ \omega\varepsilon\sqrt{\rho^{_{(1)}}/C_{1212}^{_{(1)}}}\times\Im m(k_2\varepsilon).$}
	\label{fig:taglio1abcd}
\end{figure}

\begin{figure}[h!]
	\centering
	\includegraphics[scale=0.31]{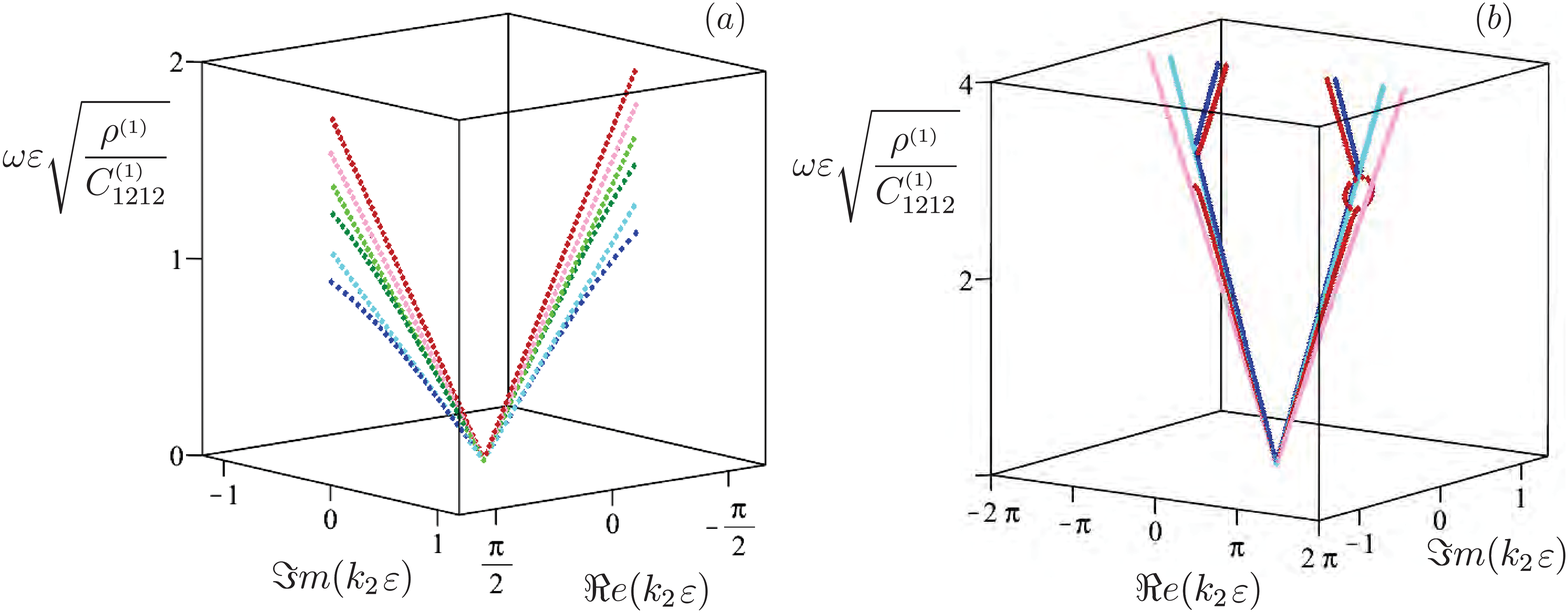}
	\caption{Dimensionless dispersion functions associated to shear waves characterized when $ k_1=0.$ Comparison between homogenized models (light curves) and their respective heterogeneous ones (dark curves) given different values of the parameters. $ (a) $ setting $ \eta=1,$ and varying the parameters as $ {C}_{1212}^{_{(2)}}/{C}_{1212}^{_{(1)}}=1$ and $ \rho^{_{(2)}}/\rho^{_{(1)}}=2,$  (red curves); $ {C}_{1212}^{_{(2)}}/{C}_{1212}^{_{(1)}}=5$ and $ \rho^{_{(2)}}/\rho^{_{(1)}}=5,$   (green curves); $ {C}_{1212}^{_{(2)}}/{C}_{1212}^{_{(1)}}=10$ and $ \rho^{_{(2)}}/\rho^{_{(1)}}=10,$ (blue curves);	$ (b) $ the non-dimensional parameters are adopted as $ {C}_{1212}^{_{(2)}}/{C}_{1212}^{_{(1)}}=1,$ $ \rho^{_{(2)}}/\rho^{_{(1)}}=2,$ and varying the thickness as $ \eta=1 $ (red curves), $ \eta=20 $ (blue curves).}
	\label{fig:taglio2ab}
\end{figure}

Likewise the previous comparative test, under the hypothesis when the wave number $ k_1$ is assumed zero and the complex frequency is taken as $ s=\mathrm{i}\omega ,$ the compressional-thermal wave function of the heterogeneous continuum via Floquet-Bloch theory and the compressional-thermal waves of the first order homogenized material are illustrated in the Fig. \ref{fig:COUPLED1} by the dark blue and red lines, and the light blue and red lines, respectively. The dimensionless parameters were chosen as $ \eta=1,$ $ \tau^{_{(2)}}/\tau^{_{(1)}}=3,$ $ {C}_{2222}^{_{(2)}}/{C}_{2222}^{_{(1)}}=3,$ $ \bar{K}_{22}^{_{(2)}}/\bar{K}_{22}^{_{(1)}}=3,$ $ p^{_{(2)}}/p^{_{(1)}}=3 ,$ $ \rho^{_{(2)}}/\rho^{_{(1)}}=2,$ $(\alpha_{22}^{_{(1)}}T_0)/{C}_{2222}^{_{(1)}}=1/100,  $ $\alpha_{22}^{_{(2)}}T_0/{C}_{2222}^{_{(2)}}=1/10,$ $\alpha_{22}^{_{(1)}}\eta\sqrt{C_{2222}^{_{(1)}}/\rho^{_{(1)}}}/\bar{K}_{22}^{_{(1)}}=1/100,$ $\alpha_{22}^{_{(2)}}\eta\sqrt{C_{2222}^{_{(1)}}/\rho^{_{(1)}}}/\bar{K}_{22}^{_{(2)}}=1/10,$  $p^{_{(1)}}T_0\eta\sqrt{C_{2222}^{_{(1)}}/\rho^{_{(1)}}}/\bar{K}_{22}^{_{(1)}}=1$ and $\tau^{_{(1)}}\sqrt{C_{2222}^{_{(1)}}/\rho^{_{(1)}}}/\varepsilon=1/10.$ In the Figs. \ref{fig:COUPLED1}$ (a)$ and \ref{fig:COUPLED1}$(c),$ both light and dark blue lines stand for the translations of their respective dispersion curves due to the periodicity of the material along $ \boldsymbol{e}_2 $ around $ \Re e(k_2\varepsilon)=2\pi n,$ which they go all along the real wave number axis $ \Re e(k_2\varepsilon) .$ Moreover, one may observe a precise estimation between the light lines and dark ones along the  compressional-thermal wave propagation. This means that the dispersion function derived from the first order homogenization process yield to a quite good matching when contrasted against the dispersion function determined with the Floquet–Bloch theory for the heterogeneous solid elucidated in the Secs. \ref{Wavepropagationheterogeneous}, \ref{heterogeneousapproach} and \ref{dispersivewaveheterogeneous} of the Supplementary Material, for the interval $ 2\pi(1-n)/3< \Re e(k_2\varepsilon)<2\pi(1+n)/3,\forall n\in\Z .$ The Figs. \ref{fig:COUPLED1}$ (a)$ and \ref{fig:COUPLED1}$(c),$ exhibit the dispersion functions when $ n=-1 ,$ $ n=0 $ and $ n=1, $ where may be noted that for a low frequency the curves of the heterogeneous continuum (dark red) are matched very well for those obtained from the homogenization process of the material (light red), in addition the same observation may be seen for the translated blue lines. Fig. \ref{fig:COUPLED1}$ (b) $ shows a zoom of the compressional-thermal wave propagation by spatial damping seen in Fig. \ref{fig:COUPLED1}$ (a) .$ A bi-dimensional view of the dimensionless angular frequency $ \omega\varepsilon\sqrt{\rho^{_{(1)}}/C_{2222}^{_{(1)}}}$ by the dimensionless real wave number $ \Re e(k_2\varepsilon) $ plane and the dimensionless angular frequency $\omega\varepsilon\sqrt{\rho^{_{(1)}}/C_{2222}^{_{(1)}}}$ by the dimensionless attenuation factor $ \Im m(k_2\varepsilon)$ plane are represented in \ref{fig:COUPLED1}$ (c)$ and \ref{fig:COUPLED1}$ (d),$ respectively. It is worthy of note that, the branches of the spectrum characterized when the real part $ \Re e(k_2\varepsilon)$ is set fixed as $ \Im m(k_2\varepsilon)$ varies, as expected, which characterize the first frequency stop-band for quasi-compressional waves and which propagate perpendicularly to the layering direction, they are unable to be approximated by the frequency spectrum obtained from the homogenized first order model, as showed in the Fig. \ref{fig:COUPLED1}$ (d).$ In order to overcome such encumbrance, it is feasible to use a perturbative technique, which will lead to a local asymptotic approximation of the compressional-thermal wave through local explicit and closed-form parametric expressions in the space of complex wave number and the angular frequency \citep{bacigalupo2016high,fantoni2020wave}.

The Fig. \ref{fig:COUPLED2} compares some cases of the compressional-thermal wave assuming different values of the dimensionless parameters, where  the dark lines represent the waves obtained for the heterogeneous material via Floquet-Bloch theory, while the light lines are for the first order homogenized model. The Figs. \ref{fig:COUPLED2}$(a)$ and \ref{fig:COUPLED2}$(b)$ were generated taking the following values for dimensionless parameters  $(\alpha_{22}^{_{(1)}}T_0)/{C}_{2222}^{_{(1)}}=1/100,  $  $\alpha_{22}^{_{(2)}}T_0/{C}_{2222}^{_{(2)}}=1/10,  $ $\alpha_{22}^{_{(1)}}\eta\sqrt{C_{2222}^{_{(1)}}/\rho^{_{(1)}}}/\bar{K}_{22}^{_{(1)}}=1/100,$ $\alpha_{22}^{_{(2)}}\eta\sqrt{C_{2222}^{_{(1)}}/\rho^{_{(1)}}}/\bar{K}_{22}^{_{(2)}}=1/10,$ $p^{_{(1)}}T_0\eta\sqrt{C_{2222}^{_{(1)}}/\rho^{_{(1)}}}/\bar{K}_{22}^{_{(1)}}=1$ and $\tau^{_{(1)}}\sqrt{C_{2222}^{_{(1)}}/\rho^{_{(1)}}}/\varepsilon=1/10.$ Assuming $ \eta=1,$ Fig. \ref{fig:COUPLED2}$(a)$ shows three zoomed different cases for the compressional-thermal wave propagation. Therefore, the red lines were plotted setting  $ \tau^{_{(2)}}/\tau^{_{(1)}}=3,$ $ {C}_{2222}^{_{(2)}}/{C}_{2222}^{_{(1)}}=3,$ $ \bar{K}_{22}^{_{(2)}}/\bar{K}_{22}^{_{(1)}}=3,$ $ p^{_{(2)}}/p^{_{(1)}}=3 ,$ $ \rho^{_{(2)}}/\rho^{_{(1)}}=2,$ the green lines were drawn choosing $ \tau^{_{(2)}}/\tau^{_{(1)}}=5,$ $ {C}_{2222}^{_{(2)}}/{C}_{2222}^{_{(1)}}=5,$ $ \bar{K}_{22}^{_{(2)}}/\bar{K}_{22}^{_{(1)}}=5,$ $ p^{_{(2)}}/p^{_{(1)}}=5 ,$ $ \rho^{_{(2)}}/\rho^{_{(1)}}=4,$ and finally the lines in blue represent the situation in which $ \tau^{_{(2)}}/\tau^{_{(1)}}=10,$ $ {C}_{2222}^{_{(2)}}/{C}_{2222}^{_{(1)}}=10,$ $ \bar{K}_{22}^{_{(2)}}/\bar{K}_{22}^{_{(1)}}=10,$ $ p^{_{(2)}}/p^{_{(1)}}=10 ,$ $ \rho^{_{(2)}}/\rho^{_{(1)}}=6.$ Taking a look closely at the curvatures of the compressional-thermal waves plotted in the Fig. \ref{fig:COUPLED2}$ (a),$ it may be affirmed that increasing the numerical values of $ \tau^{_{(2)}}/\tau^{_{(1)}},$ $ {C}_{2222}^{_{(2)}}/{C}_{2222}^{_{(1)}},$ $ \bar{K}_{22}^{_{(2)}}/\bar{K}_{22}^{_{(1)}},$ $ p^{_{(2)}}/p^{_{(1)}}$ and $ \rho^{_{(2)}}/\rho^{_{(1)}},$  the curves show a tendency of having smaller curvatures. The compressional-thermal waves behaviour is foretold in matters of different values of the thickness $ \eta, $ the Fig. \ref{fig:COUPLED2}$(b)$ displays two scenarios where is fixed $ \tau^{_{(2)}}/\tau^{_{(1)}}=3,$ $ {C}_{2222}^{_{(2)}}/{C}_{2222}^{_{(1)}}=3,$ $ \bar{K}_{22}^{_{(2)}}/\bar{K}_{22}^{_{(1)}}=3,$ $ p^{_{(2)}}/p^{_{(1)}}=3 ,$ $ \rho^{_{(2)}}/\rho^{_{(1)}}=2,$ and  varying the thickness that are $ \eta=1 $ and $ \eta=20 $ represented by the red and blue lines, respectively. Such comparisons provided by the Fig. \ref{fig:COUPLED2} shows a good agreement of the compressional-thermal wave propagation between the two models exemplified considering the interval $ 2\pi(1-n)/3< \Re e(k_2\varepsilon)<2\pi(1+n)/3,\forall n\in\Z .$

Some plots of the shear waves may be seen in Fig. \ref{fig:taglio1abcd}, where red lines stand for the central shear wave, blue lines for the translated shear waves, and the nuances in the color distinguish the shear curve of the heterogeneous continuum (dark lines) from the shear wave first order homogenized (light lines). In regards to de capability of the first order homogenization process for the shear wave, it also shows itself in a good agreement with its heterogeneous continuum from the Floquet-Bloch theory. The Fig. \ref{fig:taglio1abcd} was drawn adopting the dimensionless parameters as $ \eta=1,$ $ {C}_{1212}^{_{(2)}}/{C}_{1212}^{_{(1)}}=1$ and $\rho^{_{(2)}}/\rho^{_{(1)}}=2.$ Note that, as verified for the compressional-thermal wave, and with a cleaner comprehension looking at the Fig. \ref{fig:taglio1abcd}$ (d),$ also the frequency stop-bands of the shear wave propagating perpendicularly to the layering direction is not well described by the first order homogenization technique developed. As aforementioned, to remedy this issue the shear waves may be approached by a perturbative method through explicit and closed-form parametric expressions, in which will approximate locally the real spectrum of the non-homogeneous material \citep{bacigalupo2016high,fantoni2020wave}.

The Fig. \ref{fig:taglio2ab}$(a)$ exhibits three zoomed different cases for the shear wave propagation for a fixed thickness $ \eta=1. $ The red lines represent the choices for the parameters as $ {C}_{1212}^{_{(2)}}/{C}_{1212}^{_{(1)}}=1,$ $ \rho^{_{(2)}}/\rho^{_{(1)}}=2,$ the green lines stand for $ {C}_{1212}^{_{(2)}}/{C}_{1212}^{_{(1)}}=5,$ $ \rho^{_{(2)}}/\rho^{_{(1)}}=5,$ and lastly, the lines that approximate locally the real spectrum of the heterogeneous material in blue correspond to the case when $ {C}_{1212}^{_{(2)}}/{C}_{1212}^{_{(1)}}=10,$ $ \rho^{_{(2)}}/\rho^{_{(1)}}=10.$ From a quick analysis over the Fig. \ref{fig:taglio2ab}$ (a) ,$ it can be assured that the curvatures of the shear waves are reduced numerically due to a rise in the values of the non-dimensional parameters $ {C}_{1212}^{_{(2)}}/{C}_{1212}^{_{(1)}}$ and $ \rho^{_{(2)}}/\rho^{_{(1)}}.$ Varying the thickness $ \eta$ namely $ \eta=1 $ (red curves) and $ \eta=20 $ (blue curves), and adopting the dimensionless parameters as ${C}_{1212}^{_{(2)}}/{C}_{1212}^{_{(1)}}=1,$ $\rho^{_{(2)}}/\rho^{_{(1)}}=2,$ both shear waves for these assumptions are contrasted in the Fig. \ref{fig:taglio2ab}$(b).$ Essentially, the Fig. \ref{fig:taglio2ab} illustrates the good estimation of the shear wave propagation between the two models developed considering the interval $ 2\pi(1-n)/3< \Re e(k_2\varepsilon)<2\pi(1+n)/3,\forall n\in\Z .$

\section*{Conclusions}
\addcontentsline{toc}{chapter}{Conclusions}

The present work has formulated an asymptotic homogenization approach for describing composites that have a periodic microstructure characteristic in presence of the generalized thermoelasticity theory with a single periodic spatially dependent relaxation time \cite{lord1967generalized}. The down-scaling relations herein developed permit such technique to express the micro-fields namely, displacement and temperature, in terms of the macroscopic fields and their derivatives, through the $ \mathcal{Q} $-periodic perturbation functions. The cell problems i.e., a cascade of inhomogeneous recursive differential problems defined on the unit cell $ \mathcal{Q} $ and having zero mean values over this unit cell, give the availability to seek for those perturbation functions.

The cell problems were derived substituting series expansions for both of the micro-fields in powers of the microstructural characteristic size $ \varepsilon $ into the micro governing equations and rearranging the outcomes as asymptotic expansions in terms of $ \varepsilon$ exponents. Inserting the down-scaling relations into the microscopic thermoelastic field equations has led to the average field equations of infinite order, which are asymptotically equivalent to the initial governing equations of the heterogeneous continuum, in which the formal solutions might be derived by plugging an asymptotic infinite series of the macro fields in powers of the microstructural size $\varepsilon.$ Particularly, truncating such series of the macroscopic displacement and temperature up to the zeroth order yield to the governing field equations of the homogeneous first order (Cauchy) thermoelastic medium equivalent to the heterogeneous one, which the elastic, thermal dilatation and heat conduction overall constitutive tensors have been analytically stated. {In particular, the overall heat conduction tensor depends on the thermal conductivity tensor and the relaxation time of the phases.} Consequently, the higher the order of the truncated asymptotic expansions of the macro fields the better the estimation for solutions of the heterogeneous problem.

The reliability of such analysis has been evaluated through a benchmark test between the presented first order homogenized model and the heterogeneous continuum over an illustrative example of a bi-dimensional two-phase periodic layered material endowed with an orthotropy axis parallel to the layering direction. In what regards the investigation of the dispersion properties of the heterogeneous thermoelastic medium the Floquet-Bloch theory has been implemented. Such path has led to a sixth order eigenvector-eigenvalue problem solved over the periodic cell
subjected to Floquet-Bloch periodic boundary conditions, which throughout the comparative analysis were considered the uncoupled and coupled scenarios. The eigenproblem provided an imaginary implicit function and a real implicit function of the complex frequency and wave vector, where the intersection between them defines the frequency spectrum. In both eigenvector-eigenvalue problems from the Floquet-Bloch approach, characterizing the heterogeneous material, the thermal waves from the uncoupled problem as well as the compressional-thermal waves from the coupled problem, and the shear waves were compared against their respectives when applied the first order homogenized approximation.

Over the comparisons carried out in this work, it has been achieved a very good agreement between the dispersion curves computed from the two different methods i.e., thermal, compressional-thermal and shear waves along and normal direction to the layering. In essence, such strong connection observed between the methods confirms that the multi-field homogenization technique of the first order herein proposed has been shown itself to be a quite good tool to estimate the macroscopical elastic and thermal overall properties of the equivalent heterogeneous body having periodic microstructure under thermoelastic phenomena with sufficient accuracy.

\section*{Acknowledgments}  

The authors gratefully acknowledge financial support from National Group of Mathematical Physics (GNFM-INdAM).
AB would like to acknowledge financial support from Compagnia San Paolo, project MINIERA no.  I34I20000380007 and from University of Trento, project UNMASKED 2020. MP would like to acknowledge financial support from the Italian Ministry of Education, University and Research (MIUR) to the research project of relevant national interest (PRIN 2017) “XFAST-SIMS: Extra-fast and accurate simulation of complex structural systems” (CUP: D68D19001260001).

%
\addcontentsline{toc}{chapter}{References}
\bibliographystyle{unsrtnat}

%
\newpage
\begin{center}
	{\Large{Supplementary Material of}}\\
	\vspace{0.3cm}
	{\LARGE{\titulo}}
\end{center}
\begin{center}
	\small{\authors}
\end{center}

\renewcommand\thesection{Section \Alph{section}}
\setcounter{section}{0}
\renewcommand\thesubsection{\Alph{section}.\arabic{subsection}.}

\section{Variational-asymptotic homogenization scheme: additional mathematical details}

\subsection{Asymptotic expansions of the microscopic field equations on the Laplace domain}
\label{sec::additionaldetailshomogenization}

In order to land on the homogenized field equations set in Sec. \ref{sec:expansions}, let us consider the formulas coming from the chain rule of differentiation

\begin{subequations}
	\begin{align}
		\dfrac{D}{Dx_{k}}& \hat{\boldsymbol{u}}\left(\textbf{x},\boldsymbol{\xi}=\dfrac{\textbf{x}}{\varepsilon}\right)=\left. \left[\dfrac{\partial \hat{u}_{h}(\textbf{x},\boldsymbol{\xi})}{\partial x_{k}} + \dfrac{\partial \hat{u}_{h}(\textbf{x},\boldsymbol{\xi})}{\partial \xi_{k}} \dfrac{\partial \xi_{k}}{\partial x_{k}}\right]\right|_{\boldsymbol{\xi}= \dfrac{\textbf{x}}{\varepsilon}}=\left.\left[\dfrac{\partial}{\partial x_{k}} \hat{u}_{h}(\textbf{x},\boldsymbol{\xi})+\dfrac{1}{\varepsilon}\hat{u}_{h,k}\right]\right|_{\boldsymbol{\xi}= \dfrac{\textbf{x}}{\varepsilon}},\label{eqn:utotalderivative}\\
		\dfrac{D}{Dx_{j}}& \hat{\theta}\left(\textbf{x},\boldsymbol{\xi}=\dfrac{\textbf{x}}{\varepsilon}\right)=\left. \left[\dfrac{\partial \hat{\theta}(\textbf{x},\boldsymbol{\xi})}{\partial x_{j}} + \dfrac{\partial \hat{\theta}(\textbf{x},\boldsymbol{\xi})}{\partial \xi_{j}} \dfrac{\partial \xi_{j}}{\partial x_{j}}\right]\right|_{\boldsymbol{\xi}= \dfrac{\textbf{x}}{\varepsilon}}=\left.\left[\dfrac{\partial}{\partial x_{j}} \hat{\theta}(\textbf{x},\boldsymbol{\xi})+\dfrac{1}{\varepsilon}\hat{\theta}_{,j}\right]\right|_{\boldsymbol{\xi}= \dfrac{\textbf{x}}{\varepsilon}},\label{eqn:thetatotalderivative}
	\end{align}
\end{subequations}
which introduces the macroscopic derivatives  $\partial\hat{u}_{h}/\partial x_{k},$ $\partial\hat{\theta}/\partial x_{j} ,$ and the microscopic derivatives $\hat{u}_{h,k},$ $\hat{\theta}_{,j}$ on the transformed Laplace domain. Applying them to the asymptotic expansions $~\eqref{eqr:ulaplace}$ and $ ~\eqref{eqr:thetalaplace}, $ one leads to

\begin{subequations}
	\begin{align}
		&\dfrac{D}{Dx_{k}} \hat{\boldsymbol{u}}\left(\textbf{x},\boldsymbol{\xi} = \dfrac{\textbf{x}}{\varepsilon}\right)=\left.\left[\left(\dfrac{\partial \hat{u}^{(0)}_{h}}{\partial x_{k}} + \varepsilon \dfrac{\partial \hat{u}^{(1)}_{h}}{\partial x_{k}} + \varepsilon^{2} \dfrac{\partial \hat{u}^{(2)}_{h}}{\partial x_{k}}+\cdots\right)+\dfrac{1}{\varepsilon} \left(\hat{u}^{0}_{h,k}+ \varepsilon \hat{u}^{(1)}_{h,k} + \varepsilon^{2} \hat{u}^{(2)}_{h,k}+\cdots\right)\right]\right|_{\boldsymbol{\xi}= \dfrac{\textbf{x}}{\varepsilon}},\label{eq:gradu}\\
		&\dfrac{D}{Dx_{j}} \hat{\theta}\left(\textbf{x},\boldsymbol{\xi} = \dfrac{\textbf{x}}{\varepsilon}\right)=\left.\left[\left(\dfrac{\partial \hat{\theta}^{(0)}}{\partial x_{j}} + \varepsilon \dfrac{\partial \hat{\theta}^{(1)}}{\partial x_{j}} + \varepsilon^{2} \dfrac{\partial \hat{\theta}^{(2)}}{\partial x_{j}}+\cdots\right)+\dfrac{1}{\varepsilon} \left(\hat{\theta}^{0}_{,j}+ \varepsilon \hat{\theta}^{(1)}_{,j} + \varepsilon^{2} \hat{\theta}^{(2)}_{,j}+\cdots\right)\right]\right|_{\boldsymbol{\xi}= \dfrac{\textbf{x}}{\varepsilon}}.\label{eq:gradtheta}
	\end{align}
\end{subequations}


As mentioned in Sec. \ref{sec:expansions}, rearranging properly the terms with equal powers of $\varepsilon,$ and taking advantage of the derivative equations $~\eqref{eq:gradu} $ and $  ~\eqref{eq:gradtheta},$  yield to the following asymptotic field equations

\begin{subequations}
	\begin{eqnarray}
		\label{eqn:uhomogenized}
		\begin{split}
			&\left(\varepsilon^{-2}\left(C^{m}_{ijhk}\hat{u}^{(0)}_{h,k} \right)_{,j}+\varepsilon^{-1} \left[\left( C^{m}_{ijhk}\left( \dfrac{\partial \hat{u}^{(0)}_{h}}{\partial x_{k}}+\hat{u}^{(1)}_{h,k} \right)\right)_{,j}+\dfrac{\partial}{\partial x_{j}}\left(C^{m}_{ijhk}\hat{u}^{(0)}_{h,k} \right)-\left(\alpha^{m}_{ij}\hat{\theta}^{(0)}\right)_{,j}\right]+\right.
			\\
			&+\varepsilon^{0} \left[\left(C^{m}_{ijhk}\left( \dfrac{\partial \hat{u}^{(1)}_{h}}{\partial x_{k}}+\hat{u}^{(2)}_{h,k} \right) \right)_{,j}+\dfrac{\partial}{\partial x_{j}} \left(C^{m}_{ijhk}\left(\dfrac{\partial\hat{u}^{(0)}_{h}}{\partial x_{k}}+\hat{u}^{(1)}_{h,k}\right)\right)-\left(\alpha^{m}_{ij}\hat{\theta}^{(1)}\right)_{,j}\right.+
			\\
			&-\left.\dfrac{\partial}{\partial x_{j}} \left({\alpha}^{m}_{ij}\hat{\theta}^{(0)}\right)+\hat{b}_{i}-\rho^{m}s^{2}u^{(0)}_{h}\right]+\varepsilon\left[\left( C^{m}_{ijhk}\left(\dfrac{\partial \hat{u}^{(2)}_{h}}{\partial x_{k}}+\hat{u}^{(3)}_{h,k} \right) \right)_{,j}+\right.\\
			&\left.\left.+\dfrac{\partial}{\partial x_{j}} \left(C^{m}_{ijhk}\left(\dfrac{\partial\hat{u}^{(1)}_{h}}{\partial x_{k}}+\hat{u}^{(2)}_{h,k}\right)\right)-\left.\left({\alpha}^{m}_{ij}\hat{\theta}^{(2)}\right)_{,j}-\dfrac{\partial}{\partial x_{j}}\left({\alpha}^{m}_{ij}\hat{\theta}^{(1)}\right)-\rho^{m}s^{2}\hat{u}^{(1)}_{h}\right]+\mathcal{O}(\varepsilon^{2})\right)\right|_{\boldsymbol{\xi}= \frac{\textbf{x}}{\varepsilon}}=0,
		\end{split}
	\end{eqnarray}
	\begin{eqnarray}
		\label{eqn:thetahomogenized}
		\begin{split}
			&\left(\varepsilon^{-2}\left({K}^{m}_{ij}\hat{\theta}^{(0)}_{,j} \right)_{,i}+\varepsilon^{-1} \left[\left({K}^{m}_{ij}\left( \dfrac{\partial \hat{\theta}^{(0)}}{\partial x_{j}}+\hat{\theta}^{(1)}_{,j} \right) \right)_{,i}+\dfrac{\partial}{\partial x_{i}} \left({K}^{m}_{ij}\hat{\theta}^{(0)}_{,j}\right)-\left(\alpha^{m}_{ij}s\hat{u}^{(0)}_{i,j}\right)\right]+\right.
			\\
			& +\varepsilon^{0} \left[\left({K}^{m}_{ij}\left(\dfrac{\partial \hat{\theta}^{(1)}}{\partial x_{j}}+\hat{\theta}^{(2)}_{,j}\right) \right)_{,i}+\dfrac{\partial}{\partial x_{i}} \left({K}^{m}_{ij}\left(\dfrac{\partial\hat{\theta}^{(0)}}{\partial x_{j}}+\hat{\theta}^{(1)}_{,j}\right)\right)-\alpha^{m}_{ij}s\left( \dfrac{\partial \hat{u}^{(0)}_{i}}{\partial x_{j}}+\hat{u}^{(1)}_{i,j}\right)\right.+
			\\
			&=\left.\hat{r}-p^ms\hat\theta^{(0)}\right]+\varepsilon \left[\left({K}^{m}_{ij}\left(\dfrac{\partial \hat{\theta}^{(2)}}{\partial x_{j}}+\hat{\theta}^{(3)}_{,j} \right) \right)_{,i}+\dfrac{\partial}{\partial x_{i}} \left({K}^{m}_{ij}\left(\dfrac{\partial\hat{\theta}^{(1)}}{\partial x_{j}}+\hat{\theta}^{(2)}_{,j}\right)\right)+\right.
			\\
			&\left.\left.\left.-\alpha^{m}_{ij}s\left( \dfrac{\partial \hat{u}^{(1)}_{i}}{\partial x_{j}}+\hat{u}^{(2)}_{i,j}\right)-p^ms\hat\theta^{(1)}\right]+\left.\mathcal{O}(\varepsilon^{2})\right]\right)\right|_{\boldsymbol{\xi}= \frac{\textbf{x}}{\varepsilon}}=0,
		\end{split}
	\end{eqnarray}
\end{subequations}
where the equation $ ~\eqref{eqn:uhomogenized} $ is the homogenized micro-displacement field, and the equation $ ~\eqref{eqn:thetahomogenized} $ is the homogenized micro-temperature field.

Recalling the interface conditions $(\ref{eq:InterfaceConditionMicroDisplacement})-(\ref{eq:InterfaceConditionMicrotemperature1}),$ they may be rewritten in terms of the components, thus

\begin{subequations}
	\begin{align}
		&
		\left.
		\left[
		\left[
		\hat{u}_h
		\right]
		\right]
		\right|_{\mathbf{x}\in\Sigma}=0,\label{eq:InterfaceConditionMicroDisplacementComponents}\\
		&
		\left.
		\left[
		\left[
		\left(
		C^m_{ijhk}\dfrac{D\hat{u}_h}{Dx_k}
		-
		\alpha^m_{ij}\hat\theta
		\right)n_j
		\right]
		\right]
		\right|_{\mathbf{x}\in\Sigma}=0,
		\label{eq:InterfaceConditionMicroDisplacementComponents1}\\
		&
		\left.
		\left[
		\left[
		\hat\theta
		\right]
		\right]
		\right|_{\mathbf{x}\in\Sigma}=0,\label{eq:InterfaceConditionMicrotemperatureComponents}\\
		&
		\left.
		\left[
		\left[
		-K^m_{ij}(s)\dfrac{D\hat\theta}{Dx_j}
		n_i
		\right]
		\right]
		\right|_{\mathbf{x}\in\Sigma}=0.\label{eq:InterfaceConditionMicrotemperatureComponents1}
	\end{align}
\end{subequations}

Also, once the micro fields $\hat{\boldsymbol{u}}(\mathbf{x},\boldsymbol{\xi},s),$ and $\hat\theta(\mathbf{x},\boldsymbol{\xi},s)$ are supposed to be $\mathcal{Q}$-periodic regular functions of the variable $\mathbf{x},$ it is then possible to write the interface conditions $ (\ref{eq:InterfaceConditionMicroDisplacementComponents})$ to $(\ref{eq:InterfaceConditionMicrotemperatureComponents1})$ in terms of the fast variable $\boldsymbol{\xi}.$  \citep{Bakhvalov1984}. Therefore the asymptotic expansions $ (\ref{eqn:u}) $ and $ (\ref{eqn:theta}), $ among the derivative formulae $ ~\eqref{eq:gradu}$ and $ ~\eqref{eq:gradtheta} ,$ and the interface conditions $ (\ref{eq:InterfaceConditionMicroDisplacementComponents})$ and $(\ref{eq:InterfaceConditionMicroDisplacementComponents1})$ are rewritten in terms of $ \varepsilon $ as

\begin{eqnarray}
	\label{eq:InterfaceConditionMicroDisplacementOnSigma1}
	\begin{split}
		&
		\left.
		\left[
		\left[
		\hat{u}_h^{(0)}
		\right]
		\right]
		\right|_{\boldsymbol{\xi}\in\Sigma_1}
		+
		\varepsilon
		\left.
		\left[
		\left[
		\hat{u}_h^{(1)}
		\right]
		\right]
		\right|_{\boldsymbol{\xi}\in\Sigma_1}
		+
		\varepsilon^2
		\left.
		\left[
		\left[
		\hat{u}_h^{(2)}
		\right]
		\right]
		\right|_{\boldsymbol{\xi}\in\Sigma_1}
		+
		\mathcal{O}(\varepsilon^{3})
		=0,
		\\
		&
		\dfrac{1}{\varepsilon}
		\left.
		\left[
		\left[
		\left(
		C_{ijhk}^{m}\,\hat{u}_{h,k}^{(0)}
		\right)n_j
		\right]
		\right]
		\right|_{\boldsymbol{\xi}\in\Sigma_1}
		+
		\varepsilon^0
		\left.
		\left[
		\left[
		\left(
		C_{ijhk}^{m}
		\left(
		\dfrac{\partial \hat{u}_h^{(0)}}{\partial x_k}+
		\hat{u}_{h,k}^{(1)}
		\right)
		-
		\alpha_{ij}^m\,\hat\theta^{(0)}
		\right)n_j
		\right]
		\right]
		\right|_{\boldsymbol{\xi}\in\Sigma_1}
		+
		\\
		&+
		\varepsilon
		\left.
		\left[
		\left[
		\left(
		C_{ijhk}^{m}
		\left(
		\dfrac{\partial \hat{u}_h^{(1)}}{\partial x_k}+
		\hat{u}_{h,k}^{(2)}
		\right)
		-
		\alpha_{ij}^m\,\hat\theta^{(1)}
		\right)n_j
		\right]
		\right]
		\right|_{\boldsymbol{\xi}\in\Sigma_1}
		+
		\\
		&+
		\varepsilon^2
		\left.
		\left[
		\left[
		\left(
		C_{ijhk}^{m}
		\left(
		\dfrac{\partial \hat{u}_h^{(2)}}{\partial x_k}+
		\hat{u}_{h,k}^{(3)}
		\right)
		-
		\alpha_{ij}^m\,\hat\theta^{(2)}
		\right)n_j
		\right]
		\right]
		\right|_{\boldsymbol{\xi}\in\Sigma_1} + \mathcal{O}(\varepsilon^{3}) =0,
	\end{split}
\end{eqnarray} 
and the interface conditions $ (\ref{eq:InterfaceConditionMicrotemperatureComponents}) $ and $ (\ref{eq:InterfaceConditionMicrotemperatureComponents1}) $ become

\begin{eqnarray}
	\label{eq:InterfaceConditionMicroTemperatureOnSigma1}
	\begin{split}
		&\left.
		\left[
		\left[
		\hat\theta^{(0)}
		\right]
		\right]
		\right|_{\boldsymbol{\xi}\in\Sigma_1}
		+
		\varepsilon
		\left.
		\left[
		\left[
		\hat\theta^{(1)}
		\right]
		\right]
		\right|_{\boldsymbol{\xi}\in\Sigma_1}
		+
		\varepsilon^2
		\left.
		\left[
		\left[
		\hat\theta^{(2)}
		\right]
		\right]
		\right|_{\boldsymbol{\xi}\in\Sigma_1}
		+
		\mathcal{O}(\varepsilon^{3})
		=0,
		\\
		&
		\dfrac{1}{\varepsilon}
		\left.
		\left[
		\left[
		\left(
		K_{ij}^{m}\,\hat\theta_{,j}^{(0)}
		\right)n_i
		\right]
		\right]
		\right|_{\boldsymbol{\xi}\in\Sigma_1}
		+
		\varepsilon^0
		\left.
		\left[
		\left[
		\left(
		K_{ij}^{m}
		\left(
		\dfrac{\partial\hat\theta^{(0)}}{\partial x_j}
		+
		\hat\theta_{,j}^{(1)}
		\right)
		\right)n_i
		\right]
		\right]
		\right|_{\boldsymbol{\xi}\in\Sigma_1}+
		\\
		&+
		\varepsilon
		\left.
		\left[
		\left[
		\left(
		K_{ij}^{m}
		\left(
		\dfrac{\partial\hat\theta^{(1)}}{\partial x_j}
		+
		\hat\theta_{,j}^{(2)}
		\right)
		\right)n_i
		\right]
		\right]
		\right|_{\boldsymbol{\xi}\in\Sigma_1}
		+
		\varepsilon^2
		\left.
		\left[
		\left[
		\left(
		K_{ij}^{m}
		\left(
		\dfrac{\partial\hat\theta^{(2)}}{\partial x_j}
		+
		\hat\theta_{,j}^{(3)}
		\right)
		\right)n_i
		\right]
		\right]
		\right|_{\boldsymbol{\xi}\in\Sigma_1}+\mathcal{O}(\varepsilon^{3})=0,
	\end{split}
\end{eqnarray}
where $\Sigma_1$ denotes the interface between two different phases in the unit cell $\mathcal{Q}$.

\subsection{Recursive differential problems and their solutions}
\label{sec:recursivesection}

After some algebraic rearrangements on the asymptotic field Eqs. $\eqref{eqn:uhomogenized}$ and $\eqref{eqn:thetahomogenized},$ a hierarchical set of recursive partial differential problems, in terms of the sensitivities ${\hat{u}}^{(l)}_h,$ and $ \hat\theta^{(l)} ,$ has arisen. For this matter, at the order $\varepsilon^{-2}$, the differential problems coming from Eqs. $\eqref{eqn:uhomogenized}$ and $~\eqref{eqn:thetahomogenized}$ are, respectively,

\begin{subequations}
	\begin{align}
		&\left(C^{m}_{ijhk}\boldsymbol{\hat{u}}^{(0)}_{h,k}\right)_{,j}= f^{(0)}_{i}(\textbf{x}),\label{eq:epsilon-2f}\\
		&\left(K^{m}_{ij}\hat{\theta}^{(0)}_{,j}\right)_{,i}= g^{(0)}(\textbf{x}),\label{eq:epsilon-2g}
	\end{align} 
\end{subequations}
with interface conditions

\begin{subequations}
	\begin{eqnarray}
		\label{eq:InterfaceConditionsDisplacOrder-2}
		&&\left.
		\left[
		\left[
		\hat{u}_h^{(0)}
		\right]
		\right]
		\right|_{\boldsymbol{\xi}\in\Sigma_1}=0,\\
		&&
		\left.
		\left[
		\left[
		\left(
		C_{ijhk}^{m}\,\hat{u}_{h,k}^{(0)}
		\right)n_j
		\right]
		\right]
		\right|_{\boldsymbol{\xi}\in\Sigma_1}=0,
		\\
		\label{eq:InterfaceConditionsTermicaOrder-2}
		&&\left.
		\left[
		\left[
		\hat\theta^{(0)}
		\right]
		\right]
		\right|_{\boldsymbol{\xi}\in\Sigma_1}=0,\\
		&&
		\left.
		\left[
		\left[
		\left(
		K_{ij}^{m}\hat\theta_{,j}^{(0)}
		\right)n_i
		\right]
		\right]
		\right|_{\boldsymbol{\xi}\in\Sigma_1}=0.
	\end{eqnarray}
\end{subequations}

As matter of fact, since we are searching for solutions in the class of $\mathcal{Q}$-periodic solutions $\hat{u}^{(0)}_{h}$ and $ \hat{\theta}^{(0)}, $ it can be checked
\citep{Bakhvalov1984,SmyshlyaevCherednichenko2000} that there exists a unique solution of the equations $~\eqref{eq:epsilon-2f}$ and $~\eqref{eq:epsilon-2g},$ up to a constant, more specifically, since they are elliptic differential equations in the divergence forms, with vanishing mean values of the source terms over the unit cell $ \mathcal{Q} ,$ implies the existence of a $ \mathcal{Q}$-periodic regular solution. Such result gives rise to the also known as solvability condition for these differential problems, implying that the source terms are $f^{(0)}_{i}(\textbf{x})=0$ and $ g^{(0)}(\textbf{x})=0 ,$ therefore the differential problems $~\eqref{eq:epsilon-2f}$ and $~\eqref{eq:epsilon-2g},$ take the form

\begin{subequations}
	\begin{align}
		&\left(C^{m}_{ijhk}\hat{u}^{(0)}_{h,k}\right)_{,j}=0,\label{eqn:solmecanica-2}\\
		&\left({K}^{m}_{ij}\hat{\theta}^{(0)}_{,j}\right)_{,i}=0.\label{eqn:soltermica-2}	
	\end{align}
\end{subequations}
Hence, the solution of the first term of the macroscopic displacement expansion field transformed is given by

\begin{equation}
	\hat{u}^{(0)}_{h}(\textbf{x},\boldsymbol{\xi},s) = \hat{U}^{M}_{h}(\textbf{x},s),\label{eqn:solutiondisplac-2}
\end{equation}
and the solution for the first term of the macroscopic temperature expansion field is given by

\begin{equation}
	\hat{\theta}^{(0)}(\textbf{x},\boldsymbol{\xi},s) = \hat{\Theta}^{M}(\textbf{x},s).\label{eqn:solutiontemp-2}
\end{equation}
It is important to notice here that both solutions $\hat{U}^{M}_{h}(\textbf{x},s)$ and $ \hat{\Theta}^{M}(\textbf{x},s) $ are no longer dependent on the fast variable.

Proceeding with the succeeding terms related to $\varepsilon^{-1}$ in Eqs., $~\eqref{eqn:uhomogenized}$ and $~\eqref{eqn:thetahomogenized},$ and using recursively the two solutions above obtained, it follows that 

\begin{subequations}
	\begin{align}
		&\left(C^{m}_{ijhk}\hat{u}^{(1)}_{h,k}\right)_{,j}+C^{m}_{ijhk,j}\dfrac{\partial \hat{U}^{M}_{h}}{\partial x_{k}}- {\alpha}^{m}_{ij,j}\hat{\Theta}^{M}=f^{(1)}_{i}(\textbf{x}),\label{eq:epsilon-1f}\\
		&\left({K}^{m}_{ij}\hat{\theta}^{(1)}_{,j}\right)_{,i}+{K}^{m}_{ij,i}\dfrac{\partial \hat{\Theta}^{M}}{\partial x_{j}}=  g^{(1)}(\textbf{x}),\label{eq:epsilon-1g}
	\end{align} 
\end{subequations}
since $\hat{U}^{M}_{h,k}=0$ and $ \hat\Theta^{M}_{,j}=0. $ The interface conditions are, respectively,

\begin{subequations}
	\begin{eqnarray}
		\label{eq:InterfaceConditionsDisplacOrder-1}
		&&\left.
		\left[
		\left[
		\hat{u}_h^{(1)}
		\right]
		\right]
		\right|_{\boldsymbol{\xi}\in\Sigma_1}=0,\\
		&&
		\left.
		\left[
		\left[
		\left(
		C_{ijhk}^{m}
		\left(
		\dfrac{\partial \hat{U}_h^{M}}{\partial x_k}+
		\hat{u}_{h,k}^{(1)}
		\right)
		-
		\alpha_{ij}^m\,\hat\Theta^{M}
		\right)n_j
		\right]
		\right]
		\right|_{\boldsymbol{\xi}\in\Sigma_1}=0,
		\\
		\label{eq:InterfaceConditionsTermicaOrder-1}
		&&\left.
		\left[
		\left[
		\hat\theta^{(1)}
		\right]
		\right]
		\right|_{\boldsymbol{\xi}\in\Sigma_1}=0,\\
		&&
		\left.
		\left[
		\left[
		\left(
		K_{ij}^{m}
		\left(
		\dfrac{\partial\hat\Theta^{M}}{\partial x_j}
		+
		\hat\theta_{,j}^{(1)}
		\right)
		\right)n_i
		\right]
		\right]
		\right|_{\boldsymbol{\xi}\in\Sigma_1}=0.
	\end{eqnarray}
\end{subequations}

Likewise, the solvability condition on the class of $\mathcal{Q}$-periodic functions guarantees that  

\begin{subequations}
	\begin{align}
		f^{(1)}_{i}(\textbf{x})=&\langle f^{(1)}_{i}(\textbf{x})\rangle=\langle C^{m}_{ijhk,j} \rangle \dfrac{\partial \hat{U}^{M}_{h}}{\partial x_{k}}-\langle{\alpha}^{m}_{ij,j}\rangle\hat{\Theta}^{M},\label{mecanicameanvalue-1}\\
		g^{(1)}(\textbf{x})=&\langle g^{(1)}(\textbf{x})\rangle=\langle {K}^{m}_{ij,i} \rangle \dfrac{\partial \hat{\Theta}^{M}}{\partial x_{j}},\label{termicameanvalue-1}
	\end{align} 
\end{subequations} 
where $\langle (\cdot) \rangle=\frac{1}{|\mathcal{Q}|} \int_{\mathcal{Q}}^{} (\cdot) d\boldsymbol{\xi}$ and $|\mathcal{Q}|= \delta\gamma$ gives the mean value over the unit cell $\mathcal{Q}.$ Moreover, the $\mathcal{Q}$-periodicity of the components $C^{m}_{ijhk},$ $ \alpha^{m}_{ij} $ and ${K}^{m}_{ij}, $ along with the divergence theorem, entail both $f^{(1)}_{i}(\textbf{x})=0$ and $  g^{(1)}(\textbf{x})=0,$ thus the differential problems

\begin{subequations}
	\begin{align}
		&\left(C^{m}_{ijhk}\hat{u}^{(1)}_{h,k}\right)_{,j}+C^{m}_{ijhk,j}\dfrac{\partial \hat{U}^{M}_{h}}{\partial x_{k}}-{\alpha}^{m}_{ij,j}\hat{\Theta}^{M}=0, \quad \forall\dfrac{\partial \hat{U}^{M}_{h}}{\partial x_{k}},\hat{\Theta}^{M}\label{eqn:solmecanica-1}\\
		&\left({K}^{m}_{ij}\hat{\theta}^{(1)}_{,j}\right)_{,i}+{K}^{m}_{ij,i}\dfrac{\partial \hat{\Theta}^{M}}{\partial x_{j}}=0, \quad \forall \dfrac{\partial \hat{\Theta}^{M}}{\partial x_{j}}\label{eqn:soltermica-1}
	\end{align}
\end{subequations}
have the following solutions, respectively

\begin{subequations}
	\begin{align}
		\hat{u}^{(1)}_{h}(\textbf{x},\boldsymbol{\xi},s)=&N^{(1,0)}_{hpq_{1}}(\boldsymbol{\xi})\dfrac{\partial \hat{U}^{M}_{p}}{\partial x_{q_{1}}}+\tilde{N}^{(1,0)}_h(\boldsymbol{\xi})\hat{\Theta}^M,\label{eqn:solutiondisplac-1}\\
		\hat{\theta}^{(1)}(\textbf{x},\boldsymbol{\xi},s)=& M^{(1,0)}_{q_{1}}(\boldsymbol{\xi},s)\dfrac{\partial \hat{\Theta}^{M}}{\partial x_{q_{1}}},\label{eqn:solutiontemp-1}
	\end{align}
\end{subequations}
where $N^{(1,0)}_{hpq_{1}},$ $ \tilde{N}^{(1,0)}_h $  and $ M^{(1,0)}_{q_{1}},$ are the perturbation functions, which each of them depends on the fast variable $\boldsymbol{\xi}.$ On Sec. $ \ref{Sec::CellsProblems} ,$ in order to impose uniqueness of the homogenized solutions $\hat{u}_h $ and $ \hat\theta ,$ the perturbation functions must be supposed to have zero mean value over the unit cell $\mathcal{Q}$ and so $N^{(1,0)}_{hpq_{1}},$ $\tilde{N}^{(1,0)}_{h}$ and $M^{(1,0)}_{q_{1}}$ comply with the normalization condition, which means that

\begin{subequations}
	\begin{align}
		\label{eq:normalization1}
		&\langle N^{(1,0)}_{hpq_{1}} \rangle = \dfrac{1}{|\mathcal{Q}|} \int_{\mathcal{Q}} N^{(1,0)}_{hpq_{1}}(\boldsymbol{\xi}) d\boldsymbol{\xi}=0,\\
		\label{eq:normalization2}
		&\langle \tilde{N}^{(1,0)}_h \rangle = \dfrac{1}{|\mathcal{Q}|} \int_{\mathcal{Q}}\tilde{N}^{(1,0)}_h(\boldsymbol{\xi}) d\boldsymbol{\xi}=0,\\
		\label{eq:normalization3}
		&\langle M^{(1,0)}_{q_{1}} \rangle = \dfrac{1}{|\mathcal{Q}|} \int_{\mathcal{Q}} M^{(1,0)}_{q_{1}}(\boldsymbol{\xi},s) d\boldsymbol{\xi}=0.
	\end{align} 
\end{subequations}

At order $\varepsilon^{0},$ the differential problems are

\begin{subequations}
	\begin{align}
		\begin{split}
			\label{eq:epsilon0f}
			& \left(C^{m}_{ijhk}\left( \dfrac{\partial \hat{u}^{(1)}_{h}}{\partial x_{k}}+\hat{u}^{(2)}_{h,k}\right)\right)_{,j}+\dfrac{\partial}{\partial x_{j}}\left(C^{m}_{ijhk}\left(\dfrac{\partial \hat{u}^{(0)}_{h}}{\partial x_{k}}+\hat{u}^{(1)}_{h,k}\right)\right)-\left(\alpha^m_{ij}\hat\theta^{(1)}\right)_{,j}-\dfrac{\partial}{\partial x_j}\left(\alpha^m_{ij}\hat\theta^{(0)}\right)+\\
			&-\rho^{m}s^{2}\hat{u}^{(0)}_{i} = f^{(2)}_{i}(\textbf{x}),
		\end{split}\\
		\begin{split}
			&\left({K}^{m}_{ij}\left(\dfrac{\partial \hat{\theta}^{(1)}}{\partial x_{j}}+\hat{\theta}^{(2)}_{,j}\right)\right)_{,i}+\dfrac{\partial}{\partial x_{i}}\left({K}^{m}_{ij}\left(\dfrac{\partial \hat{\theta}^{(0)}}{\partial x_{j}}+\hat{\theta}^{(1)}_{,j}\right)\right)-\alpha^{m}_{ij}\left(\dfrac{\partial \hat{u}^{(0)}_{i}}{\partial x_{j}}+\hat{u}^{(1)}_{i,j}\right)+\\
			&-\rho^{m}s\hat{\theta}^{(0)} = g^{(2)}(\textbf{x}).\label{eq:epsilon0g}
		\end{split}
	\end{align}
\end{subequations}

Considering the solutions $~\eqref{eqn:solutiondisplac-2},$ $~\eqref{eqn:solutiondisplac-1}$ at order $\varepsilon^{-2}$ and $\varepsilon^{-1},$ respectively, from the displacement expansion, and the solutions $~\eqref{eqn:solutiontemp-2},$ $~\eqref{eqn:solutiontemp-1},$ at order $\varepsilon^{-2}$ and $\varepsilon^{-1}$, respectively, from the temperature expansion, the differential problems $~\eqref{eq:epsilon0f}$ and $~\eqref{eq:epsilon0g}$ are turned into

\begin{subequations}
	\begin{flalign}
		\begin{split}\label{eq:epsilon0fsubst} &\left(C^{m}_{ijhk}\hat{u}^{(2)}_{h,k}\right)_{,j}+\left(\left(C^{m}_{ijhk}N^{(1,0)}_{hpq_{1}}\right)_{,j}+C^{m}_{iq_{1}pk}+\left(C^{m}_{ikhj}N^{(1,0)}_{hpq_{1},j}\right)\right)\dfrac{\partial^{2}\hat{U}^{M}_{p}}{\partial x_{q_{1}}\partial x_{k}}+\\
			&+\left(\left(C^{m}_{ijhk}\tilde{N}^{(1,0)}_{h}\right)_{,j}+C^{m}_{ikhj}\tilde{N}^{(1,0)}_{h,j}-\left(\alpha^m_{ij}M^{(1,0)}_k\right)_{,j}-\alpha^m_{ik}\right)\dfrac{\partial\hat\Theta^M}{\partial x_k}-\rho^{m}s^{2}\hat{U}^{M}_{i}=f^{(2)}_{i}(\textbf{x}),\\
		\end{split}\\
		\begin{split}\label{eq:epsilon0gsubst}
			&\left({K}^{m}_{ij}\hat{\theta}^{(2)}_{,j}\right)_{,i}+\left(\left({K}^{m}_{ij}M^{(1,0)}_{q_{1}}\right)_{,i}+{K}^{m}_{q_{1}j}+\left({K}^{m}_{ji}M^{(1,0)}_{q_{1},i}\right)\right)\dfrac{\partial^{2}\hat{\Theta}^{M}}{\partial x_{q_{1}}\partial x_{j}}+\\
			&-\left(\alpha^m_{ij}N^{(1,0)}_{ipq_{1},j}+\alpha^m_{pq_{1}}\right)s\dfrac{\partial\hat{U}^M_p}{\partial x_{q_{1}}}-\left(\alpha^m_{ij}\tilde{N}^{(1,0)}_{i,j}+p^m\right)s\hat{\Theta}^{M} = g^{(2)}(\textbf{x}),
		\end{split}
	\end{flalign}
\end{subequations}
with interface conditions

\begin{subequations}
	\begin{flalign}
		\begin{split}
			&\left.
			\left[
			\left[
			\hat{u}_h^{(2)}
			\right]
			\right]
			\right|_{\boldsymbol{\xi}\in\Sigma_1}=0,
			\\
			&\left.
			\left[
			\left[
			\left(
			C_{ijhk}^{m}
			\left(
			N^{(1,0)}_{hpq_{1},j}\dfrac{\partial^{2}\hat{U}^{M}_{p}}{\partial x_{q_{1}}\partial x_{k}}+
			\tilde{N}^{(1,0)}_{h}\dfrac{\partial\hat\Theta^M}{\partial x_k}+
			\hat{u}_{h,k}^{(2)}
			\right)
			-
			\alpha_{ij}^mM^{(1,0)}_{q_1}\dfrac{\partial\hat\Theta^M}{\partial x_{q_1}}
			\right)n_j
			\right]
			\right]
			\right|_{\boldsymbol{\xi}\in\Sigma_1}=0,
			\label{eq:InterfaceConditionsDisplacOrder0}
		\end{split}\\
		\begin{split}
			&\left.
			\left[
			\left[
			\hat\theta^{(2)}
			\right]
			\right]
			\right|_{\boldsymbol{\xi}\in\Sigma_1}=0,
			\\
			&\left.
			\left[
			\left[
			\left(
			K_{ij}^{m}
			\left(
			M^{(1,0)}_{q_{1}}\dfrac{\partial^{2}\hat{\Theta}^{M}}{\partial x_{q_{1}}\partial x_{j}}
			+
			\hat\theta_{,j}^{(2)}
			\right)
			\right)n_i
			\right]
			\right]
			\right|_{\boldsymbol{\xi}\in\Sigma_1}=0.
			\label{eq:InterfaceConditionsTermicaOrder0}
		\end{split}
	\end{flalign}
\end{subequations}

Simetrizing the terms with the derivative of second order on the two differential problems $~\eqref{eq:epsilon0fsubst}$ and $~\eqref{eq:epsilon0gsubst},$ and once again, from the solvability condition of differential problem on the class of $\mathcal{Q}$-periodic functions and the divergence theorem lead to, respectively,

\begin{subequations}
	\begin{align}
		\label{mecanicameanvalue0}
		\begin{split}
			f^{(2)}_{i}(\textbf{x})=&\dfrac{1}{2}\left\langle C^{m}_{iq_{1}pk}+C^{m}_{ikhj}N^{(1,0)}_{hpq_{1},j}+C^{m}_{ikpq_{1}}+C^{m}_{iq_{1}hj}N^{(1,0)}_{hpk,j}\right\rangle\dfrac{\partial^{2}\hat{U}^{M}_{p}}{\partial x_{q_{1}}\partial x_{h}}+
			\\
			+&\left\langle C^{m}_{ikhj}\tilde{N}^{(1,0)}_{h,j}-\alpha^m_{ik}\right\rangle\dfrac{\partial\hat\Theta^M}{\partial x_k}-\Big\langle\rho^{m}\Big\rangle s^{2}\hat{U}^{M}_{i},
		\end{split}
	\end{align}
	\begin{align}
		\label{termicameanvalue0}
		\begin{split}
			g^{(2)}(\textbf{x})=&\dfrac{1}{2}\left\langle {K}^{m}_{q_{1}j}+{K}^{m}_{ji}M^{(1,0)}_{q_{1},i}+{K}^{m}_{jq_{1}}+{K}^{m}_{q_{1}i}M^{(1,0)}_{j,i}\right\rangle\dfrac{\partial^{2}\hat{\Theta}^{M}}{\partial x_{q_{1}}\partial x_{j}}-
			\\
			+&\left\langle\alpha^{m}_{ij}N^{(1,0)}_{ipq_{1},j}+\alpha^m_{pq_{1}}\right\rangle s\dfrac{\partial\hat{U}^M_p}{\partial x_{q1}}-\left\langle\alpha^m_{ij}\tilde{N}^{(1,0)}_{i,j}+p^{m}\right\rangle s\hat{\Theta}^{M}.
		\end{split}
	\end{align}
\end{subequations}
Note that, even though the interface conditions $ (\ref{eq:InterfaceConditionsDisplacOrder0}) $ and $ (\ref{eq:InterfaceConditionsTermicaOrder0}) $ are simetrized as well, they are kept hidden on this regard.

Consequently, the solution for each $ \hat{U}^M_p ,$ $ \hat{\Theta}^M $ and their derivatives of the differential problems above, at the order $\varepsilon^{0},$ are

\begin{subequations}
	\begin{equation}
		\label{eqn:solutiondisplac0}
		\hat{u}^{(2)}_{h}(\textbf{x},\boldsymbol{\xi},s) = N^{(2,0)}_{hpq_{1}q_{2}}(\boldsymbol{\xi})\dfrac{\partial^{2}\hat{U}^{M}_{p}}{\partial x_{q_{1}}\partial x_{q_{2}}}+\tilde{N}^{(2,1)}_{hq_{1}}(\boldsymbol{\xi})\dfrac{\partial\hat\Theta^M}{\partial x_{q_{1}}}+s^{2}N^{(2,2)}_{hq_{1}}(\boldsymbol{\xi})\hat{U}^{M}_{p},
	\end{equation}  
	\begin{equation}
		\label{eqn:solutiontemp0}
		\hat{\theta}^{(2)}(\textbf{x},\boldsymbol{\xi},s) = M^{(2,0)}_{q_{1}q_{2}}(\boldsymbol{\xi})\dfrac{\partial^{2}\hat{\Theta}^{M}}{\partial x_{q_{1}}\partial x_{q_{2}}}+s\tilde{M}^{(2,1)}_{pq_{1}}(\boldsymbol{\xi})\dfrac{\partial\hat{U}^M_p}{\partial x_{q_{1}}}+sM^{(2,1)}(\boldsymbol{\xi})\hat{\Theta}^{M},
	\end{equation}
\end{subequations}
where $N^{(2,0)}_{hpq_{1}q_{2}},$ $ \tilde{N}^{(2,1)}_{hq_{1}}, $ $N^{(2,2)}_{hq_{1}},$  $ M^{(2,0)}_{q_{1}q_{2}}, $  $ \tilde{M}^{(2,1)}_{pq_{1}} $  and $ M^{(2,1)} $ are the perturbation functions depending on the parameters $ \boldsymbol{\xi} $ and $s.$

\subsection{Average field equation of infinite order}
\label{sec:infinitorder}

Recalling the structure of the field equations $ (\ref{fsourceterms}) $ and $ (\ref{gsourceterms}) $ and substituting each $f_i^{(l)}(\mathbf{x})$ and $g^{(l)}(\mathbf{x}),$ obtained from each relative recursive differential problem into them, one derives the average field equations of infinite order

\begin{subequations}
	\begin{align}
		&n^{(2,0)}_{ipq_1q_2}\dfrac{\partial^2\hat{U}_h^M}{\partial x_{q_1} \partial x_{q_2}}+\tilde{n}^{(2,0)}_{iq_1} \dfrac{\partial\hat\Theta^M}{\partial x_{q_1}}-n_{ip}^{(2,2)}s^2 \hat{U}_h^M +\mathcal{O}\left( {\bf{\varepsilon}} \right)+\hat{b}_i= 0, \label{eq:averagefieldu}\\
		&w^{(2,0)}_{q_1q_2}\dfrac{\partial^2 \hat\Theta^M}{\partial x_{q_1} \partial x_{q_2}}- \tilde{w}^{(2,1)}_{pq_1} s\dfrac{\partial\hat{U}^M_h}{\partial x_{q_1}}- w^{(2,1)}s\hat\Theta^M+\mathcal{O}\left( {\bf{\varepsilon}} \right)+\hat{r} = 0,
		\label{eq:averagefieldtheta}
	\end{align}
\end{subequations}
where the constant global constitutive tensors factors are

\begin{subequations}
	\begin{align}
		&n^{(2,0)}_{ipq_1q_2} = \dfrac{1}{2} \left \langle C_{iq_1pq_2}^{m} + C_{iq_2hj}^{m} N^{(1,0)}_{hpq_1,j}+C_{iq_2pq_1}^{m} + C_{iq_1hj}^{m} N^{(1,0)}_{hpq_2,j}\right \rangle, \label{eq:componentsglobalconstitutivetensors}\\
		&\tilde{n}^{(2,0)}_{iq_1}= \left\langle C_{iq_1hj}^{m}\tilde{N}^{(1,0)}_{h,j}-\alpha^m_{iq_1}\right\rangle,\\
		& n_{ip}^{(2,2)} = \left\langle\rho^m\right\rangle\delta_{ip},\\
		& w^{(2,0)}_{q_1q_2}=\dfrac{1}{2} \left\langle K^m_{q_1q_2}+K^m_{q_2i}M^{(1,0)}_{q_1,i}+K^m_{q_2q_1}+K^m_{q_1i}M^{(1,0)}_{q_2,i}\right\rangle,\\
		&\tilde{w}^{(2,1)}_{pq_1}=\left\langle\alpha_{ij}^{m}N^{(1,0)}_{ipq_1,j}+\alpha^m_{pq_1}
		\right\rangle,\\
		&w^{(2,1)}= \left\langle\alpha_{ij}^{m}\tilde{N}^{(1,0)}_{i,j}-p^m_{iq_1}
		\right\rangle.
	\end{align}
\end{subequations}

In the interest of deriving explicit solutions of the equations $(\ref{eq:averagefieldu}) $ and $(\ref{eq:averagefieldtheta}), $ the macroscopic variables $\hat{U}^M_h$ and $\hat\Theta^M$ are asymptotically expanded as follows

\begin{subequations}
	\begin{align}
		&\hat{U}^M_h\left( {{\mathbf{x}}, s} \right) =\sum\limits_{j = 0}^{ + \infty } {{\varepsilon ^j}} {\hat{U}_h^{(j)}}\left( {{\mathbf{x}}, s} \right),\label{seriesU}\\
		&\hat\Theta^M\left( {{\mathbf{x}}, s} \right) =\sum\limits_{j = 0}^{ + \infty } {{\varepsilon ^j}} {\hat\Theta^{(j)}}\left( {{\mathbf{x}}, s} \right).\label{seriesTheta}
	\end{align}
\end{subequations}

By plugging the solutions $ (\ref{seriesU}) $ and $ (\ref{seriesTheta}), $ into the equations $(\ref{eq:averagefieldu}) $ and $(\ref{eq:averagefieldtheta}), $  the asymptotic expansions of the average field equations of infinite order become

\begin{subequations}
	\begin{align}
		\begin{split}
			&n^{(2,0)}_{ipq_1q_2} \left(\dfrac{\partial^2\hat{U}_h^{(0)}}{\partial x_{q_1} \partial x_{q_2}}+ \varepsilon \dfrac{\partial^2 \hat{U}_h^{(1)}}{\partial x_{q_1} \partial x_{q_2}}+ \varepsilon^2 \dfrac{\partial^2\hat{U}_h^{(2)}}{\partial x_{q_1} \partial x_{q_2}}+\cdots\right)+\tilde{n}^{(2,0)}_{iq_1}\left(\dfrac{\partial\hat\Theta^{(0)}}{\partial x_{q_1}}+ \varepsilon \dfrac{\partial\hat\Theta^{(1)}}{\partial x_{q_1}}+ \varepsilon^2 \dfrac{\partial\hat\Theta^{(2)}}{\partial x_{q_1}}+\cdots\right)+\\
			& -n_{ip}^{(2,2)}s^2 \left( \hat{U}_h^{(0)}+\varepsilon\hat{U}_h^{(1)}+\varepsilon^2\hat{U}_h^{(2)} \right) +\cdots+\hat{b}_i = 0, \label{eq:averagefieldassymptoticelastic}
		\end{split}
	\end{align}
	\begin{align}
		\begin{split}
			& w^{(2,0)}_{q_1q_2}\left(\dfrac{\partial^2\hat\Theta^{(0)}}{\partial x_{q_1} \partial x_{q_2}}+ \varepsilon \dfrac{\partial^2 \hat\Theta^{(1)}}{\partial x_{q_1} \partial x_{q_2}}+ \varepsilon^2 \dfrac{\partial^2\hat\Theta^{(2)}}{\partial x_{q_1} \partial x_{q_2}}+\cdots\right)-\tilde{w}^{(2,1)}_{pq_1}s\left(\dfrac{\partial \hat{U}_h^{(0)}}{\partial x_{q_1}}+ \varepsilon \dfrac{\partial\hat{U}_h^{(1)}}{\partial x_{q_1}}+\varepsilon^2 \dfrac{\partial\hat{U}_h^{(2)}}{\partial x_{q_1}}+\cdots\right)+\\
			&-w^{(2,1)}s\left(\hat{\Theta}^{(0)}+\varepsilon\hat{\Theta}^{(1)}+\varepsilon^2\hat{\Theta}^{(2)} \right) +\cdots+\hat{r} = 0.\label{eq:averagefieldassymptoticthermal}
		\end{split}
	\end{align}
\end{subequations}
Rearranging the terms at the same order of $\varepsilon$ in $ (\ref{eq:averagefieldassymptoticelastic}) $ and $ (\ref{eq:averagefieldassymptoticthermal}), $ an infinite set of macroscopic hierarchical differential problems expressed in terms of the sensitivities $\hat{U}^{(j)}_h$ and $\hat\Theta^{(j)}$ of both macroscopic displacement  $\hat{U}_h^M$ and macroscopic temperature $\hat\Theta^M$ fields can be determined. Thus,

\begin{subequations}
	\begin{align}
		\begin{split}
			&\varepsilon^0\left(n^{(2,0)}_{ipq_1q_2}\dfrac{\partial^2\hat{U}_h^{(0)}}{\partial x_{q_1} \partial x_{q_2}}+\tilde{n}^{(2,0)}_{iq_1}\dfrac{\partial\hat\Theta^{(0)}}{\partial x_{q_1}}-n_{ip}^{(2,2)}s^2\hat{U}_h^{(0)}+\hat{b}_i \right)+\\
			&+\varepsilon\left( n^{(2,0)}_{ipq_1q_2}\dfrac{\partial^2 \hat{U}_h^{(1)}}{\partial x_{q_1} \partial x_{q_2}}+\tilde{n}^{(2,0)}_{iq_1}\dfrac{\partial\hat\Theta^{(1)}}{\partial x_{q_1}}-n_{ip}^{(2,2)}s^2\hat{U}_h^{(1)}+s_i^{(1)}(\textbf{x},s)\right)+\\
			& +\varepsilon^2\left( n^{(2,0)}_{ipq_1q_2}\dfrac{\partial^2\hat{U}_h^{(2)}}{\partial x_{q_1}\partial x_{q_2}}+\tilde{n}^{(2,0)}_{iq_1}\dfrac{\partial\hat\Theta^{(2)}}{\partial x_{q_1}}-n_{ip}^{(2,2)}s^2\hat{U}_h^{(2)}+s_i^{(2)}(\textbf{x},s) \right)+\mathcal{O}(\varepsilon^{3})= 0, \label{eq:averagefieldassymptoticelastichierarchical}
		\end{split}
	\end{align}
	\begin{align}
		\begin{split}
			&\varepsilon^0\left(w^{(2,0)}_{q_1q_2}\dfrac{\partial^2\hat\Theta^{(0)}}{\partial x_{q_1} \partial x_{q_2}}-\tilde{w}^{(2,1)}_{pq_1}s\dfrac{\partial \hat{U}_h^{(0)}}{\partial x_{q_1}}-w^{(2,1)}s\hat{\Theta}^{(0)}+\hat{r}\right)+\\
			&+\varepsilon\left(w^{(2,0)}_{q_1q_2}\dfrac{\partial^2 \hat\Theta^{(1)}}{\partial x_{q_1} \partial x_{q_2}}-\tilde{w}^{(2,1)}_{pq_1}s\dfrac{\partial\hat{U}_h^{(1)}}{\partial x_{q_1}}-w^{(2,1)}s\hat{\Theta}^{(1)}+\upsilon^{(1)}(\textbf{x},s)\right)+\\
			& +\varepsilon^2\left(w^{(2,0)}_{q_1q_2}\dfrac{\partial^2\hat\Theta^{(2)}}{\partial x_{q_1} \partial x_{q_2}}-\tilde{w}^{(2,1)}_{pq_1}s\dfrac{\partial\hat{U}_h^{(2)}}{\partial x_{q_1}}-w^{(2,1)}s\hat{\Theta}^{(2)}+\upsilon^{(2)}(\textbf{x},s)\right)+\mathcal{O}(\varepsilon^{3})= 0, \label{eq:averagefieldassymptoticthermalhierarchical}
		\end{split}
	\end{align}
\end{subequations}
Therefore, the recursive problem at the macroscopic scale at order $\varepsilon^0$ reads

\begin{subequations}
	\begin{eqnarray}
		&&n^{(2,0)}_{ipq_1q_2}\dfrac{\partial^2\hat{U}_h^{(0)}}{\partial x_{q_1} \partial x_{q_2}}+\tilde{n}^{(2,0)}_{iq_1}\dfrac{\partial\hat\Theta^{(0)}}{\partial x_{q_1}}-n_{ip}^{(2,2)}s^2\hat{U}_h^{(0)}+\hat{b}_i = 0,\label{recursivemacroscopicelasticorder0} \\
		&&w^{(2,0)}_{q_1q_2} \dfrac{\partial^2\hat\Theta^{(0)}}{\partial x_{q_1} \partial x_{q_2}}- \tilde{w}^{(2,1)}_{pq_2}s\dfrac{\partial\hat{U}_h^{(0)}}{\partial x_{q_1}}-w^{(2,1)}s\hat\Theta^{(0)}+\hat{r} = 0,
		\label{recursivemacroscopicthermalorder0}
	\end{eqnarray}
\end{subequations}
while the generic recursive problem at the macroscopic scale of order $\varepsilon^r$ with $l\in\Z,$ $l\ge 1$ is found as

\begin{subequations}
	\begin{eqnarray}
		&&n^{(2,0)}_{ipq_1q_2}\dfrac{\partial^2\hat{U}_h^{(l)}}{\partial x_{q_1} \partial x_{q_2}}+\tilde{n}^{(2,0)}_{iq_1}\dfrac{\partial\hat\Theta^{(l)}}{\partial x_{q_1}}-n_{ip}^{(2,2)}s^2\hat{U}_h^{(l)}+s_i^{(l)}(\mathbf{x},s)  = 0,\label{recursivemacroscopicelasticorderl}\\
		&&w^{(2,0)}_{q_1q_2} \dfrac{\partial^2\hat\Theta^{(l)}}{\partial x_{q_1} \partial x_{q_2}}- \tilde{w}^{(2,1)}_{pq_2}s\dfrac{\partial\hat{U}_h^{(l)}}{\partial x_{q_1}}-w^{(2,1)}s\hat\Theta^{(l)}-\upsilon^{(l)}(\mathbf{x},s) = 0,
		\label{recursivemacroscopicthermalorderl}
	\end{eqnarray}
\end{subequations}
where the functions $s_i^{(l)}(\textbf{x},s)$ and $\upsilon^{(l)}(\textbf{x},s)$ are  the source terms $\mathcal{Q}$-periodic fields, which depend both on the higher order constant tensors that appear in the terms at orders equal or higher than  $\varepsilon^m$ of the equations $ (\ref{eq:averagefieldu}) $ and $ (\ref{eq:averagefieldtheta}),$ and on the sensitivities $\hat{U}_h^{(j)}$ and $\hat\Theta^{(j)}$ from macroscopic hierarchical differential problems of an order lower than $\varepsilon^m.$

\subsection{Approximation of the power-like functional through truncation of its asymptotic expansion}
\label{powerlikefunctionaltruncation}

Thus, plugging the down-scaling relations truncated at the first order \eqref{downscalingu1order}, \eqref{downscalingtheta1order} and their gradients \eqref{downscalingu1orderderivative}, \eqref{downscalingtheta1orderderivative}, into the transformed power-like functional \eqref{functionaltransformed}, it results

\begin{eqnarray}
	\label{functionaldownscalingtransf}
	\begin{split}
		\hat{\Lambda}_{m}(\hat{U}^{M}_{h},\hat\Theta^M)=&\int_{\mathfrak{L}}\langle\hat{\lambda}_{m}(\textbf{x},\boldsymbol{\xi})\rangle d\textbf{x}=\\
		=& s\left[\dfrac{1}{2} s^{2}\langle\rho^{m}\rangle\int_{\mathfrak{L}}\hat{U}^{M}_{h}\hat{U}^{M}_{h} d\textbf{x}+\left\langle\dfrac{1}{2}B^{(1,0)}_{ijlr_{1}}C^{m}_{ijhk}B^{(1,0)}_{hkpq_{1}}\right\rangle\int_{\mathfrak{L}}\dfrac{\partial\hat{U}^{M}_{l}}{\partial x_{r_{1}}}\dfrac{\partial\hat{U}^{M}_{p}}{\partial x_{q_{1}}}d\textbf{x}\quad+\right.\\
		+&\left\langle\dfrac{1}{2}B^{(1,0)}_{ijlr_{1}}C^{m}_{ijhk}\tilde{B}^{(1,0)}_{hk}\right\rangle\int_{\mathfrak{L}}\dfrac{\partial\hat{U}^{M}_{l}}{\partial x_{r_{1}}}\hat\Theta^Md\textbf{x}+   \left\langle\dfrac{1}{2}\tilde{B}^{(1,0)}_{ij}C^{m}_{ijhk}B^{(1,0)}_{hkpq_{1}}\right\rangle\int_{\mathfrak{L}}\hat\Theta^M\dfrac{\partial\hat{U}^{M}_{l}}{\partial x_{q_{1}}}d\textbf{x}\quad+\\
		+&\left\langle\dfrac{1}{2}\tilde{B}^{(1,0)}_{ij}C^{m}_{ijhk}\tilde{B}^{(1,0)}_{hk}\right\rangle\int_{\mathfrak{L}}\hat\Theta^M\hat\Theta^Md\textbf{x}  -\left\langle B^{(1,0)}_{ijlr_{1}}\alpha^{m}_{ij}\right\rangle\int_{\mathfrak{L}}\dfrac{\partial\hat{U}^{M}_{l}}{\partial x_{r_{1}}}\hat\Theta^Md\textbf{x}\quad+\\
		-&\left\langle \tilde{B}^{(1,0)}_{ij}\alpha^{m}_{ij}\right\rangle\int_{\mathfrak{L}}\hat\Theta^M\hat\Theta^Md\textbf{x}-\left.\int_{\mathfrak{L}}\hat{U}^{M}_{h}\hat{b}_{h}d\textbf{x}\right] \quad+\\
		-&\left\langle\dfrac{1}{2}A^{(1,0)}_{ir_{1}}K^{m}_{ij}A^{(1,0)}_{jq_{1}}\right\rangle\int_{\mathfrak{L}}\dfrac{\partial\hat{\Theta}^{M}}{\partial x_{r_{1}}}\dfrac{\partial\hat{\Theta}^{M}}{\partial x_{q_{1}}}d\textbf{x} -\dfrac{1}{2}s\langle p^m\rangle\int_{\mathfrak{L}}\hat\Theta^M\hat\Theta^Md\textbf{x}\quad+\\
		+&\int_{\mathfrak{L}}\hat{r}\hat\Theta^Md\textbf{x}+\mathcal{O}(\varepsilon^2).			
	\end{split}
\end{eqnarray} 

Using the concept of the variation of a functional to use the necessary condition for a functional to have an extremum, the stability condition of the transformed power-like functional $ \hat{\Lambda}_{m}$ is given by the governing equation of a non-local homogeneous continuum, in other words, one arises to the first variation of the average transformed power-like functional $\delta \hat{\Lambda}_{m}(\hat{U}^{M}_{h},\delta\hat{U}^{M}_{h},\hat\Theta^M,\delta\hat\Theta^M),$  thus

\begin{eqnarray}
	\label{firstvariation}
	\begin{split}
		\delta\hat{\Lambda}_{m}= 
		& \displaystyle\int_{\mathfrak{L}}s\left[s^{2}\langle\rho^{m}\rangle\hat{U}^{M}_{l}-\left\langle B^{(1,0)}_{ijlr_{1}}C^{m}_{ijhk}B^{(1,0)}_{hkpq_{1}}\right\rangle\dfrac{\partial^2\hat{U}^{M}_{p}}{\partial x_{q_{1}}\partial x_{r_{1}}}-\left\langle\dfrac{1}{2} B^{(1,0)}_{ijlr_{1}}C^{m}_{ijhk}\tilde{B}^{(1,0)}_{hk}\right\rangle\dfrac{\partial\hat{\Theta}^{M}}{\partial x_{r_{1}}}\quad+\right.\\
		-&\left.\left\langle\dfrac{1}{2}\tilde{B}^{(1,0)}_{ij}C^{m}_{ijhk}B^{(1,0)}_{hklq_{1}}\right\rangle\dfrac{\partial\hat{\Theta}^{M}}{\partial x_{q_{1}}}+ \left\langle B^{(1,0)}_{ijlr_{1}}\alpha^{m}_{ij}\right\rangle\dfrac{\partial\hat{\Theta}^{M}}{\partial x_{r_{1}}}-\hat{b}_l\right]\delta\hat{U}^M_ld\textbf{x}\quad+\\
		+&\displaystyle\int_{\mathfrak{L}}\left[s\left(\left\langle\dfrac{1}{2} B^{(1,0)}_{ijlr_{1}}C^{m}_{ijhk}\tilde{B}^{(1,0)}_{hk}\right\rangle\dfrac{\partial\hat{U}^{M}_{l}}{\partial x_{r_{1}}}+ \left\langle\dfrac{1}{2} \tilde{B}^{(1,0)}_{ij}C^{m}_{ijhk}B^{(1,0)}_{hkpq_{1}}\right\rangle\dfrac{\partial\hat{U}^{M}_{p}}{\partial x_{q_{1}}}\quad+ \right.\right.\\
		-&\left.\left\langle\tilde{B}^{(1,0)}_{ij}C^{m}_{ijhk}\tilde{B}^{(1,0)}_{hk}\right\rangle\hat\Theta^M -\left\langle B^{(1,0)}_{ijlr_{1}}\alpha^{m}_{ij}\right\rangle\dfrac{\partial\hat{U}^{M}_{l}}{\partial x_{r_{1}}}-2\left\langle \tilde{B}^{(1,0)}_{ij}\alpha^{m}_{ij}\hat\Theta^M\right\rangle\right)\quad+\\
		+&\left.\left\langle A^{(1,0)}_{ir_{1}}K^{m}_{ij}A^{(1,0)}_{jq_{1}}\right\rangle\dfrac{\partial^2\hat{\Theta}^{M}}{\partial x_{r_{1}}x_{q_{1}}} -s\langle p^m\rangle\hat\Theta^M+\hat{r}\right]\delta\hat\Theta^Md\textbf{x},
	\end{split}
\end{eqnarray}
where the divergence theorem has been applied in the first variation of the transformed power-like functional \eqref{firstvariation}.

\subsection{Euler-Lagrange equation via power-like functional at micro-scale}
\label{Euler-Lagrangepower-likefunctional}

Regarding the obtained governing field equations at the macro-scale, namely \eqref{eulerlagrange1macrotempo} and \eqref{eulerlagrange2macrotempo}, and the overall constitutive tensors \eqref{C} to \eqref{p} in Sec. \ref{variationalmacro}, the procedure only holds true in case that the variational approach at the micro-scale also holds. Having said that, this Section it will provide an equivalence between the governing field equations at the micro-scale $ (\ref{eq:Gov1T}),$ $(\ref{eq:Gov2T})$ and the power-like functional \eqref{functionaltransformed} through the first variation of it, a similar procedure may be seen also in \cite{fabrizio1992mathematical}.

If the functional $ \hat{\Lambda} $ in equation \eqref{functionaltransformed} attains a local minimum at $ (\hat{\boldsymbol{u}},\hat\theta), $ $ \delta $ is an arbitrary functions that has at least one derivative and vanishes at the boundary of $ \mathfrak{L} ,$ and then defining here $ \eta $ as any number close to $ 0 ,$ yields

\begin{align}
	\label{firstvariationappendixeta}
	\begin{split}
		\delta\hat{\Lambda}_{m}(\hat{\boldsymbol{u}},\delta\hat{\boldsymbol{u}},\hat\theta,\delta\hat\theta)
		&=\dfrac{d}{d\eta}\left[\int_{\mathfrak{L}}s\left(\dfrac{1}{2}\rho^{m}s^2(\hat{\boldsymbol{u}}+\eta\delta\hat{\boldsymbol{u}})\cdot(\hat{\boldsymbol{u}}+\eta\delta\hat{\boldsymbol{u}})\right.\right.\quad+\\
		&+\dfrac{1}{2}\nabla(\hat{\boldsymbol{u}}+\eta\delta\hat{\boldsymbol{u}}):(\mathfrak{C}^{m}\nabla(\hat{\boldsymbol{u}}+\eta\delta\hat{\boldsymbol{u}}))\quad+\\
		&-\left.\dfrac{1}{2}\nabla(\hat{\boldsymbol{u}}+\eta\delta\hat{\boldsymbol{u}}):(\boldsymbol\alpha^{m}(\hat\theta+\eta\delta\hat\theta))-(\hat{\boldsymbol{u}}+\eta\delta\hat{\boldsymbol{u}})\cdot\hat{\boldsymbol{b}}\right)d\mathbf{x}\quad+\\
		&-\int_{\mathfrak{L}}\left(\dfrac{1}{2}\nabla(\hat\theta+\eta\delta\hat\theta)\cdot\left(\mathbf{K}^{m}\nabla(\hat\theta+\eta\delta\hat\theta)\right)\right.\quad+\\
		&+\dfrac{1}{2}s(\hat\theta+\eta\delta\hat\theta)\left(\boldsymbol\alpha^{m}:\nabla(\hat{\boldsymbol{u}}+\eta\delta\hat{\boldsymbol{u}})\right)\quad+\\
		&+\left.\left.\left.\dfrac{1}{2}s(\hat\theta+\eta\delta\hat\theta)(p^m(\hat\theta+\eta\delta\hat\theta))-(\hat\theta+\eta\delta\hat\theta)\hat{r}\right)d\mathbf{x}\right]\right|_{\eta=0}\,,
	\end{split}
\end{align}
where it must be highlighted that $ \eta $ as defined in this Section does not stand for thickness.

Reckon that the divergence theorem, the symmetry of the tensors $ \mathfrak{C}^{m}, $ $ \mathbf{K}^{m}, $ $ \boldsymbol\alpha^{m} $ and evaluating the equation \eqref{firstvariationappendixeta} on $ \eta=0 ,$ the first variation of the power-like functional lying on the Laplace transform space at the micro-scale takes the form

\begin{align}
	\label{firstvariationappendix}
	\begin{split}
		\delta\hat{\Lambda}_{m}
		&=\int_{\mathfrak{L}}\left[s\left(\rho^{m}s^2\hat{\boldsymbol{u}}-\nabla\cdot(\mathfrak{C}^{m}\nabla\hat{\boldsymbol{u}})+\nabla\cdot(\boldsymbol\alpha^{m}\hat\theta)-\hat{\boldsymbol{b}}\right)\delta\hat{\boldsymbol{u}}\right]d\mathbf{x}\quad+\\
		&+\int_{\mathfrak{L}}\left[\left(\nabla\cdot\left(\mathbf{K}^{m}\nabla\hat\theta\right)-s\left(\boldsymbol\alpha^{m}\nabla\hat{\boldsymbol{u}}\right)-sp^m\hat\theta+\hat{r}\right)\delta\hat\theta\right] d\mathbf{x}\,,
	\end{split}
\end{align}
and since $ \delta\hat{\Lambda}_{m}(\hat{\boldsymbol{u}},\delta\hat{\boldsymbol{u}},\hat\theta,\delta\hat\theta)=0, $ the Euler-Lagrange equations corresponding to the power-like functional at the micro-scale are

\begin{subequations}
	\begin{align}
		&\rho^{m}s^2\hat{\boldsymbol{u}}-\nabla\cdot(\mathfrak{C}^{m}\nabla\hat{\boldsymbol{u}})+\nabla\cdot(\boldsymbol\alpha^{m}\hat\theta)-\hat{\boldsymbol{b}}=\boldsymbol{0},\label{microescalaapartirdaprimeiravariaçãoU}\\
		&\nabla\cdot\left(\mathbf{K}^{m}\nabla\hat\theta\right)-s\left(\boldsymbol\alpha^{m}\nabla\hat{\boldsymbol{u}}\right)-sp^m\hat\theta+\hat{r}=0,\label{microescalaapartirdaprimeiravariaçãoTheta}
	\end{align}
\end{subequations}
hence the variational approach at the micro-scale gives exactly the field equations defined over the Laplace transform space seen in equations $ (\ref{eq:Gov1T})$ and $(\ref{eq:Gov2T}).$ By consequence, the thermoelasticity governing field equations \eqref{eq:Gov1integraldifferenti} and \eqref{eq:Gov2integraldifferenti} over the time space $ t$ emerge once again after applying the inverse Laplace transform.

\section{Thermoelastic wave propagation modelling}

\subsection{Wave propagation in heterogeneous periodic material}
\label{Wavepropagationheterogeneous}

On one hand, Sec. \ref{waveprophomogenized} studied the wave propagation an homogenized continuum, taking advantage of the thermoelastic field equations on the complex frequency space $ s $ at the macro-scale, provided by the asymptotic homogenization process seen previously in the present work. Now, on the other hand, this Section is focused on investigate the heterogeneous material to determine the frequency spectrum via Floquet-Bloch theory \citep{Floquet1883,Bloch1928,brillouin1953wave}, from the thermoelastic field equations at the micro-scale over Laplace transformed space.

To this aim, firstly we recall the thermoelastic field equations \eqref{microescalaapartirdaprimeiravariaçãoU} and \eqref{microescalaapartirdaprimeiravariaçãoTheta}, without the source terms $ \hat{\boldsymbol{b}} $ and $ \hat{r} ,$ i.e.,

\begin{subequations}
	\begin{align}
		&\nabla\cdot(\mathfrak{C}^{m}\nabla\hat{\boldsymbol{u}})-\nabla\cdot(\boldsymbol\alpha^{m}\hat\theta)=\rho^{m}s^2\hat{\boldsymbol{u}},\label{microscaleUnosourceterms}\\
		&\nabla\cdot\left(\mathbf{K}^{m}\nabla\hat\theta\right)-s\left(\boldsymbol\alpha^{m}\nabla\hat{\boldsymbol{u}}\right)=sp^m\hat\theta.\label{microscaleThetanosourceterms}
	\end{align}
\end{subequations}
Secondly, from Floquet-Bloch theory, the decompositions for the thermoelastic medium  are given by

\begin{subequations}
	\begin{align}
		&\hat{\boldsymbol{u}}\left( {\mathbf{x}},s \right) = {\boldsymbol{\hat{U}}}^B({\mathbf{x}},s){\mathrm{e}^{\mathrm{i}\left( {{\boldsymbol{k}} \cdot {\mathbf{x}}} \right)}},\label{FBdecU}\\
		&\hat\theta\left( {\mathbf{x}},s \right) =\hat\Theta^B({\mathbf{x}},s){\mathrm{e}^{\mathrm{i}\left( {{\boldsymbol{k}} \cdot {\mathbf{x}}} \right)}},\label{FBdecTheta}
	\end{align}
\end{subequations}
where  ${\boldsymbol{\hat{U}}}^B({\mathbf{x}},s)$ and $\hat\Theta^B({\mathbf{x}},s)$ are $\mathcal{A}$-periodic Bloch amplitudes of the displacement field and temperature field, on the Laplace transformed space, respectively. Equations \eqref{FBdecU} and \eqref{FBdecTheta} satisfy the Floquet-Bloch periodicity boundary conditions, in which they arose by the $ \mathcal{Q}$-periodicity of the medium. Let  $\boldsymbol{k}\in\C^3$ be the wave vector, the Floquet-Bloch boundary conditions read

\begin{subequations}
	\begin{align}
		\hat{\boldsymbol{u}}(\mathbf{x}+\boldsymbol{v}_p,s) &= {\mathrm{e}^{\mathrm{i}\left( {{\boldsymbol{k}} \cdot {{\boldsymbol{v}}_p}} \right)}}\hat{\boldsymbol{u}}\left({\mathbf{x}},s\right),\label{eq:FBboundaryconditions1}\\
		\hat\theta(\mathbf{x}+\boldsymbol{v}_p,s)  &= {\mathrm{e}^{\mathrm{i}\left( {{\boldsymbol{k}} \cdot {{\boldsymbol{v}}_p}} \right)}}\hat\theta\left({\mathbf{x}},s\right),\label{eq:FBboundaryconditions2}
	\end{align}
\end{subequations}
where  $\boldsymbol{v}_p$ is the periodicity vector $ (p=1,2,3).$  

Finally, coupling the Floquet-Bloch decompositions above into the field equations \eqref{microscaleUnosourceterms} and \eqref{microscaleThetanosourceterms}, and making some simplifications, the tensorial form yields

\begin{subequations}
	\begin{align}
		&\nabla^B\cdot(\mathfrak{C}^{m}\nabla^B{\boldsymbol{\hat{U}}}^B)-\nabla^B\cdot(\boldsymbol\alpha^{m}\hat\Theta^B)-\rho^{m}s^2{\boldsymbol{\hat{U}}}^B=\boldsymbol{0},\\
		&\nabla^B\cdot\left(\mathbf{K}^{m}\nabla^B\hat\Theta^B\right)-s\left(\boldsymbol\alpha^{m}\nabla^B{\boldsymbol{\hat{U}}}^B\right)-sp^m\hat\Theta^B=0,
	\end{align}
\end{subequations}
where the differential operator $ \nabla^B $ was defined as 

\begin{subequations}
	\begin{align}
		&\nabla\hat{\boldsymbol{u}}=\Big(\nabla\boldsymbol{\hat{U}}^B+\mathrm{i}\boldsymbol{k}\otimes\boldsymbol{\hat{U}}^B\Big){\mathrm{e}^{\mathrm{i}\left( {{\boldsymbol{k}} \cdot {\mathbf{x}}} \right)}}=\nabla^B\Big({\boldsymbol{\hat{U}}}^B\Big){\mathrm{e}^{\mathrm{i}\left( {{\boldsymbol{k}} \cdot {\mathbf{x}}} \right)}},\\
		&\nabla\hat\theta=\Big(\nabla\hat{\Theta}^B+\mathrm{i}\boldsymbol{k}\otimes\hat{\Theta}^B\Big){\mathrm{e}^{\mathrm{i}\left( {{\boldsymbol{k}}\cdot{\mathbf{x}}} \right)}}=\nabla^B\Big(\hat\Theta^B\Big){\mathrm{e}^{\mathrm{i}\left( {{\boldsymbol{k}} \cdot {\mathbf{x}}} \right)}}.
	\end{align}
\end{subequations}
Particularly, we may also represent the field equations in terms of the components

\begin{subequations}
	\begin{align}
		\begin{split}
			&\left(C^m_{ijhk}\hat{U}^B_{h,k}\right)_{,j}+\mathrm{i}k_j\left[\Big(C^m_{ijhk}+C^m_{ikhj}\Big)\hat{U}^B_{h,k}+C^m_{ikhj,k}\hat{U}^B_{h}-\alpha^m_{ij}\hat\Theta^B\right]+\\
			&-\Big(k_kk_jC^m_{ijhk}+\rho^ms^2\delta_{ih}\Big)\hat{U}^B_{h}-\left(\alpha^m_{ij}\hat\Theta^B\right)_{,j}=0,\label{elasticblochcomponents}
		\end{split}\\
		\begin{split}
			&\left(K^m_{ij}\hat\Theta^B_{,j}\right)_{,i}+\mathrm{i}k_j\left[\Big(K^m_{ij}+K^m_{ji}\Big)\hat\Theta^B_{,i}+K^m_{ij,i}\hat\Theta^B-s\alpha^m_{ij}\hat{U}^B_i\right]+\\
			&-\Big(k_ik_jK^m_{ij}+sp^m\Big)\hat{\Theta}^B-s\alpha^m_{ij}\hat{U}^B_{i,j}=0,\label{thermalblochcomponents}
		\end{split}
	\end{align}
\end{subequations}
although the derivative $ (\cdot)_{,j} $ was once defined as the microscopic derivative, for convenience here it represents the partial derivative in $ x_j ,$ i.e. $ \partial/\partial x_j=(\cdot)_{,j} $

Recalling the spatial damping method in the Sec. \ref{waveprophomogenized}, let the conditions of an homogeneous wave be plugged into the field equations \eqref{elasticblochcomponents} and \eqref{thermalblochcomponents}, so given a direction $ \boldsymbol{n} $ of the wave vector $ \boldsymbol{k}=\kappa\boldsymbol{n}$ one obtains an eigenvector-eigenvalue problem where $ \kappa $ is the eigenvalue and the $\mathcal{Q}$-periodic Bloch amplitudes $ \boldsymbol{\hat{U}}^B, $ $ \hat\Theta^B $ are the eigenfunctions associated to it, while the angular frequency $ \omega $ might be a fixed parameter. Consequently to the study of harmonic waves over a layered media in the next Sec. \ref{heterogeneousapproach}, it arrives to the frequency spectrum $ \kappa{\omega} $ \eqref{surfacedispersiveheterogeneous}.

Similarly through the time damping path seen in Sec. \ref{waveprophomogenized} together with the results from the Sec. \ref{heterogeneousapproach}, it is possible to find an eigenvector-eigenvalue problem, where once resolved in this scenario it allows to determine the eigenvalues corresponding to the dispersion surfaces $ s(\boldsymbol{k}) ,$ that is writing the complex angular frequency $ s $ as functions of the wave vector $ \boldsymbol{k} $ and also the eigenfunctions corresponding to the wave polarization, with the Bloch amplitudes being its components \citep{krushynska2016visco}. 

\subsection{Frequency-band structure for periodic heterogeneous thermoelastic layered material}
\label{heterogeneousapproach}

In the present Section is outlined the procedure followed to obtain the frequency spectrum corresponding to the spatial damping over a periodic heterogeneous thermoelastic orthotropic layered material, which has led us to evaluate the benchmark test detailed in Sec. \ref{benchmark}, where the corresponding method drives into the dispersion relation \eqref{surfacedispersives(k)}. Specifically, this procedure is based on the transfer matrix method.

Initially, let us consider a body made of a given number $ n $ of overlapped layers bonded at their interfaces and stacked normal to the axis $ \boldsymbol{e}_2 $ of the plane $ (\boldsymbol{e}_1,\boldsymbol{e}_2) $ ($\boldsymbol{\xi}=\xi_1\boldsymbol{e}_1+\xi_2\boldsymbol{e}_2  $). By assumption, the line boundary of each layer must be parallel to the $\boldsymbol{e}_1 $ vector which is chosen to coincide whit the exact half of the layered plate. We also assign for each layer the number $ j $ (with $ j=1,2,\dots,n$) a local coordinate $ \mathfrak{s}_i^{_{(j)}} $ such that its origin is located in the barycentre of the layer with $ \mathfrak{s}_2^{_{(j)}} $ normal to it. Thus layer $ j $ occupies the region $ - d^{_{(j)}}/2\leq\mathfrak{s}_3^{_{(j)}}\leq d^{_{(j)}}/2 ,$ where $d^{_{(j)}} $ is its thickness, and hence $ d, $ the sum of the thickness of all individual layers $ d_2=d^{_{(1)}}+\cdots+ d^{_{(j)}}+\cdots+d^{_{(n)}}$ must be equal to the total height of the layered body, where the body occupies the region $ -d_2/2\leq\xi_2\leq d_2/2. $ 

Let us assume the same hypothesis taken in the Sec. \ref{homogenizedprocess} for the heterogeneous thermoelastic problem, this means an orthotropic bi-phase layered material with the orthotropy axes perpendicular to the layering direction $ \boldsymbol{e_2}$ and the wave vector as $ \boldsymbol{k}=(k_1,k_2)^T=(0,k_2)^T ,$ which make the field equations \eqref{elasticblochcomponents} and \eqref{thermalblochcomponents} only dependent on the variable $ \xi_2 ,$ therefore for each layer $ j $ we have a set of following governing equations

\begin{subequations} 
	\begin{align}
		\begin{split}
			&C^{_{(j)}}_{1212}\hat{U}^{{B}}_{1,22}+\mathrm{i}2k_2C^{_{(j)}}_{1212}\hat{U}^{{B}}_{1,2}-\Big(k_2^2C^{_{(j)}}_{1212}+\rho^ms^2\Big)\hat{U}^{{B}}_{1}=0,\label{elastic1}
		\end{split}\\
		\begin{split}
			&C^{_{(j)}}_{2222}\hat{U}^{{B}}_{2,22}+\mathrm{i}2k_2\left(C^{_{(j)}}_{2222}\hat{U}^{{B}}_{2,2}-\alpha^{_{(j)}}_{22}\hat\Theta^{{B}}\right)-\Big(k_2^2C^{_{(j)}}_{2222}+\rho^{_{(j)}}s^2\Big)\hat{U}^{{B}}_{2}-\alpha^{_{(j)}}_{22}\hat\Theta^{{B}}_{,2}=0,\label{elastic2}
		\end{split}\\
		\begin{split}
			&K^{_{(j)}}_{22}\hat\Theta^{{B}}_{,22}+\mathrm{i}2k_2\left(K^{_{(j)}}_{22}\hat\Theta^{{B}}_{,2}-s\alpha^{_{(j)}}_{22}\hat{U}^{{B}}_2\right)-\Big(k_2^2K^{_{(j)}}_{22}+sp^{_{(j)}}\Big)\hat{\Theta}^{{B}}-s\alpha^{_{(j)}}_{22}\hat{U}^{{B}}_{2,2}=0.\label{thermal}
		\end{split}
	\end{align}
\end{subequations}

In order to arrive at the transfer matrix, the solutions in the Floquet-Bloch form \eqref{FBdecU} and \eqref{FBdecTheta} of the specialized governing equations \eqref{elastic1}, \eqref{elastic2} and \eqref{thermal} for each layer $ j $ are obtained and are evaluated for both the upper $ (+) $ and lower $ (-)$ boundary surfaces of layer $ j .$ Among them, from the constitutive relations \eqref{sigma} and \eqref{q} transformed in the Floquet-Bloch form as well, the components $ \hat\sigma^{{B}}_{12} ,$ $ \hat\sigma^{{B}}_{22} $ and $ \hat{q}^{{B}}_2 $ are derived, which are also evaluated for the upper $ (+) $ and lower $ (-)$ boundary surfaces of layer $ j .$ Proceeding with some algebraic manipulations over them, the transformed components specialized for the upper $ (+) $ boundary of layer $ j $ can be written in terms of the same components but  specialized for the lower $ (-) $ boundary of layer $ j. $ 

In the following, by applying the above procedure to a single layer, one obtains

\begin{equation}\label{lowerandupperboundary}
	[\boldsymbol{P}]_j^+=[\boldsymbol{A}]_j[\boldsymbol{P}]_j^-, \quad j=1,2,\dots,n,
\end{equation}
where the vector $ [\boldsymbol{P}]_j^{\pm}=\left(\left[
\hat{U}^B_{1},\hat{U}^B_{2},\hat{\Theta}^B,\hat\sigma^{{B}}_{12},\hat\sigma^{{B}}_{22},\hat{q}^{{B}}_2\right]_j^{\pm}\right)^T $ ($ T $ here is for transpose) defines the column vectors of the displacement $ \boldsymbol{\hat{U}}^B, $ temperature $ \boldsymbol{\hat{\Theta}}^B, $ stress $ \boldsymbol{\hat{\sigma}}^B$ and heat flux $\boldsymbol{\hat{q}}^B,$ specialized to the upper $ (+) $ and lower $ (-)$ boundary surfaces of layer $ j ,$ and the matrix $ [\boldsymbol{A}]_j$ constitutes the local transfer matrix for layer $ j. $

Finding the equation \eqref{lowerandupperboundary} to each layer, followed by the  individual matrix multiplication $ [\boldsymbol{A}]=[\boldsymbol{A}]_n\dots[\boldsymbol{A}]_2[\boldsymbol{A}]_1$ and reminding the continuity of the thermoelastic solutions and constitutive tensors at the layer interfaces namely, $ [\boldsymbol{P}]_{j+1}^-=[\boldsymbol{P}]_j^+, $ one relates the solutions and tensors at the upper boundary, to those at its lower boundary, this results in $ [\boldsymbol{P}]^+=[\boldsymbol{A}][\boldsymbol{P}]^-, $ where $ [\boldsymbol{P}]^- $ and $ [\boldsymbol{P}]^+ $ are now the displacement, temperature and tensor column vectors specialized to the upper and lower faces of the total plate, respectively and $ [\boldsymbol{A}] $ is the global transfer matrix of the total cell. Finally, imposing  the Floquet-Bloch periodic boundary conditions namely, equations \eqref{eq:FBboundaryconditions1} and \eqref{eq:FBboundaryconditions2}, one gives $ [\boldsymbol{P}]^+={\mathrm{e}^{\mathrm{i}\left( {{\boldsymbol{k}}\cdot{\mathbf{x}}} \right)}}[\boldsymbol{P}]^- ,$ and therefore one arises the linear problem

\begin{equation}\label{linearproblemfloquetblochcondition}
	{\mathrm{e}^{\mathrm{i}\left( {{\boldsymbol{k}} \cdot {\mathbf{x}}} \right)}}[\boldsymbol{P}]^-=[\boldsymbol{A}][\boldsymbol{P}]^-.
\end{equation}

At this point, even though both the local and global transfer matrices show themselves with several properties, which are discussed and listed in \citep{achenbach2012wave}, we only exploit them computationally herein. Such properties classify the transfer matrix as being a symplectic matrix, and the characteristic polynomial attached to its respective eigenvector-eigenvalue problem i.e. $ ([\boldsymbol{A}]-{\mathrm{e}^{\mathrm{i}\left( {{\boldsymbol{k}} \cdot {\mathbf{x}}} \right)}}[\boldsymbol{I}])[\boldsymbol{P}]^-=([\boldsymbol{A}]-\lambda[\boldsymbol{I}])[\boldsymbol{P}]^-=[\boldsymbol{0}] ,$ is a palindromic polynomial \citep{Bronski2005}. 

The palindromic polynomial $ \mathscr{P}(\lambda)=\text{det}\left([\boldsymbol{A}]-\lambda[\boldsymbol{I}]\right)=0$ to our bi-layered material problem given by the field equations \eqref{elastic1}, \eqref{elastic2} and \eqref{thermal}, wrote in terms of the invariants $ I_k $ of $ [\boldsymbol{A}] $ as

\begin{equation}\label{palindromicpolynomial}
	\mathscr{P}(\lambda)=I_6+I_5\lambda+I_4\lambda^2+I_3\lambda^3+I_2\lambda^4+I_1\lambda^5+I_0\lambda^6,
\end{equation} 
it has the following symmetric relations

\begin{equation}\label{invariantsidentities}
	\quad I_0=I_6=1,\quad I_1=I_5,\quad I_2=I_4.
\end{equation}
In addition, once recognized that $ {e^{i\left( {{\boldsymbol{k}} \cdot {\mathbf{x}}} \right)}} $ is an eigenvalue of the transfer matrix $ [\boldsymbol{A}] ,$ via the transformation $ \mathfrak{z}=\lambda+ 1/\lambda$ its palindromic polynomial \eqref{palindromicpolynomial} can still be formally rephrased as

\begin{equation}\label{palindromicpolynomialthriddegree}
	\mathscr{P}(\mathfrak{z})=\mathfrak{z}^3+I_1\mathfrak{z}^2+(I_2-3)\mathfrak{z}+(I_3-2I_1).
\end{equation}

Last but not least, it is possible to invoke the transfer matrix structure and rewrite the invariants $ I_1, $ $ I_2 ,$ and $ I_3 $ in terms of the trace of $ [\boldsymbol{A}] $ as stabilised in \citep{horst1935method}, so

\begin{subequations}
	\begin{align}
		I_1 & =  -\textrm{tr}([\boldsymbol{A}]),\\
		I_2 & =  -\dfrac{1}{2}\textrm{tr}([\boldsymbol{A}]^2)+\dfrac{1}{2}(\textrm{tr}([\boldsymbol{A}]))^2,\\
		I_3 & =  -\dfrac{1}{3}\textrm{tr}([\boldsymbol{A}]^3)+\dfrac{1}{2}\textrm{tr}([\boldsymbol{A}]^2)\textrm{tr}([\boldsymbol{A}])-\dfrac{1}{6}(\textrm{tr}([\boldsymbol{A}]))^3.
	\end{align}
\end{subequations}

Still at the same assumptions made in the beginning of this Section, it is also feasible factorize the polynomial \eqref{palindromicpolynomial} into two minor factors, namely one of a second degree polynomial associated to the equation \eqref{elastic1} and another of a fourth degree polynomial associated to the equations \eqref{elastic2} and \eqref{thermal}. Each sub-polynomial will also be a palindromic polynomial and therefore governed by invariants of its sub-matrices. Furthermore, choosing the right changing of variables, after a similar process as made before, one will have factorized the third degree polynomial \eqref{palindromicpolynomialthriddegree} into two sub-polynomials a linear one and a quadratic one.

%

\subsection{Invariants of the dispersive wave propagation in the periodic heterogeneous thermoelastic layered material}
\label{dispersivewaveheterogeneous}

Throughout this Section it is presented the invariants that characterize the wave propagation along the heterogeneous material. So, for the uncoupled hypothesis of the thermoelastic problem taken in Sec. \ref{benchmark} for the comparative analysis and among the latter theories written in the Secs. \ref{Wavepropagationheterogeneous} and \ref{heterogeneousapproach}, the invariants may be obtained with the following analytical expressions

\begin{flalign}
	\begin{split}
		\tilde{I}_1^{_{1}}(s)=&-\dfrac{\sqrt{A_{1212}}B_{1212}+2A_{1212}}{4A_{1212}}\left(\mathrm{e}^{-s\left(\sum\limits^2_{r=1}\varpi_{1212}^{_{(r)}}\right)}+\mathrm{e}^{s\left(\sum\limits^2_{r=1}\varpi_{1212}^{_{(r)}}\right)}\right)+\\
		&+\dfrac{\sqrt{A_{1212}}B_{1212}-2A_{1212}}{4A_{1212}}\left(\mathrm{e}^{-s\left(\sum\limits^2_{r=1}(-1)^r\varpi_{1212}^{_{(r)}}\right)}+\mathrm{e}^{s\left(\sum\limits^2_{r=1}(-1)^r\varpi_{1212}^{_{(r)}}\right)}\right),
	\end{split}
\end{flalign}
where $ A_{1212}=C_{1212}^{_{(2)}}\rho^{_{(2)}}C_{1212}^{_{(1)}}\rho^{_{(1)}}, $ $B_{1212}=C_{1212}^{_{(1)}}\rho^{_{(1)}}+C_{1212}^{_{(2)}}\rho^{_{(2)}},$ $ \varpi_{1212}^{_{(j)}}=\mathfrak{s}_j\sqrt{C_{1212}^{_{(j)}}\rho^{_{(j)}}}/C_{1212}^{_{(j)}} $ with $ j={1,2}. $ Analogously, for the  compressional wave the associate dispersion relation provides the wave number  $ k_2 $ as an explicit function of the complex frequency $ s, $

\begin{flalign}
	\begin{split}
		\tilde{I}_1^{_{2}}(s)=&-\dfrac{\sqrt{A_{2222}}B_{2222}+2A_{2222}}{4A_{2222}}\left(\mathrm{e}^{-s\left(\sum\limits^2_{r=1}\varpi_{1212}^{_{(r)}}\right)}+\mathrm{e}^{s\left(\sum\limits^2_{r=1}\varpi_{1212}^{_{(r)}}\right)}\right)+\\
		&+\dfrac{\sqrt{A_{2222}}B_{2222}-2A_{2222}}{4A_{2222}}\left(\mathrm{e}^{-s\left(\sum\limits^2_{r=1}(-1)^r\varpi_{1212}^{_{(r)}}\right)}+\mathrm{e}^{s\left(\sum\limits^2_{r=1}(-1)^r\varpi_{1212}^{_{(r)}}\right)}\right),
	\end{split}
\end{flalign}
where $ A_{2222}=C_{2222}^{_{(2)}}\rho^{_{(2)}}C_{2222}^{_{(1)}}\rho^{_{(1)}}, $ $B_{2222}=C_{2222}^{_{(1)}}\rho^{_{(1)}}+C_{2222}^{_{(2)}}\rho^{_{(2)}},$ $ \varpi_{2222}^{_{(j)}}=\mathfrak{s}_j\sqrt{C_{2222}^{_{(j)}}\rho^{_{(j)}}}/C_{2222}^{_{(j)}} $ with $ j={1,2}. $ Lastly, for the thermal wave propagation follows that

\begin{flalign}
	\begin{split}
		\tilde{I}_1^{_{3}}(s)=&-\left(\dfrac{|s|\sum\limits^2_{r=1}\left(s\tau^{_{(r)}}+1\right)\mu^{_{(r)}}_{22}}{4s\sqrt{\left(\prod^2_{r=1}\left(s\tau^{_{(r)}}+1\right)\mu^{_{(r)}}_{22}\right)}}
		+\dfrac{1}{2}\right)
		\left(\mathrm{e}^{-\sum\limits^2_{r=1}\left(s\tau^{_{(r)}}+1\right)\varphi^{_{(r)}}_{22}}+\mathrm{e}^{\sum\limits^2_{r=1}\left(s\tau^{_{(r)}}+1\right)\varphi^{_{(r)}}_{22}}\right)+\\
		&+\left(\dfrac{|s|\sum\limits^2_{r=1}\left(s\tau^{_{(r)}}+1\right)\mu^{_{(r)}}_{22}}{4s\sqrt{\left(\prod^2_{r=1}\left(s\tau^{_{(r)}}+1\right)\mu^{_{(r)}}_{22}\right)}}
		-\dfrac{1}{2}\right)
		\left(\mathrm{e}^{-\sum\limits^2_{r=1}(-1)^r\left(s\tau^{_{(r)}}+1\right)\varphi^{_{(r)}}_{22}}+\mathrm{e}^{\sum\limits^2_{r=1}(-1)^r\left(s\tau^{_{(r)}}+1\right)\varphi^{_{(r)}}_{22}}\right),
	\end{split}
\end{flalign}
where $ \mu^{_{(r)}}_{22}=1/\bar{K}^{_{(r)}}_{22}p^{_{(r)}}, $ $ \varphi^{_{(r)}}_{22}=T_0\mathfrak{s}_{r}\sqrt{\bar{K}^{_{(r)}}_{22}p^{_{(r)}}}/\bar{K}^{_{(r)}}_{22}. $

\end{document}